\documentclass[aps,prx,twocolumn,showpacs,superscriptaddress,footinbib,longbibliography]{revtex4-2} 

\usepackage[english]{babel}
\usepackage[dvipsnames]{xcolor}
\usepackage{comment}
\usepackage{graphicx}
\usepackage{stackengine}
\usepackage{dcolumn}
\usepackage{bm}
\usepackage{bbm}
\usepackage[utf8]{inputenc}
\usepackage{amssymb}
\usepackage{amsmath}
\usepackage{bbold}
\usepackage{pstricks}
\usepackage{slashed}
\usepackage{braket}
\usepackage{hyperref}
\usepackage{natbib}
\usepackage{float}
\usepackage{verbatim}
\usepackage{siunitx}
\usepackage{lipsum}
\usepackage{slashed}
\usepackage{ulem}
\usepackage[left=1.85cm,right=1.85cm,top=1.85cm,bottom=1.85cm]{geometry}
\usepackage{verbatim}
\usepackage{xcolor}
\usepackage{soul}
\usepackage{multirow}

\def\vphi{\ensuremath\varphi}
\def\opphi{\ensuremath\hat{\varphi}}

\def\vphiss{\ensuremath \varphi_{\text{s}\bar{\text{s}}}}
\def\vphis{\ensuremath\varphi_{\bar{\text{s}}}}
\def\vertex{\ensuremath e^{i\hat{\varphi}}}
\def\ss{{\text{s}\bar{\text{s}}}}
\hypersetup{    colorlinks,    linkcolor={blue!50!black},    citecolor={blue!50!black},    urlcolor={blue!80!black}}

\graphicspath{{./Figures/}}

\begin{document}

\title{Quantum simulating continuum field theories with large-spin lattice models}

\def \IBK{Institute for Theoretical Physics, University of Innsbruck, 6020 Innsbruck, Austria}
\def \IQOQI{Institute for Quantum Optics and Quantum Information, Austrian Academy of Sciences, 6020 Innsbruck, Austria}

\author{Gabriele Calliari}\email{Gabriele.Calliari@uibk.ac.at}
\affiliation{\IBK}\affiliation{\IQOQI}

\author{Marco Di Liberto}
\affiliation{Dipartimento di Fisica e Astronomia “G. Galilei” \& Padua Quantum Technologies Research Center, Università degli Studi di Padova, 35131, Padova, Italy}
\affiliation{Istituto Nazionale di Fisica Nucleare (INFN), Sezione di Padova, 35131 Padova, Italy}

\author{Hannes Pichler}\affiliation{\IBK}\affiliation{\IQOQI}
\author{Torsten V. Zache}\email{Torsten.Zache@uibk.ac.at}\affiliation{\IBK}\affiliation{\IQOQI}

\begin{abstract}
Simulating the real-time dynamics of quantum field theories (QFTs) is one of the most promising applications of quantum simulators.
Regularizing a bosonic QFT for quantum simulation purposes typically involves a truncation in Hilbert space in addition to a discretization of space.
Here, we discuss how to perform such a regularization of scalar QFTs
by explicitly constructing suitable many-body lattice Hamiltonians using multi-level or qudit systems, and show that this enables quantitative predictions in the continuum limit by extrapolating results obtained for large-spin 
models.
With extensive matrix-product state simulations, we numerically demonstrate the sequence of extrapolations that leads to quantitative agreement of  observables for the integrable sine-Gordon (sG) QFT.
We further show how to prepare static and moving soliton excitations, and analyze their scattering dynamics in the continuum limit, in agreement with a semi-classical model and with quantitative analytical predictions.
Finally, we illustrate how a non-integrable perturbation of the sG model gives rise to dynamics reminiscent of string breaking and plasma oscillations in gauge theories.
Our methods are directly applicable in state-of-the-art analog quantum simulators, opening the door to quantitatively investigating a wide variety of scalar field theories and tackling long-standing questions in non-equilibrium QFT like the fate of the false vacuum.
\end{abstract}

\maketitle

\section{Introduction} 

Quantum field theories (QFTs) play an essential role in theoretical physics, ranging from effective low-energy descriptions in condensed matter~\cite{sachdev2023quantum} to fundamental models of nature in particle physics~\cite{weinberg1995quantum}. 
Understanding the physics of QFTs, especially in out-of-equilibrium scenarios, is often guided by computationally challenging numerical simulations, and therefore QFTs constitute an ideal target for quantum simulation~\cite{bauer2023quantum,di2024quantum}.
Recent quantum simulation experiments are starting to probe the boundary of classical simulations~\cite{daley2022practical,scholl2021quantum,joshi2023exploring,andersen2024thermalization,manovitz2024quantum}, with the exciting prospect to provide genuine insights into the equilibrium and non-equilibrium properties of quantum many-body systems.
However, it remains an outstanding challenge to faithfully simulate physical phenomena of QFTs, such as string-breaking in gauge theories~\cite{cochran2024visualizing,gonzalez-cuadra2024observation,de2024observation,crippa2024analysis}.

To clarify the underlying conceptual challenges, recall that QFTs involve continuous variables living on continuous space-time instead of, e.g., discrete spin models on lattice systems.
In condensed matter or statistical physics, QFTs often play the role of effective theories, applicable at large distances where the irrelevance of microscopic details and the emergence of a continuum description is explained by the renormalization group.
Conversely, space-time lattice discretizations exploit this interpretation of QFTs and provide not only a rigorous non-perturbative definition, but also a computational paradigm  that underlies, e.g., lattice QCD~\cite{montvay1994quantum}.
For quantum simulations, which are often set up in a Hamiltonian formulation, these well-established classical techniques need to be fundamentally adapted~\cite{hackett2019digitizing,carena2021lattice,singh2022qubit,davoudi2021search,kreshchuk2021light}. An important open problem is to find practically useful truncations of field variables which is unavoidable to represent them on finite-dimensional Hilbert spaces that can be controlled by synthetic quantum systems~\cite{jordan2012quantum,klco2019digitization}.
How to perform controlled truncations of continuous field variables in order to obtain quantitative results in the continuum limit of QFTs is the central question of this work.
 
Recent research has mainly addressed this question in the context of universal \emph{qubit}-based quantum computers~\cite{tong2022provably,ingoldby2024enhancing,hardy2024optimized}. Despite the rapid progress in experimental quantum computing, simulating QFTs in the continuum limit currently remains out of reach of \emph{digital} approaches due to the large local Hilbert spaces and system sizes required.
A promising direction to save resources employs \emph{qudit} (multi-level) systems~\cite{illa2024qu8its,popov2024variational,calajo2024digital,zache2023quantum} as realized with trapped ions~\cite{ringbauer2022universal,meth2023simulating} or proposed with Rydberg atoms~\cite{gonzalez-cuadra2022hardware,maskara2023programmable,kruckenhauser2022highdimensional}.
At the same time, a limited class of field theories is accessible in \emph{analog} quantum simulators based on continuous quantum gases~\cite{tajik2023verification,viermann2022quantum,frolian2022realizing}.

\begin{figure}[hbt!]
    \centering
    \includegraphics[width=0.9\linewidth]{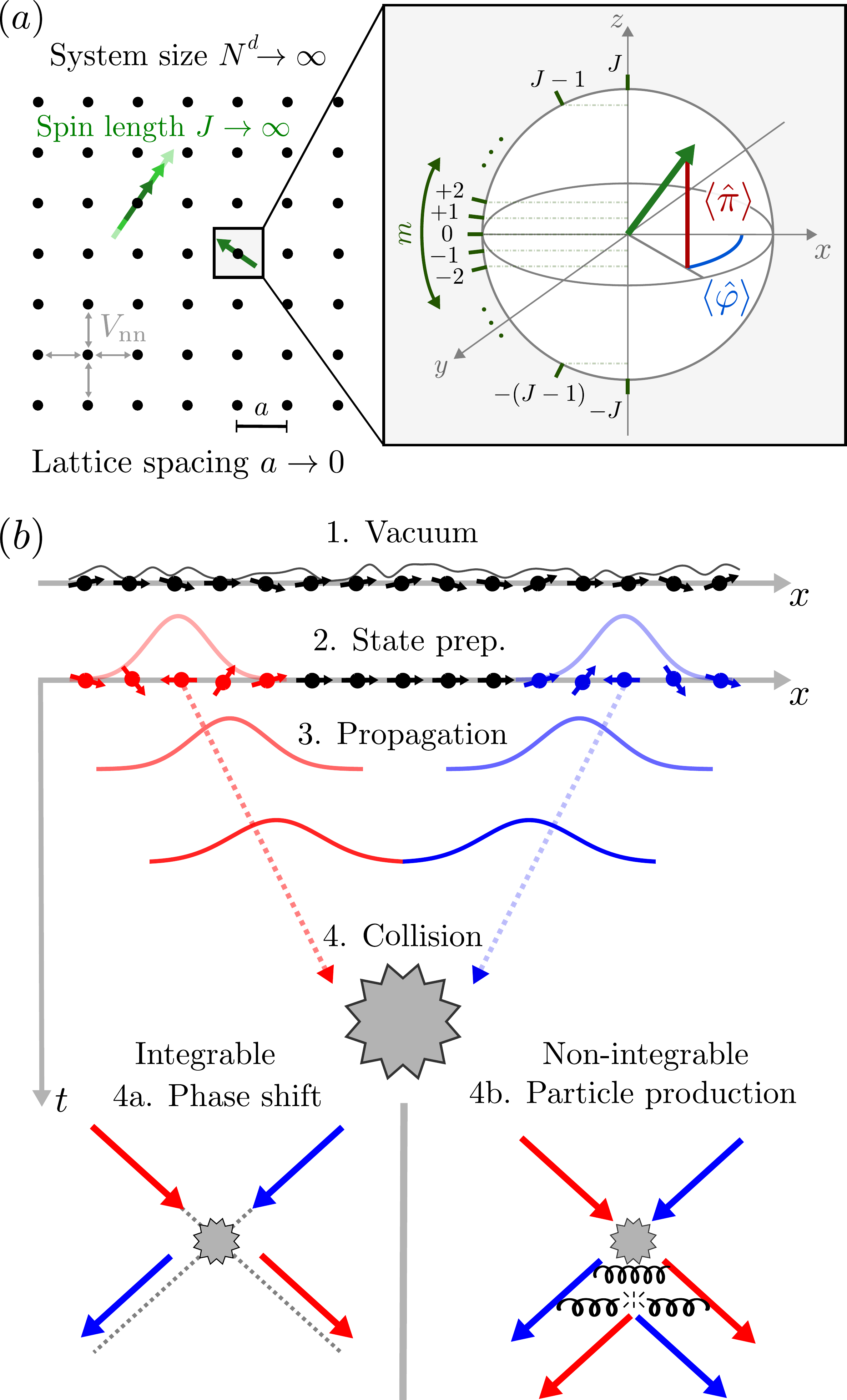}
    \caption{
    \textbf{Overview.}
    (a) $d$-dimensional square lattice of particles with nearest-neighbor interactions. We assume each particle to be described by a large spin of length $J$, with $\langle \hat{J}_+ \rangle\sim \langle e^{i\hat{\varphi}} \rangle $ and $\langle \hat{J}_z \rangle \sim \langle \hat{\pi} \rangle$ identifying field theory and lattice operators, as described in Sec.~\ref{sec:general_implementation}.
    Continuum physics is extracted by a sequence of extrapolations in spin length $J$, system size $N$ and lattice spacing $a$ [see Fig.~\ref{fig:Fig2}].
    (b) Sketch of real-time scattering dynamics from large-spin models. We show how to faithfully implement 1.~the vacuum (Sec.~\ref{sec:sG_vacuum}), 2.~quasi-particle wave-packets and 3.~their propagation (Sec.~\ref{sec:sG_solitons}), as well as 4.~collisions (Sec.~\ref{sec:sG_scattering} resp.~\ref{sec:beyond_sG}), resulting in phase shifts resp. particle production for the example of an integrable resp. non-integrable 1D QFT.}
    \label{fig:Fig1}
\end{figure}

In this work, we lay out a viable path
to quantitatively recover the continuum limit of scalar quantum field theories with an explicit truncation of the bosonic field variables.
Motivated by a recent proposal to employ large-spin Rydberg atom arrays~\cite{kruckenhauser2022highdimensional}, our approach is based on a family of generalized large-spin Heisenberg lattice Hamiltonians.
In a bottom-up approach, we show how appropriate extrapolations utilizing models of increasing spin length [illustrated in Fig.~\ref{fig:Fig1}(a)] enable the accurate implementation of general scalar field theories, including variants of the paradigmatic sine-Gordon (sG) model.
Our results conclusively show that the large-spin regularization provides a clean mapping between microscopic and continuum observables with a precise and systematic understanding how the continuum limit emerges, which enables quantum simulations of QFTs in regimes where no classical benchmarks are available.
We emphasize that the conceptual aspects of this regularization are applicable to any quantum simulation that can realize Hamiltonian dynamics, although the specific form of our model Hamiltonian makes it especially well suited for an analog implementation.

To demonstrate our proposal, we perform extensive benchmark simulations employing tensor-network methods for the example of the 1D sG model where exact analytical results are available.
In equilibrium, we find quantitative agreement for ground and excited state correlation functions in the continuum limit, validating the correct identification of microscopic degrees of freedom with QFT variables.
Out of equilibrium, we discuss how to create solitonic excitations as a feasible alternative to preparing single-particle wave-packets, and find that their spreading induced by quantum fluctuations is well described by a phenomenological semi-classical model. 
We then simulate the non-equilibrium dynamics of scattering pairs of soliton wave-packets, in agreement with predictions from the known S-matrix. 
To the best of our knowledge this constitutes one of the first numerically exact simulations of non-equilibrium scattering dynamics in the continuum limit of a bosonic QFT (see also Ref.~\cite{jha2024realtime} where  scattering in a strongly-coupled Ising field theory is simulated and Ref.~\cite{wybo2022quantum} which focused on the sG model).
As an outlook, we consider a perturbed sG model, where a quantum simulator would provide access to a spatially and temporally resolved picture of scattering dynamics in regimes that potentially lie beyond the reach of traditional methods.

Before delving into the details, the following Section~\ref{sec:general_implementation} provides an overview of our approach, including a discussion of its broad applicability and relevance for emulating condensed matter and high-energy physics. We also provide a brief summary of our main results, and the structure of the rest of this paper, which is also illustrated in Fig.~\ref{fig:Fig1}. Taken together, our results provide a concrete roadmap towards quantitative quantum simulations of field theories.

\section{Scalar field theories from large-spin models\label{sec:general_implementation}}
Our target are $d$-dimensional QFTs described by the Hamiltonian
\begin{align}\label{eq:general_HV}
    \hat{H}_V=\int {\rm d}^dx \left\{ \frac{1}{2}\left[\hat{\pi}(x)\right]^2 +\frac{1}{2} \left[\boldsymbol{\nabla } \hat{\vphi}(x)\right]^2  +V(\hat{\varphi}(x))\right\} \;,
\end{align}
defined in terms of the continuum field operators $\hat{\varphi}(x)$ and their conjugate momentum $\hat{\pi}(x)$ which fulfill the canonical commutation relation \mbox{$[\hat{\varphi}(x), \hat{\pi}(y)]=i\delta(x-y)$}. Different models are distinguished by the form of self-interactions as described by the potential $V$.

Below in Sec.~\ref{sec:microscopic_Rydbergs}, we introduce our prototypical large-spin lattice model that will be used as a regularization of $\hat{H}_V$.
We then discuss the theoretical sequence of limits that leads from this microscopic model to scalar quantum field theories in Sec.~\ref{sec:general_scalars}. 
Sec.~\ref{sec:other_QFTs_and_implementations} elaborates on the range of accessible QFTs and interactions, where we also comment on their relevance.
In Sec.~\ref{sec:overview}, we summarize our main results, pointing the interested reader to the corresponding subsequent sections.

\subsection{Large-spin lattice 
models \label{sec:microscopic_Rydbergs}}
Throughout this manuscript, we study quantum many-body models whose degrees of freedom are large spins of length $J$. More precisely, our microscopic regularization of scalar QFTs is based on a generalized Heisenberg lattice Hamiltonian of the form
\begin{align}
\label{eq:LatticeHamiltonian}
\hat{H}_\text{latt} &= \sum_i \left[ \Omega \hat{J}_{x}^{(i)} - \Delta \hat{J}_{z}^{(i)} + \chi \left(\hat{J}_{z}^{(i)}\right)^2 \right] \nonumber\\
&\qquad-  \sum_{i,\kappa} \left[ \lambda_\kappa \left(\hat{J}_{+}^{(i)}\right)^\kappa + \mathrm{H. c.}   \right] 
\\ 
&\qquad+  \sum_{\langle ij \rangle} V_\text{nn} \left[ \hat{J}_{z}^{(i)} \hat{J}_{z}^{(j)} - \frac{1}{4} \left( \hat{J}_{+}^{(i)}\hat{J}_{-}^{(j)} + \mathrm{H. c.} \right) \right]\;,\nonumber
\end{align}
with $\hat{J}_{x/y/z}^{(i)}$ spin-$J$ operators and   $\hat{J}_{\pm}^{(i)} = \hat{J}_{x}^{(i)} \pm i \hat{J}_{y}^{(i)}$ the corresponding ladder operators, i.e., they obey SU(2) commutation relations $\left[\hat{J}_{x}^{(i)}, \hat{J}_{y}^{(i)}\right] = i \hat{J}_{z}^{(i)}$ and have a fixed length $J(J+1) \hat{1}^{(i)} = \sum_{\alpha=x,y,z}\left(\hat{J}_{\alpha}^{(i)}\right)^2$. We assume the spins to be located on the sites $i$ of a regular (square) lattice in $d$ spatial dimensions.

In Eq.~\eqref{eq:LatticeHamiltonian}, we combine several terms that arise naturally for spin-$J$ models, including
$(i)$ terms linear in single spins, such as transverse $(\propto \Omega)$ and longitudinal fields $(\propto \Delta)$, as well as direct raising/lowering operators $(\propto \lambda_\kappa$ for $\kappa = 1)$;
$(ii)$ single-spin non-linearities, in particular one-axis twisting $(\propto \chi)$ and higher-order raising/lowering terms $(\propto \lambda_\kappa$ with $\kappa =2,3,4, \dots)$; $(iii)$ two-spin interactions $(\propto V_\text{nn})$, here in a specific ``XXZ'' form typical for dipolar interactions, truncated to nearest-neighbor pairs 
$\langle ij \rangle$ at sites $i$ and $j$.

In App.~\ref{sec:AppendixRydbergHamiltonian}, we briefly review a possible experimental implementation of the above Hamiltonian using the SO(4)-symmetric high-dimensional manifold of Rydberg atoms recently discussed in Ref.~\cite{kruckenhauser2022highdimensional}. In this context the spin length $J=\frac{n-1}{2}$ can be controlled by the principal quantum number $n$ of the employed Rydberg excitation, and all coupling parameters $\Omega, \Delta, \chi, \lambda_\kappa, V_\text{nn}$ can be experimentally tuned by controlling atom-light interactions or rearranging the lattice geometry. 
Other experimentally relevant setups that give rise to related large-spin models include dipolar molecules~\cite{bao2023dipolar}, magnetic atoms~\cite{chomaz2022dipolar,claude2024optical} and hybrid digital-analog proposals using qubit clusters of Rydberg atoms~\cite{maskara2023programmable}. More generally, large-spin models can be implemented with general qudit computers~\cite{wang2020qudits} such as those realized with trapped ions~\cite{ringbauer2022universal} as well as hybrid qubit-oscillator processors based on cavity QED~\cite{crane2024hybrid}.

\subsection{
From the lattice to continuum field theories
\label{sec:general_scalars}}

We now show how to tune the microscopic Hamiltonian to obtain scalar field theories. For quantitative predictions, this requires a sequence of controlled extrapolations [see Fig.~\ref{fig:Fig1}(a) and Fig.~\ref{fig:Fig2}]: large spins ($J \rightarrow \infty$), scaling to the thermodynamic (system size $N\rightarrow \infty$), and the continuum limit (lattice spacing $a \rightarrow 0$).
Below we summarize how to take these limits in theory, which we later repeatedly carry out in our numerical simulations.

\subsubsection{Large-spin limit}
For a square lattice regularization of $\hat{H}_V$ with lattice spacing $a$, one typically considers dimensionless fields ${\hat{\varphi}}_i$ and ${\hat\pi}_i$, that obey $[{\hat{\varphi}}_i, {\hat{\pi}}_j]=i\delta_{ij}$.
As a key ingredient to represent scalar fields by spins, we instead employ compactified fields where $\hat{\varphi}_i\in [0, 2\pi]$ is restricted periodically. This means we work with the well-defined vertex operator $\vertex$, which obeys \mbox{$[\hat{\pi}_i, e^{i{\hat\varphi}_j}]= \delta_{ij} e^{i{\hat\varphi}_j}$} 
and thus acts as a raising operator on the eigenstates $\ket{m}_\pi$ of $\hat{\pi}$ (at every lattice site). These operators can be regularized on the finite-dimensional Hilbert space of a spin-$J$ by identifying~\cite{haldane1983continuum,zache2022continuum}
\begin{align}\label{eq:fields_to_spins}
e^{\pm i \hat{\varphi}_i} \leftrightarrow \frac{\hat{J}^{(i)}_\pm}{\sqrt{J(J+1)}} \;, &&  \hat{\pi}_i \leftrightarrow \hat{J}^{(i)}_z \;.
\end{align}
Here the $2J+1$ eigenstates $\ket{m}$  of $\hat{J}_z$ ($m\in [-J, \dots, J]$) provide a natural restriction $|m| \leq J$ for the momentum eigenstates $\{\ket{m}_\pi\}$ of  $\hat{\pi}$ at every lattice site. 

The identification becomes formally exact on states that occupy sufficiently small $|m|\ll J$. This can be seen by comparing the action of the rescaled raising operators $\hat{J}_\pm/\sqrt{J(J+1)}$ with that of the vertex operators, i.e., 
\begin{align}
    \frac{\hat{J}_\pm}{\sqrt{J(J+1)}}\ket{m}&=\frac{\sqrt{J(J+1)-m(m\pm 1)}}{\sqrt{J(J+1)}}\ket{m\pm 1} \nonumber\\
&= \left[1 - \mathcal{O}\left(\frac{m(m\pm 1)}{J(J+1)}\right)\right] \ket{m\pm 1}\\
    \longleftrightarrow e^{\pm i\opphi}\ket{m}_\pi&=\ket{m\pm 1}_\pi \;. \nonumber
\end{align}
Sending $J \rightarrow \infty$ therefore produces the desired lattice field operators. Note that this scaling renders the Ising term ($\propto V_\text{nn} \hat{J}_{z}^{(i)} \hat{J}_{z}^{(j)}$) in $\hat{H}_\text{latt}$ insignificant in comparison to the ``flip-flop'' interactions ($\propto V_\text{nn} \hat{J}_{+}^{(i)}\hat{J}_{-}^{(j)}$), see App.~\ref{sec:AppsG_compare}. 

\subsubsection{Numerical extrapolations}
Conceptually, the large-$J$ limit of the $d$-dimensional lattice Hamiltonian $\hat{H}_\text{latt}$ can be translated to a continuum model by increasing the number of lattice sites $N$ and decreasing a fictitious lattice spacing $a$, while keeping the total volume $V = L^d$ with $L = N a$ fixed. Afterwards one can also remove the IR regulator by sending $L\rightarrow\infty$.

\begin{figure*}
    \centering
    \includegraphics[width=2\columnwidth]{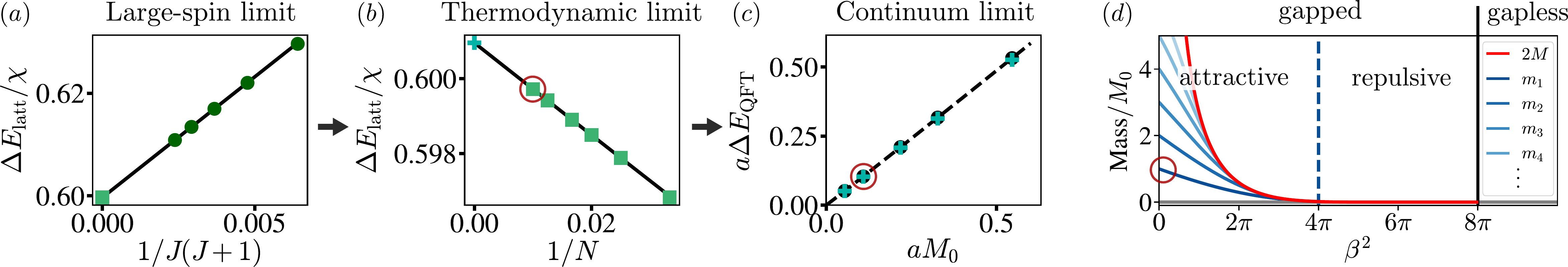}
\caption{\textbf{Extrapolating to the continuum limit of the simulated sine-Gordon model.} 
    (a)-(c) Extrapolation to the continuum value of the mass gap $\Delta E$. We consider the $d=1$ lattice Hamiltonian $\hat{H}_\text{latt}$ Eq.~\eqref{eq:LatticeHamiltonian} and determine the mass gap $\Delta E$ in the continuum limit by simulating the model at several finite $J=12,...,20$ and $N=30,...,100$. 
    (a) For fixed $N=100, V_\text{nn}'/\chi=40.53, \lambda_\kappa'/\chi=0.20$ we perform a linear fit (solid black line) of the energy gap  $\Delta E_\text{latt}$ (dark green dots, in lattice units) against $1/J(J+1)$, and extrapolate the value in the large spin-length limit $J\rightarrow \infty$ (light green square). (b) The values of the first extrapolation (light green squares) are then fitted against $1/N$ to extrapolate the result in the thermodynamic limit $N\rightarrow \infty$ (light blue plus). The red circle identifies the asymptotic value extrapolated in (a). (c) The data obtained from the fit in $1/N$ corresponds to the continuum limit value $\Delta E_\text{QFT}$ (in QFT units) as evidenced by the correct scaling behaviour w.r.t. the bare mass $a M_0$. The dashed line represents the theoretical prediction  [Eq.~\eqref{eq:breathermass}] for  the mass gap in the simulated QFT, the sine-Gordon model [Eq.~\eqref{eq:QFTHamiltonian} for $d=1$]. The red circle identifies the asymptotic value extrapolated in (b).
    (d) The phase diagram of the sine-Gordon model. The coupling $\beta^2$ distinguishes different regimes of the QFT. A BKT transition at $\beta^2=8\pi$ separates a gapless ($\beta^2>8\pi$) from a gapped region ($\beta^2<8\pi$). In the latter the fundamental excitations are (quantum) solitons of mass $M$. Two solitons interact repulsively for $\beta^2>4\pi$, while they attract for $\beta^2<4\pi$, giving rise to bound states (breathers) with mass $m_n$ Eq.~\eqref{eq:breathermass}. The maximal number $n$ of breathers is fixed by $\beta^2$ [Eq.~\eqref{eq:breathermass}]. The mass gap extrapolations from Figs.~(a)-(b) correspond to the lowest breather mass $m_1$ deep in the attractive regime with $\beta^2 \approx 0.1 \times \pi$, as highlighted by the red circle.}
    \label{fig:Fig2}
\end{figure*}

In practice, we instead implement a three-step procedure, where we take the thermodynamic limit directly after the large-spin limit and extract continuum observables in the very end. The last step requires a precise parameter matching as discussed in the next subsection. As an illustration of the whole process, consider Fig.~\ref{fig:Fig2}, where we extract the mass gap of the sG model.

To obtain the results shown in Fig.~\ref{fig:Fig2}, we start by fixing the microscopic lattice parameters of $\hat{H}_\text{latt}$ at finite system size $N$, and compute the expectation value of the relevant observables for varying $J$.
The numerical results are then fitted with a linear function in $1/J(J+1)$ [see Fig.~\ref{fig:Fig2}(a)], from which we retrieve the results in the asymptotic limit $1/J(J+1)\rightarrow 0$, equivalent to the large spin-length limit $J\rightarrow \infty$ [see App. \ref{sec:AppendixFittingDiscussion} for further details].
In a second step, the asymptotic values obtained for different $N$ are fitted with a quadratic function in $1/N$ [see Fig.~\ref{fig:Fig2}(b)].
From this additional fit we retrieve the thermodynamic limit $1/N \rightarrow 0$, equivalent to $N\rightarrow\infty$.
For small enough lattice spacing $a\rightarrow 0$, we expect scaling behaviour indicative of the continuum limit, which can be compared to theoretical expectations.

Note that we do not tune the lattice spacing $a$ as part of the physical geometry, but we instead vary a dimensionless control parameter $aM_0$ -- which is a function of the microscopic parameters as discussed below.
This parameter sets the mass scale $M_0$ of the simulated QFT at the UV cutoff $1/a$ determined by our microscopic lattice model.
When all extrapolations are done correctly for sufficiently small values of $aM_0$, the system develops a large correlation length $\xi \sim 1/{M_0} \gg a$, such that the extrapolated results become insensitive to the lattice regularization, and indeed agree with the continuum predictions.
We emphasize that extrapolating $J\rightarrow \infty$ in the first step is an essential ingredient that allows us to perform this analysis. This is in contrast to a fixed small spin length, where the accessible QFT spectrum would be severely truncated to very low energies.

\subsubsection{Matching microscopic and QFT parameters}
In general, the couplings of the quantum field theory arising in the continuum limit are related to the microscopic parameters -- defined in the UV by the lattice regularization -- through renormalization.
To gain a better intuition, we now show how to estimate this relation by dimensional analysis, although we will later match these parameters directly from our numerical simulations without further input.

From Eq.~\eqref{eq:fields_to_spins}, together with the freedom to rescale $\hat\varphi_i \rightarrow \alpha \hat\varphi_i, \hat\pi_i \rightarrow \frac{1}{\alpha} \hat\pi_i$ by an arbitrary constant $\alpha$, and reinserting appropriate powers of the lattice spacing, we identify continuum expressions as
\begin{align}\label{eq:vertex_opertor_identification}
\hat{J}_\pm^{(i)} &\rightarrow \sqrt{J(J+1)}\exp\left[i \frac{\beta}{\kappa} \hat\varphi(x) \right] \;,
\\
\hat{J}_z^{(i)} &\rightarrow  \frac{\kappa}{\beta'}a^{\frac{d+1}{2}} \hat{\pi}(x)\;,\\
a^d\sum_i &\rightarrow \int {\rm d}^d x \;.
\end{align}
Note that this introduces the parameter $\beta = \beta' a^{(d-1)/2}$ with (at this point arbitrary) dimensionless coupling $\beta'$, such that $\hat{\varphi}(x)$ is $(2\pi\kappa/\beta)$-periodic.

These identifications are chosen such that $\hat{H}_\text{latt}$ with $\Delta = \Omega = 0$, and a single fixed $\kappa$ results in (for $a \rightarrow 0$, and up to an irrelevant energy shift) 
\begin{align}
\label{eq:QFTHamiltonian}
    &\frac{\left(\beta'\right)^2}{2 \kappa^2 \chi a} \hat{H}_\text{latt}  \rightarrow \hat{H}_\text{QFT}\\
    &\quad =\int {\rm d}^d x \left\{ \frac{1}{2}\left[\hat{\pi}(x)\right]^2 +\frac{1}{2} \left[\boldsymbol{\nabla }\hat{\vphi}(x)\right]^2  -\frac{M_0^2}{\beta^2}\cos{\left[\beta \hat{\varphi} (x)\right]}\right\} \;. \nonumber
\end{align}
This result is obtained by matching the parameters according to
\begin{align}\label{eq:parameter_ident}
\left(\beta'\right)^4 = 4\kappa^4 \frac{\chi}{V'_\text{nn}} \;, && \left(M_0'\right)^2 = 4 \kappa^2 \frac{\lambda_\kappa'}{V'_\text{nn}} \;,
\end{align}
which provides the required direct relation of the bare dimensionless parameters $\beta'$ and $M_0' = aM_0$ of the QFT to the microscopic parameters of the lattice model [see App.~\ref{sec:AppendixMappingHamiltonian} for more details].
Note that this identification requires tuning 
\begin{align}
\lambda_\kappa' = \lambda_\kappa [J(J+1)]^{\kappa/2} \;, && V'_\text{nn} = J(J+1) V_\text{nn} \;,
\end{align}
upon increasing $J$.
For the example of the mass gap $\Delta E$ shown in Fig.~\ref{fig:Fig2}, Eq.~\eqref{eq:QFTHamiltonian} also provides the correct relation of lattice and continuum units, i.e., \mbox{$\Delta E_\text{latt}/\chi=2\kappa^2 a\; \Delta E_\text{sG}/(\beta')^2$} \footnote{Here $\Delta E_\text{sG}$ denotes the energy gap in ``continuum'' units, while $\Delta E_\text{latt}$ corresponds to the same quantity in ``lattice'' units. In practice, we set $\chi=1$, and either insert powers of $\chi$ or $a$ to form dimensionless quantities}.

\subsection{Accessible QFTs\label{sec:other_QFTs_and_implementations}}
Generalizing from the discussion of the previous section, the general lattice Hamiltonian Eq.~\eqref{eq:LatticeHamiltonian} with multiple simultaneous higher-order terms $\propto \lambda_\kappa$ leads to Eq.~\eqref{eq:general_HV}
with a tunable potential
\begin{align}
V( \hat{\varphi} ) = -\frac{M_0^2}{\beta^2} \sum_{\tilde{\kappa}} \frac{\lambda'_{\tilde{\kappa}}}{\lambda'_\kappa} 
\cos \left( \beta \frac{\tilde{\kappa}}{\kappa} \hat{\varphi} \right) \;,
\end{align}
where we have singled out a specific $\kappa$ as a reference that sets the scale $M_0$. This shows that by choosing, e.g., $\tilde{\kappa}$ an integer multiple of $\kappa$, we can construct arbitrary symmetric $(2\pi \kappa/\beta)$-periodic potentials by adjusting suitable ratios $\lambda'_{\tilde{\kappa}}/\lambda'_\kappa$, which now play the role of Fourier expansion coefficients. In this sense, we can indeed access general scalar field theories in $d$ spatial dimensions.
In the following, we briefly discuss a few examples of the wide range of models that could be studied in this way.

In one spatial dimension ($d=1$), for a single cosine potential $V(\opphi)=-M_0^2/\beta^2 \cos{(\beta \opphi)}$, we obtain the standard quantum sG model, which is a paradigmatic integrable QFT~\cite{zamolodchikov1995mass,lukyanov1997exact,pallua2001uv,daviet2019nonperturbative,roy2021quantum}. It describes the low-energy regime of a variety of experimental setups, such as tunnel-coupled superfluids~\cite{gritsev2007linear,schweigler2017experimental}, strongly interacting Bose gases~\cite{haller2010pinning}, superconducting quantum circuits~\cite{roy2021quantum}, or spin chains~\cite{wybo2023preparing,wybo2022quantum}. As an integrable model, many of its properties are known exactly [see Fig.~\ref{fig:Fig2}(d) and Sec.~\ref{sec:sG_continuum}]. In the remainder of this paper we will largely focus on this one-dimensional case for simplicity, and perform thorough numerical analyses in order to demonstrate that a faithful quantitative realization of the sG model is indeed possible.

An interesting non-integrable perturbation of the 1D sG model is a simple quadratic potential $\sim \hat{\varphi}^2$. The resulting theory is the bosonized version of the massive Schwinger model, i.e., quantum electrodynamics in 1D~\cite{coleman1976more,abdalla1991nonperturbative,jentsch2022physical,batini2024particle}, a model that has attracted a lot of recent attention as a testbed for quantum simulation experiments of lattice gauge theories~\cite{martinez2016realtime,klco2018quantumclassical,kokail2019selfverifying,nguyen2022digital,farrell2024quantum}.

For larger spatial dimensions ($d>1$), scalar fields have a non-vanishing mass dimension, namely $[\hat{\vphi}]=\frac{d-1}{2}$ in units of $[M_0]=1$. According to textbook arguments~\cite{peskin1995introduction}, renormalizable field theories in $d=2$ and $d=3$ include interaction potentials up to $\sim\hat\varphi^6$ and $\sim\hat\varphi^4$, respectively. These define two interesting target models, where quantum simulators could tackle long-standing questions in non-equilibrium QFT about the fate of the false vacuum~\cite{coleman1977fate}. For recent alternative approaches to this problem, we refer to the literature~\cite{lagnese2021false,lagnese2023detecting,batini2024realtime,batini2024particle}.

Moreover, our scalar fields inherit a fundamental periodicity from the microscopic quantum spins, which can have important consequences for their dynamics. A famous example in two spatial dimensions ($d=2$) is compact U(1) lattice gauge theory, where topological configurations (monopoles) give a mass to the ``photon'', effectively prohibiting a Coulomb phase, such that the model always remains confined~\cite{polyakov1977quark}. This compact U(1) gauge theory with monopoles turns out to be dual to the 2D sG model~\cite{sachdev2023quantum, kaplan2020gausss,bender2020gauge}, such that the former also becomes accessible within our approach. A related approach based on Josephson junction arrays has recently been proposed in Ref.~\cite{pardo2024truncationfree}.

As a final example of a compact scalar field in three spatial dimensions ($d=3$), we mention the ``axion'' -- a hypothetical periodic scalar field that has been introduced as a potential resolution of the so-called strong CP problem~\cite{peccei1977cp,weinberg1978new,wilczek1978problem}. In the original proposal~\cite{peccei1977cp}, the axion field is naturally compact as a consequence of the Higgs mechanism. Although a direct observation of real-world axions is still lacking~\cite{sikivie2021invisible}, analogous phenomena have recently been observed experimentally in condensed matter systems~\cite{gooth2019axionic}, and could also be explored in quantum simulation setups based on large-spin models following our approach.

\subsection{\label{sec:overview}Overview of results}

We close this section with an overview of the different results obtained in this work. Our results can be grouped in several qualitatively distinct regimes of a QFT simulation, as illustrated in Fig.~\ref{fig:Fig1}(b).

In Sec.~\ref{sec:sG_vacuum}, we show with detailed numerical tensor-network simulations that our proposed large-spin regularization faithfully captures equilibrium properties of the 1D sine-Gordon model in the continuum limit. 
Focusing first on the underlying conformal field theory (CFT), the Luttinger liquid model, we simulate several system sizes $N$ and spin lengths $J$, and perform the sequence of extrapolations discussed in Sec.~\ref{sec:general_scalars} [see also Fig.~\ref{fig:Fig2}], to identify the parameter regime of the predicted gapless phase with associated central charge $c=1$.
We further determine the Luttinger parameter $K$ of the field theory and clarify its renormalization w.r.t. the classical microscopic mapping from the lattice Hamiltonian.

We then consider the sine-Gordon model where we first determine the values of the renormalized coupling strengths $\beta^2, M_0'$, confirming the theoretical prediction $\beta^2=\pi \kappa^2 K$ between renormalized parameters of the sG and the CFT.
We emphasize that our parameter matching procedures and the identification of microscopic with continuum variables does not rely on the solvability of the sG QFT, but is also directly applicable to the quantum simulation of other target models, without requiring input from theory simulations.
By performing the sequence of limits in $J$ and $N$, we extract the scaling behaviour of several vacuum and first-excited state observables in the continuum limit, including the mass gap and vertex operators, in the numerically accessible regime $\beta^2\lesssim1$. 
Our numerical results demonstrate quantitative agreement with theoretical QFT predictions for the renormalized coupling parameters, which we further use to test the unproven Lukyanov-Zamolodchikov conjecture.
Overall, we find that the investigated system sizes of $N\lesssim 100$ lattice sites and spin lengths $J\lesssim20$ -- both within reach of state-of-the-art quantum simulators -- are adequate for the equilibrium quantum simulation of the sG model.
These results form the basis for the following sections where we apply our approach to the investigation of localized excitations, their dynamics and collisions in the sG model. 

In Sec.~\ref{sec:sG_solitons}, we simulate the real-time quantum dynamics of individual solitons for small $\beta^2$, as a warm-up for studying collisions of single-particle wave-packets. We propose a simple preparation scheme of ``semi-classical solitons'', i.e., quantum states whose phase profile agrees with the classical soliton solution, but are subject to the same fluctuations dictated by quantum correlations in the true ground states.
Assuming local control over the microscopic spins, our preparation protocol applies local spin rotations on the ground state of our lattice regularization of the quantum sG model to adjust the one-point field expectation values while preserving the two-point functions.
Starting from such initial states, we find that the time evolution displays the expected classical linear motion in the mean position of the soliton, with a velocity that agrees with the one imprinted by our protocol. 
On top of this classical motion, the presence of quantum fluctuations induces other observables -- such as the topological charge density -- to spread and smear out over time. 
This observation can be explained by interpreting the prepared states as wave-packets of solitons, whose initial velocities and positions are stochastically distributed around the imposed classical values. 
We make this interpretation precise by constructing an appropriate semi-classical phenomenological model, essentially a truncated Wigner approximation restricted to the single-soliton sector, from which we derive analytical predictions that quantitatively agree with the observed dynamics.

In Sec.~\ref{sec:sG_scattering}, we further leverage the state preparation tools developed in the previous section and simulate the scattering of a soliton and antisoliton. 
Within the parameter regimes accessible to our classical numerical simulations, we observe how the two excitations initially counter-propagate, with the expected linear motion, followed by transmissive scattering, after which the individual excitations continue to  propagate linearly. The considered type of collision is expected to introduce a position shift of the outgoing trajectories compared to the incoming ones, which we extract from our simulation. 
Our results agree with predictions from the known S-matrix of the sG model for several different initial conditions, verifying that also the non-equilibrium dynamics is faithfully captured by the large-spin regularization. 
In contrast to most previous works, which have focused on qualitative aspects of scattering dynamics, here we demonstrate \emph{quantitative} agreement of real-time observables with continuum QFT results. A notable exception is the recent work Ref.~\cite{jha2024realtime}, where the authors study real-time dynamics of scattering in a 1D Ising field theory, which is based on the well known Ising phase transition in a spin-1/2 chain. Similarly, Ref.~\cite{wybo2022quantum} studies a realization of the sG model with two coupled spin-1/2 chains. In contrast, our approach using large spins enables the realization of general scalar fields in arbitrary dimensions.

As an outlook illustrating the generality of our approach, we consider a variant of our proposed regularization that corresponds to a non-integrable perturbation of the sG model in Sec.~\ref{sec:beyond_sG}.
The form of the perturbation is motivated by quantum electrodynamics and is expected to introduce a confining potential between soliton and antisoliton, which in turn leads to the formation of a meson-like bound state.
We classically simulate scattering dynamics of this modified model, which reveals qualitatively distinct regimes depending on the perturbation strength. Small integrability breaking leads to repeated scattering of the individual quasi-particles, consistent with the formation and oscillations of a bound state.
Upon increasing the perturbation strength, we observe dynamics reminiscent of string breaking, as suggested by the pairwise production of quasi-particles of opposite topological charge, as well as ``plasma'' oscillation of observables akin to the electric field strength.
While a quantitative analysis of this dynamics at long times is prohibitively costly with classical means, this setup is a prime example for the application of our approach in a quantum simulation experiment.

Finally, we conclude in Sec.~\ref{sec:discussion} with an outlook on future directions.

\section{Simulating the sine-Gordon vacuum\label{sec:sG_vacuum}}
In this section, we focus on the ground and first excited state of the sG model. After briefly reviewing known facts used for benchmarking (Sec.~\ref{sec:sG_continuum}), we present our numerical results in Sec.~\ref{sec:sG_from_lattice}. Our findings demonstrate quantitative agreement of the lattice regularization with large spins and the anticipated continuum physics.

\subsection{The sG model in the continuum\label{sec:sG_continuum}}
The sine-Gordon (sG) model, defined by Eq.~\eqref{eq:QFTHamiltonian} for $d=1$, is a well-studied integrable relativistic 1+1D quantum field theory~\cite{zamolodchikov1995mass,lukyanov1997exact,pallua2001uv,daviet2019nonperturbative,roy2021quantum}. 
The dimensionless coupling $\beta^2=(\beta')^2\geq 0$ identifies different regions of the model: at $\beta^2=8\pi$ a BKT transition separates a gapless phase  ($ \beta^2 > 8\pi $) from a gapped phase ($\beta^2~<~8\pi$) [see Fig.~\ref{fig:Fig2}(d)]. The gapped phase  supports single-particle excitations known as (quantum) solitons whose mass $M$ is known analytically \cite{daviet2019nonperturbative}
\begin{equation}\label{eq:solitonmass}
    M =b\Lambda\frac{2\Gamma\left(\frac{\xi}{2}\right)}{\sqrt{\pi}\Gamma\left(\frac{1+\xi}{2}\right)}\left[\frac{M_0^2(1+\xi)\Gamma\left(\frac{1}{1+\xi}\right)}{16\xi \Gamma\left(\frac{\xi}{1+\xi}\right)(b\Lambda)^2}\right]^{\frac{1+\xi}{2}},
\end{equation}
where $\xi=\beta^2/(8\pi -\beta^2)$, $\Lambda=\pi/a$ the UV cutoff, and \mbox{$b=e^C/2$} a scale parameter dependent on the cutoff implementation.
For the intermediate region $4\pi ~<~\beta^2~<~8\pi$ two solitons repel each other, while for $0~<\beta^2~<~4\pi$ they interact attractively and form bound states (so-called breathers) of mass $m_n$ given by
\begin{equation}
    m_n=2M\sin{\left(\frac{n\pi\xi}{2}\right)}, \qquad  n=1,2,\dots, \left\lfloor \frac{1}{\xi} \right\rfloor \;.
    \label{eq:breathermass}
\end{equation}
Here $n$ is the maximal number of breathers, which is set by $\xi$.

For our purposes, it is useful to regard the sG model as an integrable deformation of the free, compactified boson conformal field theory (CFT)~\cite{zamolodchikov1995mass, roy2021quantum}.
The perturbation is given by the vertex operator $e^{i\beta \opphi}$ with scaling dimension $\Delta_\varphi = \frac{\beta^2}{4\pi}$~\cite{pallua2001uv}.
Its expectation value in the ground state of the sG model is given by~\cite{lukyanov1997exact}
\begin{multline}
    \langle e^{i\beta \opphi}\rangle=\frac{(1+\xi)\pi \Gamma\left(\frac{1}{1+\xi}\right)}{16\sin{(\pi\xi)}\Gamma\left(\frac{\xi}{1+\xi}\right)}
    \\
    \times \left[\frac{\Gamma\left(\frac{1+\xi}{2}\right)\Gamma\left(1-\frac{\xi}{2}\right)}{4\sqrt{\pi}}\right]^{-\frac{2}{1+\xi}}m_1^{\frac{2\xi}{1+\xi}} \;.
    \label{eq:vertextheory}
\end{multline}
Measured with respect to the CFT, also the ground-state energy density of the sG model is finite and given by
\begin{equation}
        E_0(M)=\frac{\langle \hat{H} \rangle_\text{sG} - \langle \hat{H} \rangle_\text{CFT} }{V}=-\frac{M^2}{4}\tan{\left(\frac{\pi\xi}{2}\right)}
    \label{eq:gsenergydensity} \;.
\end{equation}

\subsection{The sG model from the lattice\label{sec:sG_from_lattice}}

We now turn to numerical simulations of the lattice regularization with finite-length spins.  
Throughout this work, we employ matrix product states (MPS) techniques using ITensor~\cite{fishman2022itensor}. In order to obtain continuum results, we need to carry out the extrapolations discussed in the previous section. 
In practice, we do this by working with a fixed $\chi = 1$ and scanning several values of system size $N = 30, \dots, 100$, spin length $J = 12, \dots, 20$ and couplings $V_\text{nn}, \lambda_\kappa$. 
Overall, we find our simulations to be well described by first fitting the numerical data with a linear function in $1/J(J+1)$, and then with a quadratic function in $1/N$, see Fig.~\ref{fig:Fig2} and the discussion in App. \ref{sec:AppendixFittingDiscussion}.
Unless stated otherwise, we present observables obtained after this two-step extrapolation ($J \rightarrow \infty$ followed by $N \rightarrow \infty$) and only highlight the more delicate continuum scaling from now on.

\begin{figure}
    \centering
    \includegraphics[width=\linewidth]{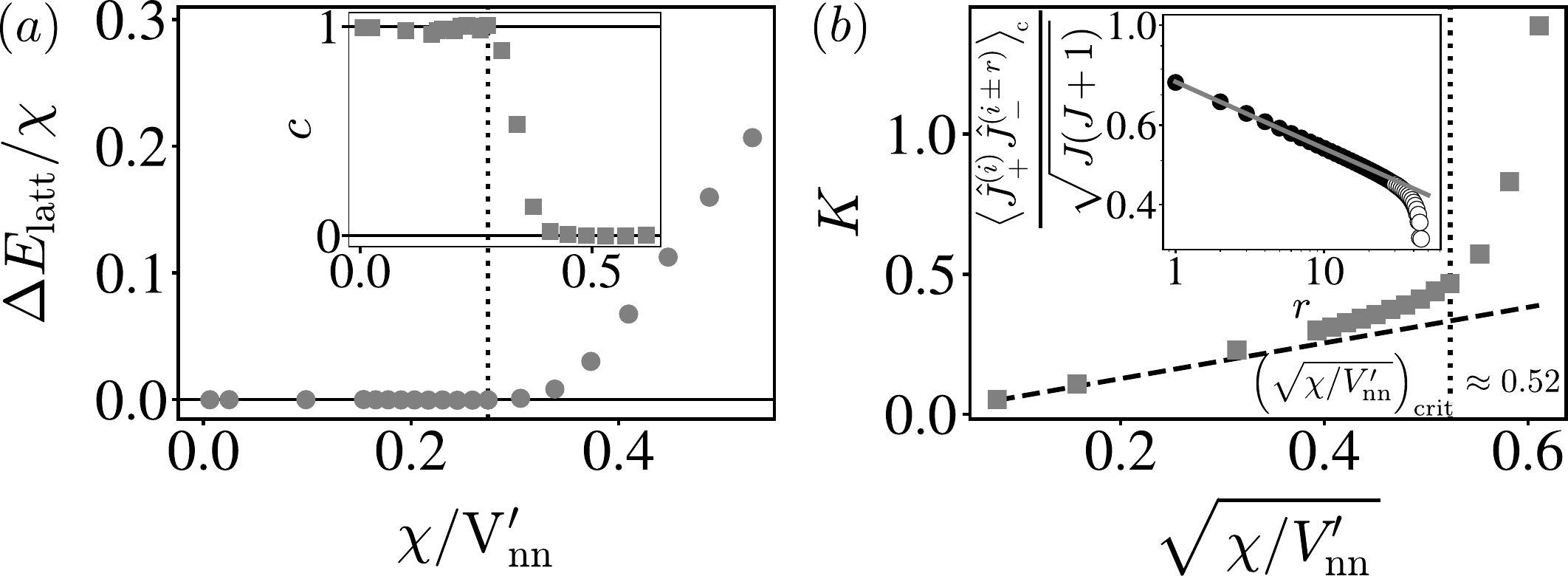}
    \caption{\textbf{Characterization of the CFT.} (a) The energy gap $\Delta E_\text{latt}/\chi$ in lattice units extracted from our lattice regularization $\hat{H}_\text{latt}$ clearly shows a gapless phase at small $\chi/V'_\text{nn}$. In the same region, the central charge $c$ (inset) is compatible with $1$. The dotted black line delimitates the simulability region.
    (b) The Luttinger parameter $K$ of the CFT as a function of the microscopic parameters $\sqrt{\chi/V_\text{nn}'}$ (gray squares) follows the simple microscopic prediction (dashed black line) for small $K$ and deviates at larger $K$ as expected. The value of $K$ is determined from the power-law decay of the connected vertex-vertex correlation function $\langle \hat{J}_+^{(i)} \hat{J}_-^{(i\pm r)} \rangle_\text{c} /\sqrt{J(J+1)}$ with distance $r$ (inset, for parameters $N=90, J=18, \sqrt{\chi/V_\text{nn}'}=0.393, i=N/2=45$).}
    \label{fig:CFT}
\end{figure}

\subsubsection{The underlying CFT}
Calculating both the ground and first excited state using DMRG~\cite{fishman2022itensor} for $\lambda_\kappa = 0$, we indeed find a gapless phase in the regime [see Fig.~\ref{fig:CFT}(a)]
\begin{align}\label{eq:range_gapless_phase}
0<\frac{\chi}{V'_\text{nn}} < \left(\frac{\chi}{V'_\text{nn}}\right)_\text{crit} \approx 0.27.
\end{align}
The fact that a massive phase appears for large $\chi$ is expected as the $\left(\hat{J}_{z}^{(i)}\right)^2$- term favors the trivial product state $\bigotimes_{i}|m=0\rangle_i$. 

To test the nature of the anticipated CFT at small $\chi/V'_\text{nn}$, we further calculate the von Neumann entanglement entropy 
$S_\text{vN}$ in the ground state for a bipartition of $n$ sites and their complement of $N-n$ sites and fit the central charge $c$ according to
\begin{equation}
    S_\text{vN}
    =\frac{c}{6}\log{\left(\frac{2N}{\pi}\sin{\left(\frac{\pi n}{N}\right)}\right)} + c',
    \label{eq:centralchargeEE}
\end{equation}
with $c'$ a non-universal constant \cite{calabrese2004entanglement} [see App.~\ref{sec:AppendixCFT}]. Our results confirm $c=1$ as expected for a Luttinger liquid [see inset of Fig.~\ref{fig:CFT}(a)].

According to the previous section, we expect that the perturbing operators $\left(\hat{J}_\pm / \sqrt{J(J+1)} \right)^\kappa$ provide lattice regularizations of the vertex operators $e^{\pm i \beta \hat \varphi (x)}$. In the gapless phase of a Luttinger liquid, we expect power law connected correlations of the form
\begin{align}\label{eq:Luttinger_power_law}
    \frac{1}{(J(J+1))^\kappa} \left\langle \left(\hat{J}_+^{(i)}\right)^\kappa \left(\hat{J}_-^{(j)}\right)^\kappa   \right\rangle_\text{c} &\leftrightarrow  \langle e^{i\beta \opphi(x_i)} e^{-i\beta \opphi(x_j)} \rangle \nonumber
\\
&\propto |i-j|^{-\kappa^2 K/2},
\end{align}
which defines the Luttinger parameter $K > 0$. From Eq.~\eqref{eq:parameter_ident}, together with $\kappa^2 K/2 = 2\Delta_\varphi$,  i.e., $\Delta_\varphi = \frac{\beta^2}{4\pi}$, and the numerical result in Eq.~\eqref{eq:range_gapless_phase}, we expect
\begin{align}
    0 < K = \frac{2}{\pi} \sqrt{\frac{\chi}{V'_\text{nn}}} \lesssim 0.33 \;.
    \label{eq:K_microscopic}
\end{align}
These expectations are also confirmed by our simulations as shown in Fig.~\ref{fig:CFT}(b), at least for sufficiently small $K$.
For larger $\sqrt{\chi/V'_\text{nn}} \gtrsim 0.4$, we find quantitative deviations, which we attribute to higher derivative terms appearing in the expansion of
\mbox{$\cos\left[\beta(\opphi_i - \opphi_j)/\kappa\right]=1-\beta^2(\opphi_i - \opphi_j)^2/(2\kappa^2) + \dots$} from the dipole-dipole interaction $\hat{J}_{+}^{(i)}\hat{J}_{-}^{(j)} + \mathrm{H. c.} \;$.
Our numerics thus shows that these neglected higher-order terms can be absorbed in a renormalization of the Luttinger parameter $K$. For quantitative predictions in a potential experimental realization, one could similarly extract $K$ (which determines the coupling $\beta$ in the sG model through $\beta^2 =\pi\kappa^2 K$) by a fit of Eq.~\eqref{eq:Luttinger_power_law} instead of relying on the above theoretical estimates. We also emphasize that larger values of $\beta^2$ become accessible by realizing interactions with $\kappa >1$.
In App.~\ref{sec:AppendixCFT} we further investigate the impact of next-nearest-neighbors (NNN) terms, and show that they lead to the same Luttinger liquid model, yet with reduced deviations in the Luttinger parameter $K$.

\subsubsection{The sine-Gordon model}
Having established the underlying CFT with $c=1$ and extracted $K$, we now test analytical predictions
for our lattice regularization of the sG model.
As in the CFT case, we focus on the ground and first excited state.

\begin{figure}
    \centering
    \includegraphics[width=\linewidth]{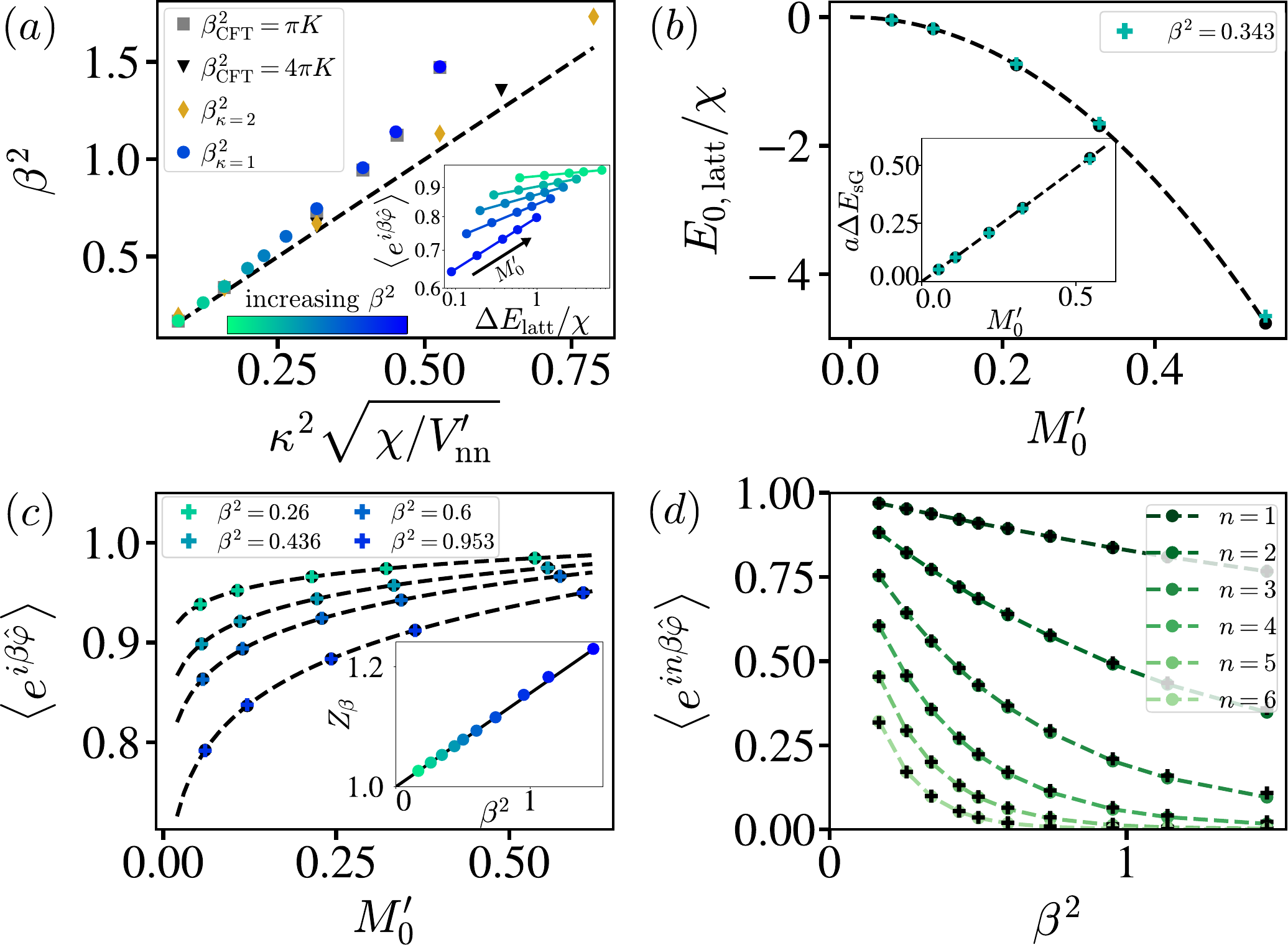}
    \caption{\textbf{Renormalized sG QFT.} (a) Coupling $\beta^2$ as a function of the microscopic parameters $\kappa^2 \sqrt{\chi/V_\text{nn}'}$.
    The colored dots resp. the yellow diamonds represent the $\beta^2$ values extracted in the sG model for $\kappa=1$ resp. $\kappa=2$ with a fit according to Eq.~\eqref{eq:sG_beta_fit} [see inset]. The extracted coupling parameters agree with the  CFT relation $\beta^2=\pi\kappa^2 K$ (gray squares for $\kappa=1$, and black triangles for $\kappa=2$), but show deviations from the naive estimate from the microscopic parameters [Eq.~\eqref{eq:parameter_ident}, dashed black line], signalling a renormalization of $\beta^2$.
    (b) Ground-state energy density $E_{0, \text{latt}}$ (in lattice units) measured with respect to the CFT. For the same value of  $\beta^2$ we display the energy gap $\Delta E_\text{sG}$ (in sG units) in the inset.
    (c) Expectation value of the vertex operator $\langle e^{i \beta \hat{\varphi}} \rangle$ at $\kappa=1$ versus $M_0'$. Numerical results (colored plus) and corresponding theoretical predictions (dashed black line). 
    As discussed in the main text, the bare numerical result is renormalized with a factor $Z_\beta \approx 1 + \mathcal{O}(\beta^2)$ [see inset].
    (d) The expectation value of higher powers of the vertex operators $\langle e^{i n \beta \hat{\varphi}} \rangle$ (dark plus)  at $\lambda'_\kappa/V_\text{nn}'=0.0025$ agree with the Lukyanov-Zamolodchikov conjecture (green circles; dashed lines are guides to the eye)  upon rescaling the numerical results with the factor $(Z_\beta)^{n^2}$ [see main text and App.~\ref{sec:AppsG_conjecture} for more details].}
    \label{fig:Equilibrium}
\end{figure}

Our main observables are the expectation value of the vertex operator $\langle e^{i\beta \opphi}\rangle \leftrightarrow \left\langle \left(\hat{J}_+/\sqrt{J(J+1)}\right)^\kappa\right\rangle$ in the ground state and the mass gap $\Delta E = m_1$.
From Eqs.~\eqref{eq:breathermass} and \eqref{eq:vertextheory} for fixed $\beta^2$, we expect a proportionality 
\begin{align}\label{eq:sG_beta_fit}
\langle e^{i\beta \opphi}\rangle \propto \left(\Delta E\right)^{\alpha} \;, &&  \alpha = 2\xi/(1+\xi)=\beta^2/4\pi = \Delta_{\varphi}  \;, 
\end{align}
which we test in our numerics by a linear fit in log-log scale. More precisely, we fit data points corresponding to a given $\beta^2$, while we vary over $M_0'$, thus resulting in different values of the observables. The data points considered are obtained by the double extrapolation in $J$ and $N$. The result of the linear fit is shown in Fig.~\ref{fig:Equilibrium}(a), where we compare the fit results with the scaling dimensions extracted from the CFT. The observed agreement corroborates our identification of the renormalized coupling $\beta^2$.
In App.~\ref{sec:AppsG_compare}, we further investigate the impact of next-nearest-neighbor interaction terms on the  value of $\beta^2$, and show that they lead to a reduced renormalization of the coupling, as already observed for the Luttinger parameter $K$.

Since the sG model has no further coupling parameters, we are left with the renormalization scale, which is set by the bare mass $M_0'$ on the microscopic lattice. Given the observed renormalization of $\beta^2$ due to higher derivatives from the term $\propto \cos \left[\beta \left(\hat{\varphi}_i - \hat{\varphi}_{j}\right)/\kappa\right]$ [see App.~\ref{sec:AppsG_renormbeta}], we depart from the microscopic identification given in Eq.~\eqref{eq:parameter_ident}. Instead, we fix $M_0'$ through the formally equivalent relation
\begin{align}
\left(M_0'\right)^2 = \left(\beta'\right)^4 \frac{\lambda'_\kappa}{\chi \kappa^2} \;,
\label{eq:M0_renormalization}
\end{align}
which is, however, independent of the microscopic parameter $V'_\text{nn}$. We obtain an adjusted prediction of $M_0'$ by inserting the value of $\beta' = \beta$ from the previously extracted $\beta^2$ [either through $\beta^2 = \pi \kappa^2  K$ or Eq.~\eqref{eq:sG_beta_fit}]. As shown in Fig.~\ref{fig:Fig2}(c), the numerically obtained mass gap $\Delta E$ as a function of mass scale set by this procedure shows excellent agreement with the analytical predictions from Eqs.~\eqref{eq:solitonmass} and \eqref{eq:breathermass}. This clearly indicates that our numerical results have reached the scaling limit and indeed realize the sG QFT with correctly extracted parameters $M_0'$ and $\beta^2$. Further details about the matching of the mass scale can be found in App.~\ref{sec:AppsG_renormM0}.

We also calculate the ground-state energy density $E_0$ with respect to the CFT, which provides an independent check of the procedure to match $\beta^2$ and $M_0'$. As shown in Fig.~\ref{fig:Equilibrium}(b) our simulations are in perfect agreement with the prediction of Eq.~\eqref{eq:gsenergydensity} at small $M_0'$. 

Our results for the vertex operators [see Fig.~\ref{fig:Equilibrium}(c)] are also consistent with Eq.~\eqref{eq:vertextheory} when taking into account a rescaling of the absolute value.
We attribute this to an operator renormalization of the lattice regularization, i.e.,  the identification in Eq.~\eqref{eq:vertex_opertor_identification} requires a multiplicative correction with a factor $Z_\beta \overset{\beta^2 \rightarrow 0}{\longrightarrow} 1$ that depends on $\beta$, but is independent of $M_0'$ in the scaling regime [see the inset of Fig.~\ref{fig:Equilibrium}(c)]. 

So far, we have compared our numerics to exact results of the sG QFT. Although the model is integrable, not all quantities are known exactly. In particular, there is an unproven conjecture about the expectation values of $\langle e^{i n \beta \hat{\varphi}}\rangle$ for generic $n$ [see App.~\ref{sec:AppsG_conjecture} for the explicit formula]. Here, we test this conjecture in our simulations for integer $n=1, \dots, 6$ [see Fig.~\ref{fig:Equilibrium}(d)]. We find quantitative agreement with the conjectured functional form when taking into account a multiplicative factor $Z_{\beta}^{(n)} = \left(Z_\beta\right)^{n^2}$, which relates the higher powers of the vertex operators to their corresponding lattice regularization $\propto \left(\hat{J}_+ \right)^n$.

Altogether, our findings confirm the expected simulability of the sG model regularized with large spins on a finite lattice, although we are restricted to a coupling region close to the classical limit $\beta^2\rightarrow 0$.
Similarly to the renormalization of the parameter $\beta^2$, we attribute this restriction to undesired higher-derivatives terms in the Taylor expansion of the term $ \cos[\beta(\hat{\varphi}_i-\hat{\varphi}_j)/\kappa]$ [see App.~\ref{sec:AppsG_renormbeta}]. 
While the accessible couplings for $\kappa=1$ are limited to the massive regime with $\beta^2 \ll 4\pi$, the comparison with our results for $\kappa=2$ [see Fig.~\ref{fig:Equilibrium}(a)] also illustrates that the accessible region can be systematically enlarged by increasing $\kappa$. 
Specifically, according to Eq.~\eqref{eq:parameter_ident}, for $\kappa=5$ we expect to reach the sG model with large $\beta^2$ up to $\beta^2 \gtrsim 8\pi$. 
While we have not investigated this regime classically because the extrapolations for larger $\kappa\gg 1$ require simulations at increasingly large spin length $J$, an experimental implementation could nevertheless probe this regime.

\section{Time evolution of solitons\label{sec:sG_solitons}}
We now turn to real-time dynamics of excitations in the sG model, first focusing on individual semi-classical soliton wave-packets. In Sec.~\ref{sec:classical_solitons}, we briefly review their description in the classical limit and then discuss how to prepare them in our regularized model (Sec.~\ref{sec:soliton_preparation}). Quenching away from the classical limit (Sec.~\ref{sec:soliton_quench}), we find that the corresponding localized initial states are long-lived, but eventually spread out due to quantum fluctuations.

\subsection{Individual solitons in the classical limit\label{sec:classical_solitons}}
In the limit $\beta^2 \rightarrow 0$ the quantum sG model reduces to a classical field theory. 
Elementary solutions to the classical field equation $\partial_t^2 \vphi - \partial_x^2 \vphi + (M_0^2/\beta) \sin{(\beta \vphi)}=0$, so-called solitons, are given by
\begin{align}
    \vphi_\text{s}(x,t)=\frac{4}{\beta}\arctan{\left(e^{\gamma M_0 (x-v t + \delta)}\right)}\;.
    \label{eq:classicalsoliton}
\end{align}
They move with constant speed $v$ and unchanged shape, with $\gamma=(1-v^2)^{-1/2}$ the relativistic Lorentz factor and $\delta$ a displacement in position space~\footnote{Note that we set the speed of light $c=1$.}. The corresponding conjugate momentum reads
\begin{equation}
    \pi_\text{s}(x,t)=\partial_t \vphi_\text{s}(x,t)=-\frac{2\gamma M_0 v }{\beta}\frac{1}{\cosh{(\gamma M_0 (x-v t + \delta))}} \;.
    \label{eq:momentum_class_sol}
\end{equation}

The classical soliton is a topological excitation; the topological charge $Q(t)=\int_{-\infty}^{+\infty} {\rm d}x \rho(x,t)$, which counts the number of solitons, is a conserved quantity. Here $\rho(x,t)=\beta \partial_x \vphi(x,t)/2\pi$ is the corresponding charge density, which for the classical soliton $\vphi_s(x,t)$ is given by
\begin{align}
    \rho_\text{s}(x,t)&=\frac{\gamma M_0}{\pi}\frac{1}{\cosh(\gamma M_0(x-v t + \delta))} \;.
    \label{eq:topo_charge_class_sol}
\end{align}
The sG model also features multi-soliton configurations, including bound states (``breathers'') as well as scattering solutions, which we address further below in Sec.~\ref{sec:sG_scattering}.

\subsection{Preparing a semi-classical soliton\label{sec:soliton_preparation}}
We are interested in the fate of semi-classical solitons when evolving with the quantum many-body Hamiltonian at finite $\beta^2$. More precisely, we aim to prepare a state matching the classical phase profile $\langle \hat{\varphi} \rangle = \vphi_\text{s}$, but including quantum fluctuations. 

\begin{figure}
    \centering
    \includegraphics[width=\linewidth]{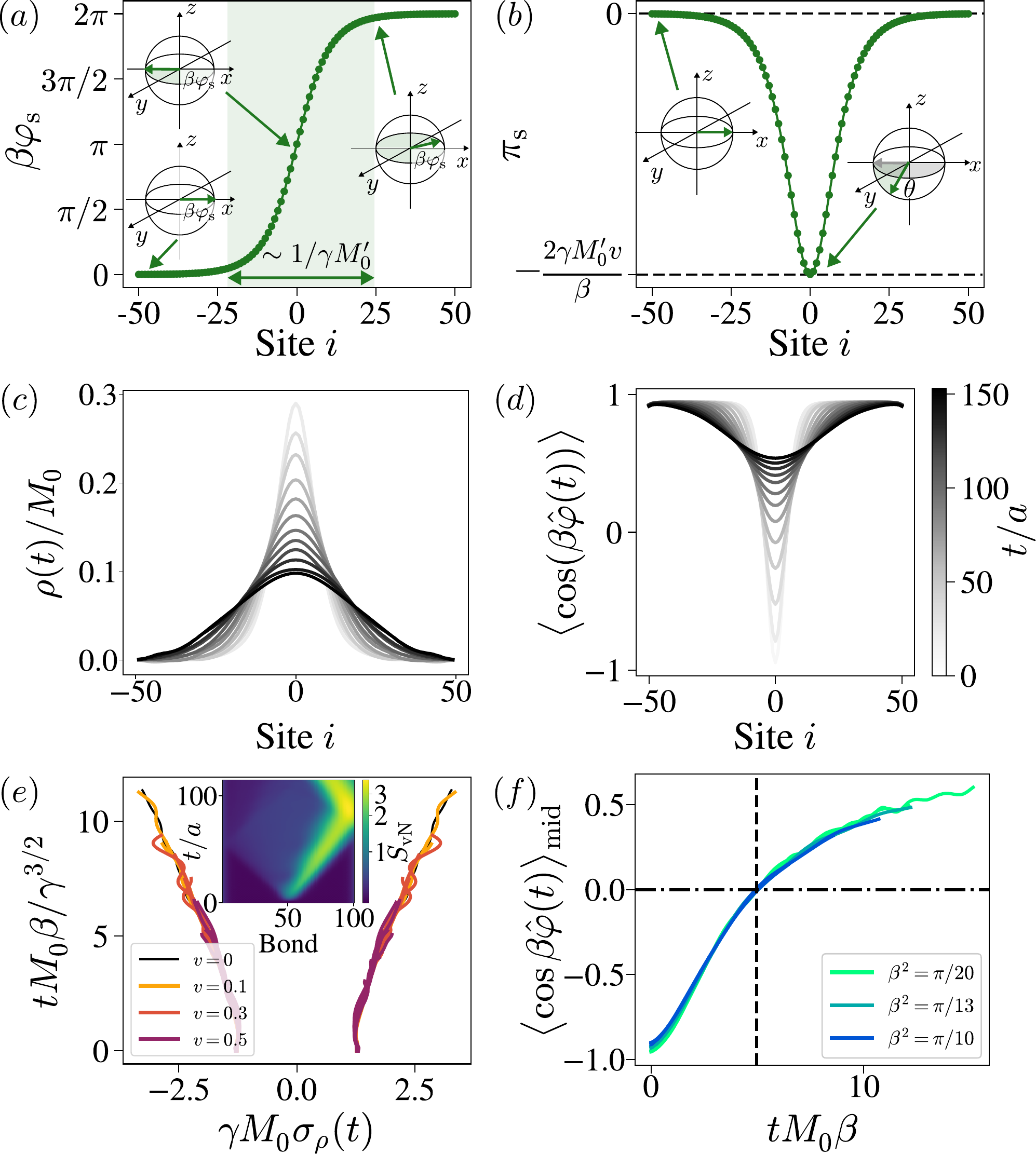}
    \caption{\textbf{Soliton preparation and dynamics.} (a)-(b) Preparation of the semi-classical soliton state: (a) we first apply site-dependent $z$-rotations with an angle $\beta \vphi_\mathrm{s}^{(i)}$ (green shaded area) to rotate in the equatorial plane the spins (initially pointing in $+x$-direction). (b) We then apply rotations around axes in the equatorial plane with a site-dependent angle $\theta^{(i)}$ (green shaded area) to imprint the momentum profile of a moving soliton.
    (c)  resp. (d) Charge density profile $\rho(t)/M_0$  resp. $\langle \cos{(\beta \opphi)} \rangle$  as a function of time for $N=101, J=20, \beta^2=\pi/20, M_0'=0.2$, and $v=0$.
    (e) Collapse of the width $\sigma_\rho(t)$ extracted from a Gaussian fit of the charge distribution $\rho$, by rescaling space and time as discussed in the main text. Here we  plot $\sigma_\rho(t)$
    for several $\beta^2, M_0'$ and $v=0$. (Inset) Von Neumann entanglement entropy $S_\text{vN}$ as a function of time $t$ and the bond that defines the entanglement bipartition for $N=101, J=20, \beta^2=\pi/20, M_0'=0.2$, and $v=0.5$. 
    (f) Collapse of the real part of the vertex operator at half-chain $\langle \cos(\beta \opphi(t))\rangle_\text{mid}$ for several $\beta^2, M_0'$ at $v=0$. The vertical dashed line indicates the ``decay" time $t_\text{d}$ of the initial soliton, defined by the zero crossing of $\langle \cos(\beta \opphi(t))\rangle_\text{mid}$ (dash-dotted line).}
    \label{fig:TimeEvolFig}
\end{figure}

To achieve the desired initial state, we first calculate the ground state of the sG Hamiltonian $\hat{H}_\text{sG}$ at finite $\beta^2$. For our microscopic implementation (at $\kappa=1$), this state corresponds to a correlated chain of spins pointing in $+x$-direction, as $\langle\beta \opphi \rangle = \arg{(\langle \hat{J}_+\rangle/\sqrt{J(J+1)})}=0$. For a static soliton with $v=0$ and $\gamma=1$, it is now sufficient to imprint the desired phase profile by applying site-dependent $z$-rotations $\hat{U}_\text{p}^{(i)}=\exp{\left( -i \beta \vphi_\text{s}^{(i)}\hat{J}_z^{(i)}\right)}$,
with $\vphi_\text{s}^{(i)}=\vphi_\text{s}(x=i \cdot a, t=0)$ the value of the phase profile at site $i$ [see Fig.~\ref{fig:TimeEvolFig}(a)].

To prepare a moving soliton of finite speed $v\neq 0$, we instead first imprint the phase profile and afterwards also the corresponding momentum $\pi_\text{s}$.
We achieve this by applying site-dependent gates $\hat{U}_\text{m}^{(i)}=\exp{( -i \theta^{(i)} \hat{J}_\varphi^{(i)})}$,
with  $ \hat{J}_\varphi^{(i)} = \sin{(\beta \vphi_\text{s}^{(i)})} \hat{J}_x^{(i)} -\cos{(\beta \vphi_\text{s}^{(i)})}\hat{J}_y^{(i)}$ and angle $\theta^{(i)}=\arcsin{(\pi^{(i)}_\text{s} \kappa a/(\beta |\langle \hat{J}_+^{(i)}\rangle|))}$, with $|\langle \hat{J}_+^{(i)}\rangle|$ the ``classical'' length of the spin [see Fig.~\ref{fig:TimeEvolFig}(b)]. Here, $\pi^{(i)}_\text{s}$ is the conjugate momentum at site $i$ and time $t=0$, i.e.,  $\pi_\text{s}^{(i)}=\pi_\text{s}(x=i\cdot a, t=0)$, and the applied gates correspond to a rotation with angle $\theta^{(i)}$ around an axis $w=(\sin{(\beta \vphi_\text{s}^{(i)})}, -\cos{(\beta \vphi_\text{s}^{(i)})})$ in the equatorial plane, orthogonal to the direction in which the expectation value of the spin points. 

From here on and in the rest of this work, we perform simulations deep in the gapped phase, such that the values of $\beta^2, M_0'$ can be determined from the microscopic parameters $\chi, \lambda_\kappa', V_\text{nn}'$, and $\kappa=1$ according to Eq.~\eqref{eq:parameter_ident} \footnote{We estimate the maximal relative error in neglecting the $\beta^2$-renormalization to $\approx 9\%$.}.
We also note that the non-commutativity of the two sets of unitaries induces a small systematic effect on the desired profiles in the central region, of size  proportional to $(\gamma M_0')^{-1}$, where the phase profile connects two degenerate minima of the cosine potential [for more details see App.~\ref{sec:AppSol_preparation}.] We notice that the magnitude of this effect increases with $M_0'$ and $v$, while decreasing with $\beta^2$, which is consistent with the dependence of the momentum profile on the sG parameters.
We find that the expectation value of $\hat{J}_z$ displays instead a perfect agreement with the theoretical prediction for $\pi_\text{s}$.

We emphasize that after applying the gates, in both cases the resulting state displays the same vertex-vertex correlation function (upon accounting for the correct phase of the spin operators $\hat{J}_\pm\rightarrow \hat{J}_\pm e^{\mp i\beta \vphi_\text{s}}$), $\langle \opphi^{(i)} \opphi^{(j)} \rangle \propto  \langle \hat{J}_+^{(i)} \hat{J}_-^{(j)} \rangle_\text{c} = \langle \hat{J}_+^{(i)} \hat{J}_-^{(j)} \rangle - \langle \hat{J}_+^{(i)} \rangle \langle \hat{J}_-^{(j)} \rangle$, as the original ground state of the lattice Hamiltonian. In other words, we have effectively changed the one-point functions -- allowing us to interpret the resulting state as a semi-classical soliton -- while the two-point correlators remain untouched and display the correct vacuum quantum fluctuations. Note that this preparation protocol assumes local programmability of the quantum simulator, similar to Ref.~\cite{wybo2023preparing}.
We further comment on experimentally achievable numerical values of the required parameters in App.~\ref{sec:AppSol_StaticEvol}.

\subsection{Evolving a semi-classical soliton\label{sec:soliton_quench}}
We continue by investigating the time evolution of the semi-classical soliton state for both $v=0$ and $v\neq0$ by evolving the system with the sG Hamiltonian Eq.~\eqref{eq:QFTHamiltonian} for $d=1$. Numerically, the time evolution is performed using time-evolving block decimation (TEBD), i.e.,  applying discrete Trotter steps and updating the MPS representation of the state. 
We find that TEBD with a moderate maximal bond dimension of $\chi=128$, truncation cutoff $10^{-9}$, and time step $\tau=0.01$ leads to quantitatively converged results 
[see the discussion in App.~\ref{sec:AppSol_StaticEvol} resp.~\ref{sec:AppSol_MovingEvol} for more details about static resp.~moving soliton simulations]. The following results are obtained from spin lengths $J\in\{ 16, 18, 20\}$ for a sufficiently large system size of $N=101$. 
In the presented parameter regimes our observables are essentially converged in $N$, but require an extrapolation in $J$ to retrieve quantitative results. 

We track the evolution of the initial soliton by monitoring the topological charge density $\rho=\langle \hat\rho \rangle$, as well as the expectation value of the interaction potential $1-\langle\cos{(\beta \opphi)\rangle}$.
As time progresses, we find that both observables are increasingly  well-approximated by an appropriately normalized Gaussian function characterized by a mean position $\bar{x}$ and a width $\sigma$, 
measured in lattice units
[see App.~\ref{sec:AppSolitonPrepTimeevol}], which we discuss next. 

\subsubsection{Mean position: classical motion}

First, we compute the instantaneous position $\bar{x}(t)$ 
\cite{wybo2022quantum}
\begin{equation}
    \bar{x}(t)=\frac{a}{\mathcal{N}}\sum_{j=1}^N j\bar{\varepsilon}(j,t)^2+\frac{a}{2}, \quad \mathcal{N}=\sum_{j=1}^N \bar{\varepsilon}(j,t)^2,
    \label{eq:quasi-particle_position}
\end{equation}
with $\bar{\varepsilon}(j,t)$ the excess energy density at the bond between sites $j$ and $j+1$ and time $t$. 
In $\bar{\varepsilon}(j,t)$ we subtract ground-state contributions from the bare energy density $\varepsilon(j,t)$. 
From $\bar{x}(t)$ we extract the speed of the soliton by a linear fit, which is in agreement with the imposed classical velocity $v$ within $3.5\%$, corroborating our state preparation protocol.
We attribute the deviation, which grows with increasing $\beta^2$, to the imposed classical momentum profile determined from the microscopic parameter identification rather than the renormalized $\beta^2$.

\subsubsection{Delocalization: impact of quantum fluctuations}
On top of the ``classical" motion of the mean soliton position,
we find that both the charge density $\rho$ and the cosine term $1-\langle\cos{(\beta \opphi)}\rangle$ delocalize over time. This is quantified by the width $\sigma(t)$ of a Gaussian fit to both observables.
Fig.~\ref{fig:TimeEvolFig}(e) shows the spreading of the width $\sigma_\rho(t)$ of the charge density $\rho$ over time. We find the evolution to be self-similar, which is revealed by rescaling the axes $\sigma_\rho \rightarrow \sigma_\rho \cdot \gamma M_0$ and $t \rightarrow t \cdot \beta M_0/\gamma^{3/2}$, such that the trajectories of several $\beta^2, M_0',$ and $v$ collapse.
Here we only present results for moving solitons up to a speed of $v=0.5$. For larger velocities ($v\geq 0.7$), our state preparation creates high-energy excitations at the lattice scale, which propagate with $v\approx c=1$, and interfere with the targeted soliton dynamics. We note that a small admixtures of these excitations is visible for all velocities, but their effect is negligible for small $v$.

To gain a better understanding of the spreading, consider the static soliton case with $v=0, \gamma=1$.
When approximating the topological charge density $\rho$ with a Gaussian, we obtain an initial width $\sigma_\rho(0)=\sigma_{\rho,0}\approx 1.237/ M_0$,
which we observe to evolve as 
\begin{align}
    \sigma_\rho(t)= \sqrt{\sigma_{\rho,0}^2 + (v_\text{spr} \cdot t)^2} \;.
\end{align}
Here, $v_\text{spr} \approx 0.273 \cdot \beta$ is the spreading velocity, which we find to exhibit a dominant dependence on $\beta$, with a prefactor estimated from the smallest $M_0'=0.15$ simulated.
The dynamics of $1-\langle\cos{(\beta \opphi)}\rangle$ shows a similar behaviour with $\sigma_\text{cos}(0)=\sigma_{\text{cos},0} \approx 0.794 / M_0$. Given this evolution, together with the 
(approximate) conservation of the spatial integral of both quantities in time, the maximum of $1-\langle\cos{(\beta \opphi)}\rangle$ -- which coincides with the value in the middle of the chain -- decreases over time [see Fig.~\ref{fig:TimeEvolFig}(f)], and eventually becomes smaller than 1 at some time $t_\text{d}$. 
In other words, after $t_\text{d}$ the expectation value  $\langle \cos{(\beta \opphi)}\rangle=\Re{\langle \hat{J}_+ \rangle /\sqrt{J(J+1)}}$ assumes only positive values in the middle of the chain and the extracted classical phase profile $\langle \beta \opphi \rangle = \arg{\langle \hat{J}_+ \rangle}$ gets restricted to the interval $[-\pi/2,\pi/2]$. This is incompatible with a classical soliton profile $\beta \varphi_\text{s}\in[0,2\pi]$ that is expected to smoothly connect the vacua at $0$ and $2\pi$. In this sense, the interpretation as a single classical soliton becomes inapplicable and $t_\text{d}\approx 4.96/(M_0 \beta)$ indicates a time where the profile has ``decayed" into fluctuations.

To make this interpretation more precise, note that the ``decay" time $t_\text{d}\propto \beta^{-1}$ depends on $\beta$, which measures the overall strength of quantum fluctuations. 
By construction, our chosen initial state indeed exhibits the classical soliton profile (formally valid only in the classical limit $\beta^2 \rightarrow 0$) as a non-vanishing one-point function, while retaining the two-point correlations of the ground state of the quantum sG Hamiltonian prepared at a small, but finite $\beta^2$.
Therefore, we expect the impact of quantum fluctuations to increase with $\beta$, leading to a faster ``decay" of the classical soliton profile. 

We emphasize that the observed ``decay" only concerns the classical soliton profile, while the topological charge $Q(t)$ 
is conserved at all times. The initial excitation therefore does not vanish. 
In view of the duality between sine-Gordon and massive Thirring model~\cite{coleman1975quantum}, where a single  (anti-)soliton corresponds to an (anti-)fermion and the topological charge density $\rho$ can be identified with the fermion number density, our results are consistent with a delocalizing \emph{superposition} of single-fermion states, i.e., the spreading of a wave-packet.

\subsubsection{Semi-classical model}
We conclude this section with a semi-classical model that accurately captures the observed dynamics, namely a truncated-Wigner approximation~\cite{polkovnikov2010phase}, restricted to the single-soliton sector and with appropriately chosen initial conditions.
In this model, we consider a wave-packet of single solitons, whose velocities $v$ and position displacements $\delta$ are sampled according to Gaussian functions.
Observables are evaluated along the individual classical soliton trajectories, with an average compatible with the spreading of a wave-packet.

More precisely, we approximate quantum expectation values  of an operator $\mathcal{O}$ at evolution time $t$ as \begin{align}
\langle \mathcal{O}(t) \rangle \approx \int D\varphi_\text{s}(0) D\pi_s(0) \, W[\varphi_\text{s}(0), \pi_s(0)] \, \mathcal{O}_\text{cl} \left[\varphi_\text{s}(t)\right] \;.
\label{eq:semi-classical_evoleq}
\end{align}
Here, $\mathcal{O}_\text{cl}\left[\varphi_\text{s}(t)\right] $ represents the operator evaluated along the classical time-dependent single-soliton trajectory $\varphi_\text{s}(t)$, i.e., the solution [Eq.~\eqref{eq:classicalsoliton}] of the classical equations of motion. The distribution $W$  instead captures the different initial conditions $\varphi_\text{s}(0)$ and $\pi_s(0) = \partial_t \varphi_\text{s}(0)$. That is, the RHS of Eq.~\eqref{eq:semi-classical_evoleq} implements a truncated-Wigner approximation as a statistical average of observables evaluated on trajectories of classical single-soliton solutions  $\varphi_\text{s}(t)$ and averaged over the initial conditions distribution $W$ in the single-soliton sector. 

We choose the distribution $W$  as two independent normalized Gaussian distributions in $\delta$ and $v$, centered at $\mu_\delta~=~\mu_v~=~0$ and with widths $\sigma_\delta, \sigma_v$, which account for the initial quantum fluctuations. 
To obtain analytical results, we further approximate both observables $\rho_\text{s}(x,t)$ and $1-\cos{(\beta \varphi_\text{s}(x,t))}$ at $t=0$ with a Gaussian distribution of width $\sigma_0$. 
Assuming $v~\ll~1, \gamma~\approx~1$, these choices lead to evolved observables that remain well-approximated by a Gaussian distribution with a time-dependent width $\sigma(t)=\sqrt{(\sigma_0^2 + \sigma_\delta^2) + (\sigma_v \cdot t)^2}$ [for more details see  App.~\ref{sec:AppSolitonPrepTimeevol}].

This model thus reproduces the numerically observed spreading of the width upon identifying $\sigma_v$ with $v_\text{spr}$, and upon assuming that quantum fluctuations are proportional to $\beta$. 
In App.~\ref{sec:AppSol_Fluctuations} we further show that the direct proportionality $\sigma_v \propto \beta$ can be attributed to the $zz$-correlations of the quantum ground state.
The additional $\sigma_\delta$ term in the time-dependent width $\sigma(t)$ can be neglected, as it would simply modify the static profile of the observables, which we instead impose to agree with the classical predictions. 

The case $v\neq 0$ can be understood analogously to the static $v=0$ case by interpreting the moving soliton as being static in an inertial frame of reference, which is moving with velocity $v$ with respect to the observer [for further details see App.~\ref{sec:AppSol_semi-classical}]. Thus, upon taking into account the relativistic Lorentz contraction of space and dilation of time, we obtain additional Lorentz factors $\gamma$. Moreover, we find that the standard deviation of the velocity for the moving soliton is given by  $\sigma_v\propto \beta/\sqrt{\gamma}$. Altogether, our semi-classical model also reproduces the $\gamma$-dependence of the rescaled space and time axes in Fig.~\ref{fig:TimeEvolFig}(e), giving further credence to the interpretation of the initial state as a single-soliton wave-packet with the desired velocity.

\section{Soliton-antisoliton scattering\label{sec:sG_scattering}}

In the following section, we investigate the real-time dynamics of a counter-propagating soliton-antisoliton pair. 
After briefly reviewing the expected scattering shifts, we present our numerical results in quantitative agreement with the theoretical predictions.

\subsection{Quantum and classical scattering shifts}
Beyond a single soliton, the classical sine-Gordon model admits several other exact solutions, including the antisoliton $\vphis$,
with topological charge $Q=-1$, as well as a scattering soliton-antisoliton pair, with topological charge $Q=0$. 
The latter is given by
 \begin{align}
     \vphi(x,t)&=-\frac{4}{\beta}\arctan{\left( \frac{\sinh( M_0 t \sinh\theta)}{\tanh(\theta)\cosh{(M_0 x \cosh\theta)}}\right)}
     \nonumber
     \\ &=-\frac{4}{\beta}\arctan{\left( \frac{\sinh(\gamma M_0 v t)}{v\cosh{(\gamma M_0 x)}}\right)},
     \label{eq:pairsgeq}
 \end{align}
where $\theta$ is the rapidity ($\cosh\theta=\gamma, \tanh\theta=v$).
In the limit of infinitely late or early time $t \rightarrow \pm \infty$, the pair solution $\vphi$ reduces to an independent soliton and antisoliton~\cite{koch2023exact}, which upon scattering are displaced by a position shift
\begin{align}
\delta x_\text{cl}=\frac{\phi_\text{cl}(2\theta)}{M_\text{cl} \cosh{\theta}},
\label{eq:classposshift}
\end{align}
with classical soliton mass $M_\text{cl}=8\gamma M_0/\beta^2$~\footnote{This classical mass $M_\text{cl}$ can be derived as the $\beta^2 \rightarrow 0$ limit of Eq.~\eqref{eq:solitonmass}.} and classical scattering phase shift 
\begin{align}
    \phi_\text{cl}(\theta)=\frac{8}{\beta^2}\log\left( \frac{\cosh\theta +1 }{\cosh \theta -1}\right).
\end{align}

In the quantum sG model, asymptotic scattering of a soliton and antisoliton with rapidities $\pm\theta$ leads to a phase shift $\phi(\theta)=-i \log{S(\theta)}$ 
determined by the scattering matrix $S(\theta)$ \cite{wybo2022quantum}. 
In the limit $\beta^2\rightarrow0$, the scattering becomes completely transmissive and we restrict our attention to the transmissive part $S_T (\theta)$ \cite{koch2023exact}
\begin{multline}
    S_T(\theta)=-\frac{\sinh{(\xi^{-1}\theta)}}{\sinh{((i\pi - \theta)\xi^{-1})}}\times \\
    \exp{\left[-i\int_0^\infty \frac{{\rm d}t}{t}\frac{\sinh(\pi t(1-\xi)/2)}{\sinh{(\pi\xi t/2)}\cosh{(\pi t/2)}}\sin(\theta t)\right]},
\end{multline}
with $\xi=\beta^2/(8\pi - \beta^2)$.
The resulting position shift $\delta x$ can be then determined according to~\cite{doyon2018soliton}
\begin{equation}
    \delta x(\theta_\text{L}, \theta_\text{R})=-\frac{\partial_\theta \phi(\theta)}{M \cosh{\theta_\text{L}}}\Big|_{\theta = \theta_\text{L}-\theta_\text{R}},
    \label{eq:positionshift}
\end{equation}
with $M$ the mass of the soliton, and $\theta_\text{L}$ resp. $\theta_\text{R}$ the rapidity of the incoming quasi-particle from the left resp. right~\footnote{
One can check that the quantum position shift of a soliton-antisoliton scattering with $\theta_\text{L}~=~-\theta_\text{R}~=~\theta$ reduces to the classical prediction $\delta x~\rightarrow~\delta x_\text{cl}$ by taking the classical limit $\beta^2\rightarrow 0$.}.

\subsection{Numerical results}
In analogy to the moving soliton, we prepare an initial semi-classical soliton-antisoliton pair by applying on top of the ground state of the sG Hamiltonian two sets of gates: one to imprint the phase profile and one to imprint the momentum. We then evolve the system with the sG Hamiltonian.

\begin{figure}[t!]
    \centering
    \includegraphics[width=\linewidth]{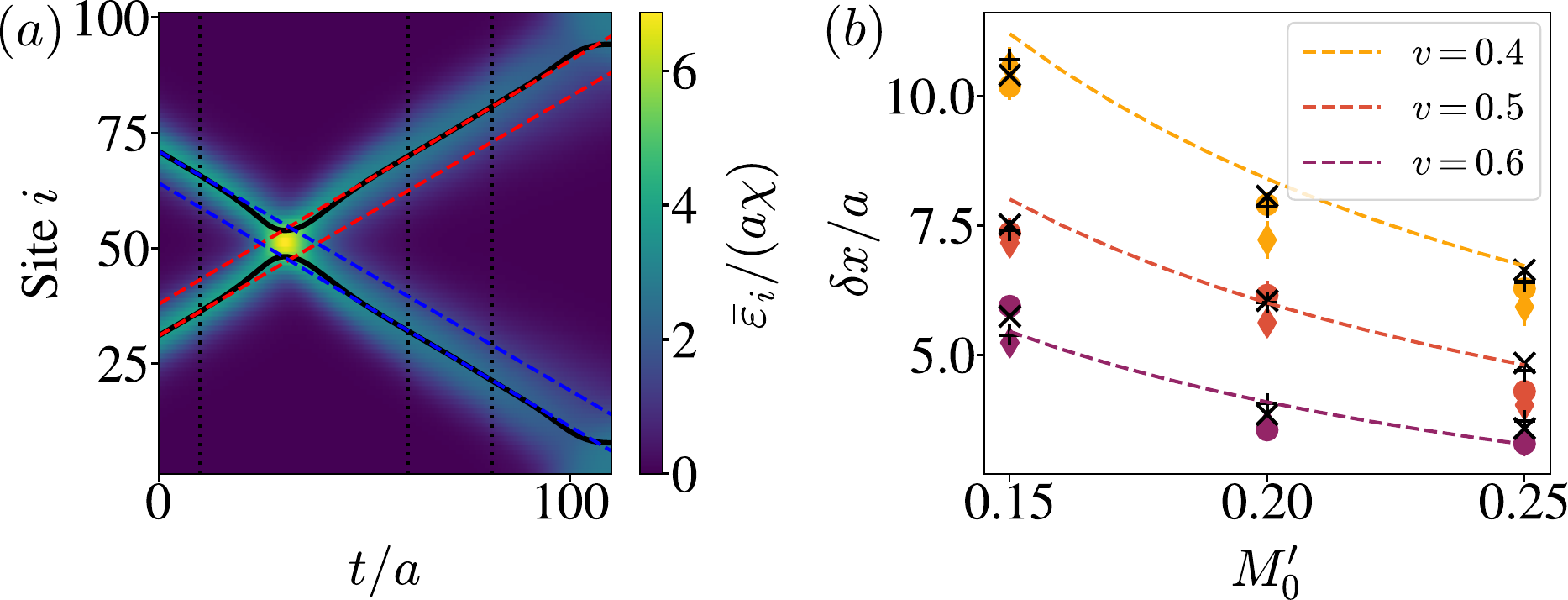}
    \caption{\textbf{Soliton-antisoliton scattering.} 
    (a) Heatmap of the excess lattice energy density $\bar{\varepsilon}_i$ as a function of lattice site $i$ and time $t$, with quasi-particle trajectories indicated by the black solid lines. To estimate the scattering position shift, we fit the trajectories with a linear function in the pre-/post-collision time intervals delimited by the vertical dashed black lines. The resulting fits are represented by the dashed red line (soliton) and the blue line (antisoliton). 
    The simulation parameters are $N=101, J=20, \beta^2=\pi/13, M_0'=0.15, v=0.5, \Delta x_\ss/a =40$.
    (b) Scattering position shift $\delta x/a$ as a function of $M_0'$ for $\beta^2=\pi/13$. 
    The circles resp. diamonds are the numerical results for an initial separation $\Delta x_\ss/a=40$ resp. $\Delta x_\ss/a=50$ sites. The dashed color lines represent the classical position shift according to Eq.~\eqref{eq:classposshift}. The black crosses resp. plus represent the theoretical predictions of Eq.~\eqref{eq:positionshift} for $\Delta x_\ss/a=40$ resp. $\Delta x_\ss/a=50$, using as input the renormalized $\beta^2, M_0'$ values from the equilibrium analysis and the effective velocity extracted from the trajectories after the scattering event. The numerical results are extrapolated to the continuum limit with the procedure described in Sec.~\ref{sec:general_scalars}.}
    \label{fig:ScatteringFig}
\end{figure}

To determine the position shift in our simulations, we first extract the time-dependent position of each \mbox{(anti-)}soliton from the excess lattice energy density $\bar{\varepsilon}$
\begin{align}
\bar{x}_\text{L/R}(t)=\frac{a}{\mathcal{N}}\sum_{j\in \text{L/R}} j \bar{\varepsilon}(j,t)^2 +\frac{a}{2}, \quad \mathcal{N}=\sum_{j\in \text{L/R}}  \bar{\varepsilon}(j,t)^2 , 
\end{align}
with $\bar{x}_\text{L}$ resp. $\bar{x}_\text{R}$ the (anti-)soliton position in the left resp. right half of the chain.
We find that this approach captures well the (anti-)soliton positions as localized quasi-particles for the essentially free evolution before and after the scattering event. Conversely, we can identify the scattering region in space-time as a single region of high energy density when soliton and antisoliton cannot be distinguished [see Fig.~\ref{fig:ScatteringFig}(a)].
We thus fit the trajectories in the regions of linear motion and extrapolate the position shift $\delta x$ as the distance between soliton and antisoliton at the scattering time, i.e.,  the time at which the linear fits cross [see App.~\ref{sec:AppScattering} for more details, including a systematic error analysis of this method].

The resulting value of the position shift $\delta x$ extrapolated to the continuum limit is shown in Fig.~\ref{fig:ScatteringFig}(b), which exhibits only slight variations (up to $\mathcal{O}(1)$ lattice site) between different initial soliton-antisoliton separations $\Delta_\ss/a=40, 50$. We attribute these deviations to the difficulty in precisely locating the quasi-particles.
When comparing our numerical results to the theoretical quantum prediction $\delta x_\text{theo}=\delta x(\theta, -\theta)$ [Eq.~\eqref{eq:positionshift}], we use the renormalized values of $\beta^2$ and $M_0'$ determined from the equilibrium analysis in Sec.~\ref{sec:sG_vacuum} and the quasi-particle velocities extracted from the trajectories after the scattering event. 
Overall, we find good agreement of the numerically obtained position shifts $\delta x$ with the theoretical predictions. 

\section{ Scattering beyond integrability\label{sec:beyond_sG}}
In this section, we introduce a non-integrable perturbation to the sine-Gordon model and study how this changes the scattering dynamics of the soliton-antisoliton pair. We observe phenomena reminiscent of particle production and plasma oscillations in quantum electrodynamics.

\subsection{Choice and interpretation of the perturbation}
Since the sine-Gordon model is an integrable quantum field theory, analytic predictions are available for a plethora of observables. So far, we showed an agreement between the theoretical predictions and the results of the proposed large-spin lattice implementation.

As a first step toward more non-trivial quantum simulation, we now consider a more general interaction potential of the type discussed in Sec.~\ref{sec:other_QFTs_and_implementations}. To be explicit, we reconsider the scattering of the previous section when perturbing the standard sG Hamiltonian $\hat{H}_\text{sG}$ by an additional cosine term with different frequency,
\begin{equation}
    \hat{H}_\text{sG,p}=\hat{H}_\text{sG}-p\cdot \int {\rm d}x \frac{M_0^2}{\beta^2} \cos{\left(\frac{\beta\opphi}{2}\right)},
    \label{eq:psGHamiltonian}
\end{equation}
with $p$ the strength of the perturbation. 
The corresponding lattice version of the perturbed Hamiltonian $\hat{H}_\text{latt,p}$ could be realized, e.g., in a Rydberg setup using both a microwave coupling with $\kappa=1$ and a ponderomotive coupling with $\kappa=2$ [see App. \ref{sec:AppendixRydbergHamiltonian}]. The microscopic parameter $\lambda'_2$ is then fixed according to  Eq.~\eqref{eq:parameter_ident}, while $\lambda'_1=p\cdot \lambda_2'$.

\begin{figure}
    \centering
        \includegraphics[width=\linewidth]{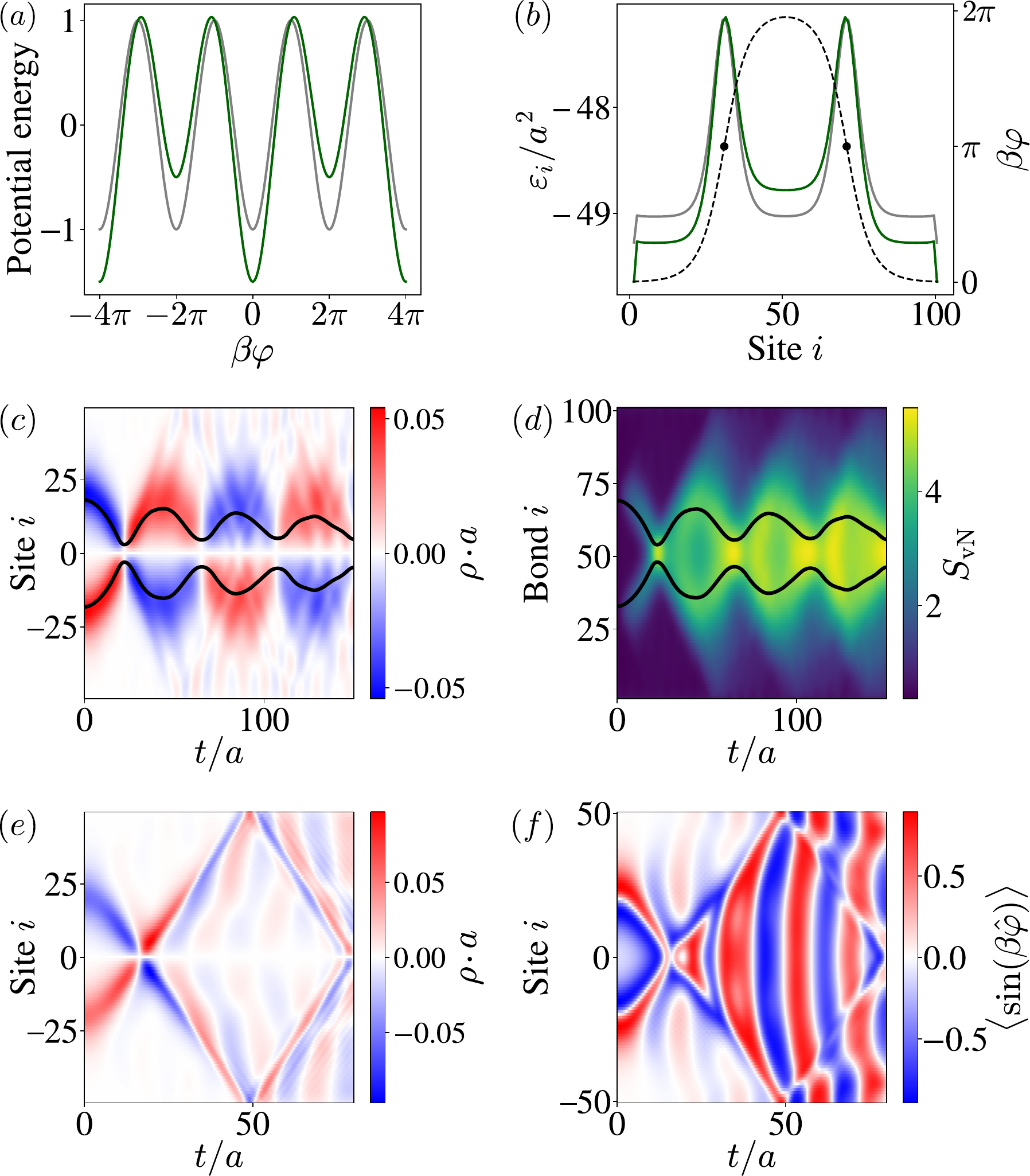}

    \caption{\textbf{Beyond integrability.} (a) Interaction potential $V(\opphi)\beta^2/M_0^2$ of Eq.~\eqref{eq:cosinepotentialperturbed} for $p=0$ (gray line) and $p=0.5$ (green line).  (b) Energy density $\varepsilon_i$ (in sine-Gordon units) of the soliton-antisoliton static pair (dashed black) for the unperturbed $p=0$ (solid gray) and the perturbed $p=0.5$ (solid green). The dashed black line represents the classical soliton-antisoliton profile, while black dots capture the quasi-particle position in the unperturbed case. The simulation parameters are $N=101, J=22, \beta^2=\pi/20, M_0'=0.2, \Delta_\ss/a=40$. (c) resp.~(e) Topological charge density $\rho$ with quasi-particles trajectories (solid black) as a function of time $t$ and site $i$ for $p=0.5$ resp.~$p=1.5$. (d) Von Neumann entanglement entropy, in units of $\log(2)$, with quasi-particles trajectories for $p=0.5$. (f) Expectation value of $\sin(\beta \opphi)$ as a function of time $t$ and site $i$ for $p=1.5$. This observable can be interpreted as an ``electric'' field [see main text].}
    \label{fig:PerturbationFig}
\end{figure}

Consider the profile $\vphiss=\vphi_\text{s} + \vphis$ of a classical static soliton and antisoliton (of the unperturbed model) separated by a distance $\Delta x_\ss$. Due to the presence of the perturbation, the interaction potential $V(\opphi)=-M_0^2/ \beta^2  \cos{(\beta \opphi)} $ becomes
\begin{equation}
    V(\opphi)=-\frac{M_0^2}{\beta^2} \left( \cos{(\beta \opphi)} + p\cdot  \cos{\left(\frac{\beta\opphi}{2}\right)} \right) \;.
\label{eq:cosinepotentialperturbed}
\end{equation} 
This doubles the periodicity [see Fig.~\ref{fig:PerturbationFig}(a)], i.e., the phase profile $\beta\varphi$ is now defined over a period of $4\pi$ (instead of $2\pi$). The perturbed potential further modifies the energy density of the soliton-antisoliton state, which increases in the center of the chain, where $\beta \vphiss=2\pi$, while it is lowered at the ends, where $\beta \vphiss=0$ [see Fig.~\ref{fig:PerturbationFig}(b)]. The total energy of such a configuration therefore grows approximately linearly with the distance $\Delta x_\ss$. In other words, this classical analysis predicts a confining potential for the soliton-antisoliton pair.

From this perspective, we view the double-frequency sine-Gordon model Eq.~\eqref{eq:psGHamiltonian} as a rough approximation of the massive sG model \cite{kruckenhauser2022highdimensional} with a potential $V(\opphi)~\sim~+\frac{M^2}{2} \opphi^2~-\frac{M_0^2}{\beta^2}\cos (\beta \opphi)$. As mentioned in Sec.~\ref{sec:other_QFTs_and_implementations}, the latter is known to be the dual bosonized version of the massive Schwinger model, i.e., quantum electrodynamics in one spatial dimension. Motivated by this relation, we will interpret the scattering dynamics qualitatively in terms of ``electric fields'' and ``electron-positron'' pair production (= excitations with negative/positive charge).

\subsection{Numerical results}
We initialize our system by imprinting the desired phase profile, this time on the ground state of the quantum sine-Gordon Hamiltonian realized with $\kappa=2$. To deal with the new $4\pi$-symmetry, we first check that the spin is pointing in $+x$-direction (if instead it is pointing in $-x$-direction, we rotate it accordingly), before applying the site-dependent $z$-rotations with angle $\beta \vphiss^{(i)}/2$, such that the expectation value of the phase profile $\langle\beta \opphi\rangle=\arg\langle (\hat{J}_+/\sqrt{J(J+1)})^2 \rangle$ coincides with the desired classical solution. We then evolve the prepared semi-classical soliton-antisoliton superposition under the perturbed Hamiltonian $\hat{H}_\text{sG,p}$ for several perturbation strengths $0.5~\leq~p~\leq~ 3.5$.

\begin{figure*}[hbt!]
    \centering
\includegraphics[width=\linewidth]{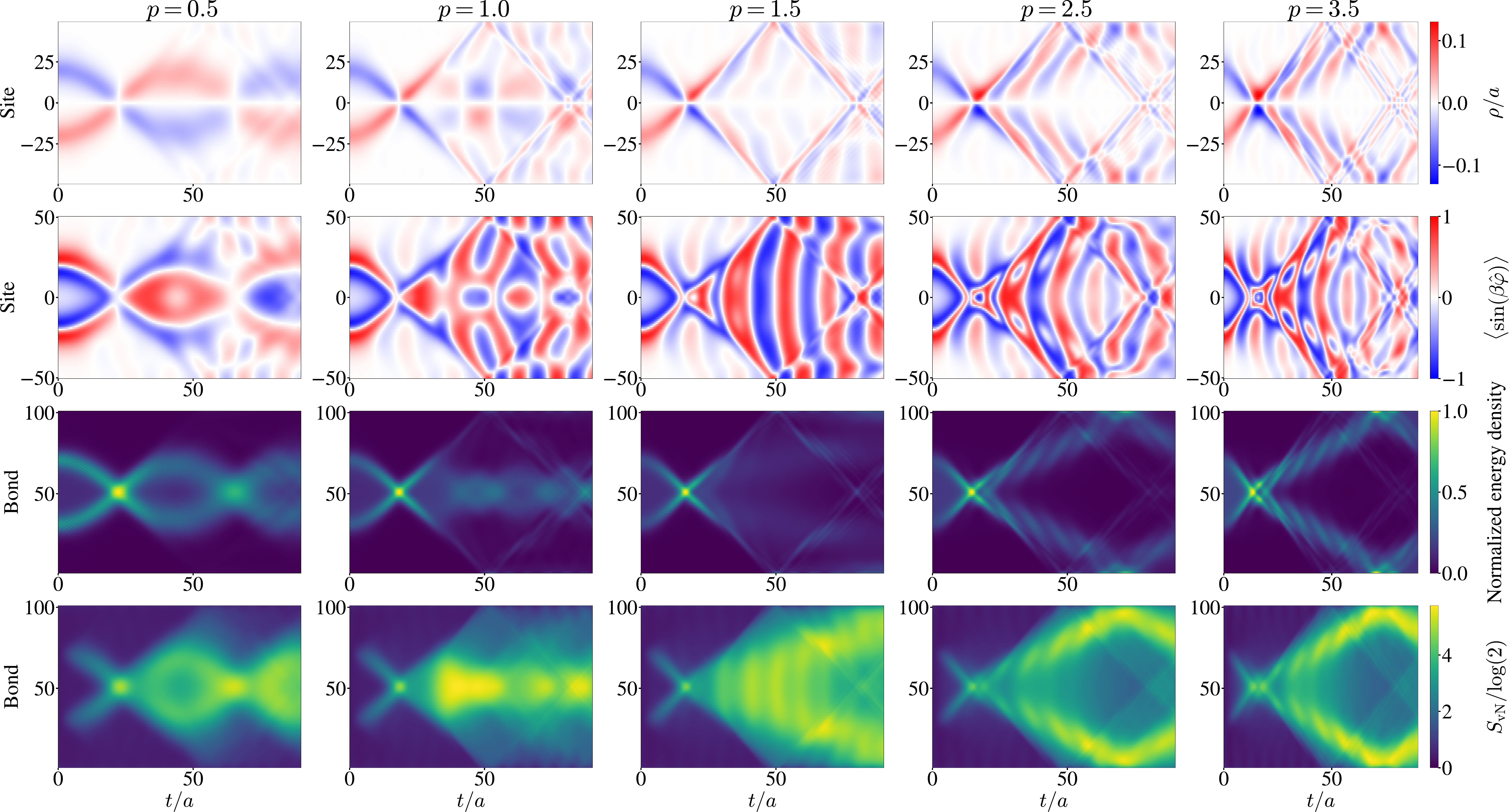}
    \caption{\textbf{Dynamics of a soliton-antisoliton pair in the perturbed sine-Gordon model.} We investigate the dynamics of an initially static soliton and antisoliton for the parameters $J=22, \beta^2=\pi/20, M_0'=0.2, \Delta_\ss/a=40$. The perturbation strength $p$ of the perturbed sine-Gordon model is increasing from left to right: $p=0.5, 1.0, 1.5, 2.5, 3.5$.
    (First row) Topological charge density $\rho$: the red color represents a positive density (soliton), the blue a negative one (antisoliton). (Second row) Expectation value of the operator $\sin(\beta \opphi)$, as a proxy for the ``electric'' field. (Third row) Normalized energy density, obtained by subtracting the ground-state energy density under the perturbed Hamiltonian. For each perturbation strength the color map goes from the respective minimum to the maximal value. (Fourth row) Von Neumann entanglement entropy $S_\text{vN}$ for each bond along the chain.}
\label{fig:PerturbationCompareFig}
\end{figure*}

The resulting dynamics under a relatively small perturbation $p=0.5$ is illustrated in Fig.~\ref{fig:PerturbationFig}(c)-(d). 
As expected due to the confining potential, both quasi-particles of the initially separated pair are accelerated and collide. After the collision, the quasi-particles separate again, but the confinement eventually decelerates their outward motion, such that they eventually invert their direction and scatter again. This process repeats over time and we observe an oscillatory behaviour of repeated scattering events.
We find that the von Neumann entanglement entropy $S_\text{vN}$ mimics these oscillations in time [see Fig.~\ref{fig:PerturbationFig}(d)]. Except for a small amount of free excitations produced at each scattering event, the entanglement reaches an approximately constant value ``inside'' the pair after the first scattering event. In this sense, we can interpret the whole process as the formation of a composite ``meson'' from two initially free wave-packets of one ``electron'' and one ``positron''.

This picture radically changes upon increasing the perturbation strength as shown in Fig.~\ref{fig:PerturbationCompareFig}. At $p\sim 1.5$ the periodic oscillations and collisions disappear. Instead, we observe the production of new particles in the middle of the chain, as evidenced by the topological charge density $\rho$ in Fig.~\ref{fig:PerturbationFig}(e). Moreover, after the scattering event the observable $\langle \sin(\beta \opphi) \rangle$ displays pronounced oscillations in the middle of the chain [see Fig.~\ref{fig:PerturbationFig}(f)]. 
To put this into context, we note that in the Schwinger model strong electric fields lead to repeated particle pair-production events inducing plasma oscillations of the electric field \cite{hebenstreit2013realtime}.
In the dual massive sine-Gordon model, it is precisely the observable $\sin(\beta \opphi)\approx \beta \opphi$ that plays the role of the electric field. 
The observed behaviour is therefore qualitatively similar to a breaking of the initial electric flux string and the expected subsequent plasma oscillations. 
Moreover, the half-chain von Neumann entanglement entropy at $p=1.5$ displays recurring local maxima, which we may interpret as a signature of the creation of new particle pairs. 
Note that for all observables the late-time dynamics is clearly affected by the reflection of excitations at the boundaries of the finite system.

Upon further increasing the perturbation strength $p$ [see Fig.~\ref{fig:PerturbationCompareFig}], our numerics indicates a secondary scattering event right after the initial soliton-antisoliton collision.
We associate this additional event with extra initial particles due to boundary or finite-size effects, and the breakdown of the perturbative interpretation of our initial state as two well-separated wave-packets.
Moreover, at larger perturbations both the von Neumann entanglement entropy and the energy density display periodic maxima along two outgoing branches, in correspondence with the expected charges produced due to string breaking. 
The nearby presence of two oppositely charged excitations along the same branch leads to the emergence of a local ``electric'' field pointing in a direction opposite to the background field from plasma oscillation. This interpretation is also compatible with the visible islands of opposite value in the observable $\sin{(\beta \opphi)}$ for $p\geq 2.5$.

\section{Conclusions and outlook\label{sec:discussion}}
In summary, we have developed an approach to (quantum) simulate scalar quantum field theories (QFTs) based on large-spin models. For the (perturbed) 1D sine-Gordon QFT, we explicitly demonstrated that its large-spin realization quantitatively captures the continuum physics of $(i)$ the vacuum, $(ii)$ the preparation of quasi-particle wave-packets, $(iii)$ their real-time propagation, and $(iv)$ the scattering in the integrable case and beyond. We have discussed how to implement general interaction potentials and how to realize higher-dimensional models with a single real scalar field. An extension to complex or multi-component fields is straightforward, and it would also be interesting to couple to other types of fields, in particular fermions~\cite{rad2024analog}.

Our results form the basis for studying non-equilibrium real-time dynamics of QFTs in the continuum limit with large-spin models. Relevant examples for future work include the decay of the false vacuum~\cite{manovitz2024quantum,lagnese2021false,lagnese2023detecting,batini2024realtime,batini2024particle}, or a more extensive study of scattering in non-integrable models~\cite{milsted2022collisions,jha2024realtime,pichler2016realtime,papaefstathiou2024realtime,bennewitz2024simulating,zhu2024probing}. While such studies are possible in one spatial dimension using classical tensor-network techniques, an experimental implementation of our approach in higher spatial dimensions has the potential to reach uncharted territory.

From an experimental perspective, the required spin lengths are not too large ($J\lesssim20$), making it interesting to further investigate the feasibility of implementing our proposal with Rydberg atom arrays~\cite{kruckenhauser2022highdimensional,claude2024optical}. Such a setup could also make use of quantum state verification protocols~\cite{elben2020crossplatform} to perform a self-consistent convergence test with respect to the spin length within a single experiment. Another important point in this atomic setup is the (ir)relevance of long-range tails of the dipole-dipole interactions for the continuum physics~\cite{defenu2024outofequilibrium}. 

While our approach was motivated by analog quantum simulations, our results are also of interest for digital quantum simulations based on qudits~\cite{wang2020qudits,ringbauer2022universal}, as well as hybrid digital-analog approaches~\cite{davoudi2021simulating,maskara2023programmable}. 
In this context, an important outstanding question is the resource trade-off between the local Hilbert space dimension (spin length or qudit size) and the total system size (i.e. number of spins or qudits). 
In the case of QFTs, this question is especially intriguing because we are only interested in universal continuum physics independent of irrelevant microscopic details. As the numerical results in this work have been achieved with remarkably low bond dimensions, comparable with analogous spin-$1/2$ models, we speculate that the large-spin approach provides some level of encoding, which could imply a natural robustness against errors. This conjecture could be tested in a detailed resource comparison of, e.g., a spin-$1/2$ XXZ realization of the sine-Gordon model vs. the approach presented here, including the effects of experimental imperfections, such as finite coherence or gate fidelities, which we leave for future work.

Throughout this work we demonstrated quantitative agreement of our approach by comparing the extrapolated numerical results to exact analytical QFT prediction.
In the case of an unknown QFT to be simulated, it will be essential to perform an error-bounding analysis, especially in the most interesting regime where classical simulations are infeasible~\cite{daley2022practical}.
While for continuum QFTs a naive analysis based on operator norms is not possible due the unboundedness of the involved operators, we note that this issue could be avoided by projecting to finite-energy subspaces~\cite{burgarth2024strong,tong2022provably}.
A possible verification of an analog experimental implementation of the desired QFT could be achieved using Hamiltonian learning techniques, which were recently adapted to the field theory setting~\cite{ott2024hamiltonian}.

\section*{Acknowledgements}
The authors thank Rick van Bijnen, Andreas Kruckenhauser, Vincent Liu, Robert Ott, and Peter Zoller for discussions and collaborations on related works.
This publication has received funding under Horizon Europe programme HORIZON-CL4-2022-QUANTUM-02-SGA via the project 101113690 (PASQuanS2.1) and the EU-QUANTERA project TNiSQ (N-6001). Work in Innsbruck is supported by the ERC Starting grant QARA (Grant No.~101041435) and by the Austrian Science Fund (FWF): COE1 (Grant No. DOI 10.55776) and quantA.
M.D.L. acknowledges support from the Italian Ministry of University and Research via the Rita Levi-Montalcini program.
The computational results presented here have been achieved using the LEO HPC infrastructure of the University of Innsbruck.

\bibliography{bibliography.bib}

\cleardoublepage
\appendix

\onecolumngrid
\section{\label{sec:tableobservablesfigures} Overview table}

In order to quickly locate plots in the main text, below we provide a  table of the investigated observables and their corresponding figures  for the three studied models. 

\begin{table}[h!]
\caption{Investigated observables for each QFT model and corresponding figure.}
\setlength{\tabcolsep}{20pt}
\begin{tabular}{c|c|c}
       \textbf{QFT Model           }& \textbf{Observable} & \textbf{Figure}  \\ \hline\hline
\multirow{4}{*}{\shortstack{Luttinger liquid\\ CFT}} & Central charge $c$ &  Inset \ref{fig:CFT}a \\
                  &  Energy gap $\Delta E_\mathrm{latt} $& \ref{fig:CFT}a  \\
                  & $\langle \hat J_+ ^{(i)} \hat J_- ^{(i\pm r)} \rangle/\sqrt{J(J+1)}$ & Inset \ref{fig:CFT}b  \\
                  & Luttinger parameter $K$  & \ref{fig:CFT}b \\ \hline
\multirow{10}{*}{sine-Gordon model} & Renormalized $\beta^2$ & \ref{fig:Equilibrium}a \\
                  & Ground-state energy density $E_{0, \mathrm{latt}}$ &  \ref{fig:Equilibrium}b\\
                  &  Energy gap $\Delta E_\mathrm{sG}$&  Inset \ref{fig:Equilibrium}b\\
                  & Vertex operator $\langle e^{i\beta\hat\varphi}\rangle$  & \ref{fig:Equilibrium}c  \\
                  & Higher powers of the vertex operator $\langle e^{in\beta\hat\varphi}\rangle$  & \ref{fig:Equilibrium}d  \\
                  & Topological charge density $\rho$ & \ref{fig:TimeEvolFig}c \\
                  & $\langle \cos{\beta\hat\varphi}\rangle$  &  \ref{fig:TimeEvolFig}d \\
                & von Neumann entanglement entropy $S_\mathrm{vN}$& Inset \ref{fig:TimeEvolFig}e \\
                & Excess lattice  energy density $\bar{\varepsilon}$ & \ref{fig:ScatteringFig}a \\
                  & Position shift $\delta x$ & \ref{fig:ScatteringFig}b \\\hline
                
\multirow{4}{*}{\shortstack{Perturbed \\ sine-Gordon model}} & Topological charge density $\rho$ & \ref{fig:PerturbationFig}c / \ref{fig:PerturbationFig}e / \ref{fig:PerturbationCompareFig} (1st row) \\
& von Neumann entanglement entropy $S_\mathrm{vN}$ & \ref{fig:PerturbationFig}d / \ref{fig:PerturbationCompareFig} (4th row) \\
                  & $\langle \sin{\beta\hat\varphi}\rangle$  & \ref{fig:PerturbationFig}f / \ref{fig:PerturbationCompareFig} (2nd row) \\
                & Normalized energy density & \ref{fig:PerturbationCompareFig} (3rd row)\\

\end{tabular}
\end{table}

\twocolumngrid

\section{\label{sec:AppendixRydbergHamiltonian} Description of large-spin Rydberg atom arrays}
In this Appendix we recall how a more general version of lattice Hamiltonian $\hat{H}_\text{latt}$ Eq.~\eqref{eq:LatticeHamiltonian} can be engineered with Rydberg atom arrays exploiting the SO(4)-symmetry \cite{kruckenhauser2022highdimensional}.

Consider a square array of Rydberg atoms in $d=1,2$ or $3$ spatial dimensions. Motivated by the SO(4)-symmetric high-dimensional Rydberg manifolds introduced in Ref.~\cite{kruckenhauser2022highdimensional}, we focus on Rydberg states with a single principal quantum number $n$ for parallel electric and magnetic fields in the regime of linear Stark and Zeeman shifts. In this case, the state space of a single Rydberg atom is well described by two coupled angular momentum operators $\hat{J}_a, \hat{J}_b$, both having spin length $J = \frac{n-1}{2}$.  We further restrict our attention to a configuration where $\hat{J}_b$ is fully polarized, isolating a set of states $\{|m\rangle\}$ with $m=-J,\dots, J$ which are eigenstates of  $\hat{J}_z=\hat{J}_{a,z}$. We also note that in the relevant regime of $|m| \ll J$ the quantum defects discussed in Ref.~\cite{kruckenhauser2022highdimensional} are irrelevant.

As discussed in detail in Ref.~\cite{kruckenhauser2022highdimensional}, any pair of Rydberg atoms naturally experiences dipole-dipole interactions of strength $V_{ij}=V_\text{nn} a^3/|\mathbf{r}_i-\mathbf{r}_j|^3$, where $i,j$ label the individual atoms at positions $\mathbf{r}_{i/j}$ and $V_\text{nn}$ denotes the interaction strength of nearest neighbors $\langle ij \rangle$ at distance $|\mathbf{r}_i-\mathbf{r}_j| = a$. 
In order to simulate the desired scalar QFTs, we assume control over the following three experimental ingredients~\cite{kruckenhauser2022highdimensional}.
First, one can drive single-atom transitions using microwaves with Rabi frequency $\Omega$ and detuning $\Delta$. Second, off-resonant coupling to higher lying states $n'>n$ allows one to engineer a one-axis twisting term with squeezing strength $\chi$. Third, we also include a ponderomotive  drive that transfers $\kappa$ quanta of orbital angular momentum with coupling  strength $\lambda_\kappa$. Collecting everything, the dynamics of the system can be described by the many-body Hamiltonian
\begin{align}
\label{eq:RydbergLatticeHamiltonian}
\hat{H}_\text{Ryd} &= \sum_i \left[ \Omega \hat{J}_{x}^{(i)} - \Delta \hat{J}_{z}^{(i)} + \chi \left(\hat{J}_{z}^{(i)}\right)^2 \right] \\
&\qquad-  \sum_i \left[ \lambda_\kappa \left(\hat{J}_{+}^{(i)}\right)^\kappa + \mathrm{H. c.}   \right] 
\nonumber\\ 
&\qquad+ \frac{1}{2} \sum_{i \neq j} V_{ij} \left[ \hat{J}_{z}^{(i)} \hat{J}_{z}^{(j)} - \frac{1}{4} \left( \hat{J}_{+}^{(i)}\hat{J}_{-}^{(j)} + \mathrm{H. c.} \right) \right]\;,\nonumber
\end{align}
where $\hat{J}_{z}^{(i)}$ resp. $\hat{J}_{\pm}^{(i)} = \hat{J}_{x}^{(i)} \pm i \hat{J}_{y}^{(i)}$ denote  the $z$-component resp. raising/lowering operators of the spin-$J$ degree of freedom at every lattice site $i$.
Upon restricting to nearest-neighbors only $\langle ij \rangle$, we retrieve the lattice Hamiltonian $\hat{H}_\text{latt}$ Eq.~\eqref{eq:LatticeHamiltonian} discussed in the main text. 

\section{\label{sec:AppendixMappingHamiltonian} Mapping the lattice Hamiltonian to the continuum scalar QFT}
Here we discuss the mapping between the lattice and the scalar QFT Hamiltonian, as well as the necessary identifications between discrete and continuum degrees of freedom and parameters.

 The $d$-dimensional lattice Hamiltonian 
\begin{align}
\hat{H}_\text{latt} &= \sum_i \left[ \Omega \hat{J}_{x}^{(i)} - \Delta \hat{J}_{z}^{(i)} + \chi \left(\hat{J}_{z}^{(i)}\right)^2 \right] \\
&\qquad-  \sum_i \left[ \lambda_\kappa \left(\hat{J}_{+}^{(i)}\right)^\kappa + \mathrm{H. c.}   \right] 
\nonumber\\ 
&\qquad+  \sum_{\langle ij \rangle} V_\text{nn} \left[ \hat{J}_{z}^{(i)} \hat{J}_{z}^{(j)} - \frac{1}{4} \left( \hat{J}_{+}^{(i)}\hat{J}_{-}^{(j)} + \mathrm{H. c.} \right) \right]\;,\nonumber
\end{align}
allows the simulation of the continuum scalar QFTs Hamiltonian $\hat{H}_\text{QFT}$ in  Eq.~\eqref{eq:QFTHamiltonian} upon proper choice of the parameters. 
We begin by considering a single $\kappa$
and choose $\Omega=\Delta=0$. In order to employ the identification between spin and QFT operators 
Eq.~\eqref{eq:fields_to_spins}, we further rescale the lattice parameters as $V_\text{nn}'=V_\text{nn} J(J+1)$ and  $\lambda_\kappa'=\lambda_\kappa[J(J+1)]^{\kappa/2}$. Upon rescaling, the Ising term $\hat{J}_{z}^{(i)} \hat{J}_{z}^{(j)}$ acquires an additional factor $1/J(J+1)$ and becomes negligible in the large spin-length limit $J \rightarrow \infty$ [see App.~\ref{sec:AppsG_compare}].
We thus obtain the discrete lattice version of the scalar QFTs Hamiltonian
\begin{multline}
    \label{eq:Ham_lattice_QFT}
    \hat{H}_\text{QFT}^\text{latt}=\chi \sum_i \left\{(\hat{\pi}_i)^2 - 2\frac{\lambda'_\kappa}{\chi}\cos{\left(\kappa \hat{\vphi}_i \right) } \right\} \\  - \chi \sum_{\langle ij \rangle} \frac{V_\text{nn}'}{2\chi}\cos{\left( \hat{\vphi}_i-\hat{\vphi}_{j}\right)}.
\end{multline}
To reach the continuum we then identify 
\begin{equation}
\frac{\lambda'_\kappa}{\chi}=\frac{(M_0')^2 \kappa^2}{(\beta')^4}, \qquad \frac{V_\text{nn}'}{\chi}=\frac{4\kappa^4 }{(\beta')^4},
\end{equation}
and rescale the field operators
\begin{equation}
\kappa  \hat{\vphi}_\text{latt} \rightarrow \beta \hat{\vphi}_\text{cont},\quad \hat{\pi}_\text{latt} \rightarrow \hat{\pi}_\text{cont} \frac{\kappa a^{\frac{d+1}{2}}}{\beta'} , 
\label{eq:AppMappingOperators}
\end{equation}
as well as the Hamiltonian $\hat{H}_\text{latt} \cdot \beta'^2/(2 \chi \kappa^2 a ) \rightarrow \hat{H}_\text{QFT}$.
Moreover, we Taylor expand the cosine term $\cos{\left(\hat{\vphi}_i-\hat{\vphi}_{j} \right)}$ in $\delta\opphi=\opphi_i - \opphi_{j}$ and truncate it to second order: this term accounts for the spatial derivative term $(\partial_x \opphi)^2$.
The continuum scalar QFT is thus described by the Hamiltonian 
\begin{multline}
    \hat{H}_\text{QFT}^\text{cont}=\int {\rm d}^d x \left\{ \frac{1}{2}[\hat{\pi}(x)]^2 +\sum_{k=1}^d\frac{1}{2} [\partial_{x_k} \hat{\vphi}(x)]^2 \right. \\
    \left. -\frac{M_0^2}{\beta^2}\cos{(\beta \hat{\varphi})} \right\},
\end{multline}
where we sum over the spatial derivatives  $[\partial_{x_k} \opphi(x)]^2$ in each of the $d$ directions.
The implementation of the continuum QFT from the Rydberg lattice Hamiltonian is only valid in the limit of large spin length $J \rightarrow \infty$, large atom number $N\rightarrow \infty$, as well as in the continuum limit $M_0' \rightarrow 0$ [see discussion in the main text].

\section{\label{sec:AppendixFittingDiscussion}
Estimation of the continuum limit of observables}
In this Appendix we provide additional details on the extrapolations performed to retrieve the continuum limit value of the observables, as well as their uncertainty.

As discussed in the main text, the analog quantum simulation of scalar QFTs on a lattice is only valid in the continuum limit and thus, we need to extrapolate the value and uncertainty of relevant observables in the limit $J, N \rightarrow \infty $.
To this end, we run numerical simulations for fixed QFT parameters $\beta^2, M_0'$ at several $J$ and $N$, thus obtaining a set of data points $\{O_i\}$. We then fix $N$ and consider a minimal subset of data points, corresponding to the numerical results at the largest $J$ values simulated. We fit this minimal subset with a function $h(x)$ linear in $x=1/J(J+1)$, thus obtaining a first estimate $O_\alpha$ for the asymptotic limit $J\rightarrow \infty$. We repeat this procedure for increasingly larger subsets of $\{O_i\}$ by adding data points at smaller $J$ values, and thus obtain new predictions for the asymptotic value $O_\alpha$ ($\alpha=1, \dots, N_\text{fits}$, with $N_\text{fits}$ the number of subsets considered).
If we do not have an estimate for the error $\Delta O_i$ of the data point $O_i$, we choose as our best guess $\bar{O}_\alpha$ for the asymptotic limit the median value of the set $\{O_\alpha\}$, and we estimate the uncertainty $\Delta O_\alpha$ as half of the difference between the largest and smallest values $\{O_\alpha\}$.
Instead, if we have already an estimate for the error $\Delta O_i$, the best guess $\bar{O}_\alpha$ is determined according to a cumulative distribution function $f_\alpha$, introduced in Ref.~\cite{banuls2013mass},
\begin{equation}
    f_\alpha=\frac{\sum_{j=1}^\alpha\exp{(-\chi_j^2/N_\text{d.o.f})}}
    {\sum_{j=1}^{N_\text{fits}}\exp{(-\chi_j^2/N_\text{d.o.f})}},
\end{equation}
with $\chi_j^2=\sum_{i: x_i \in \text{fit} j} (h(x_i)-O_i)^2/(\Delta O_i)^2$ the reduced $\chi_j^2$ of the $j$-th estimated value $O_j$ and $N_\text{d.o.f}$ the corresponding number of degrees of freedom. Note that we sort $\{O_j\}$ in increasing order. Our best guess $\bar{O}_\alpha$ is then the first $O_\alpha$ for which $f_\alpha\geq 0.5$, while the uncertainty is defined as half of the difference between the first value for which $f_\alpha \geq 0.8415$ and the first value for which $f_\alpha\geq 0.1585$.
Once we obtain an estimate for $J\rightarrow \infty$ for each fixed $N$, we repeat the same procedure by fitting the data points $\{\bar O_\alpha\}$ with a quadratic function in $1/N$ to determine the asymptotic limit $N\rightarrow \infty$.

\section{\label{sec:AppendixCFT} Conformal field theory and next-nearest-neighbor interaction}

In this Appendix we discuss additional results regarding the quantum simulation of the conformal field theory (CFT). We consider the so far neglected next-nearest-neighbor term of the dipole-dipole interaction and investigate how this affects the numerical results.

To carry out this analysis we look at two paradigmatic observables: the energy gap $\Delta E$ resp. the central charge $c$, which should be equal to 0 resp. 1 for the considered CFT.
\begin{figure}
    \centering
    \includegraphics[width=\linewidth]{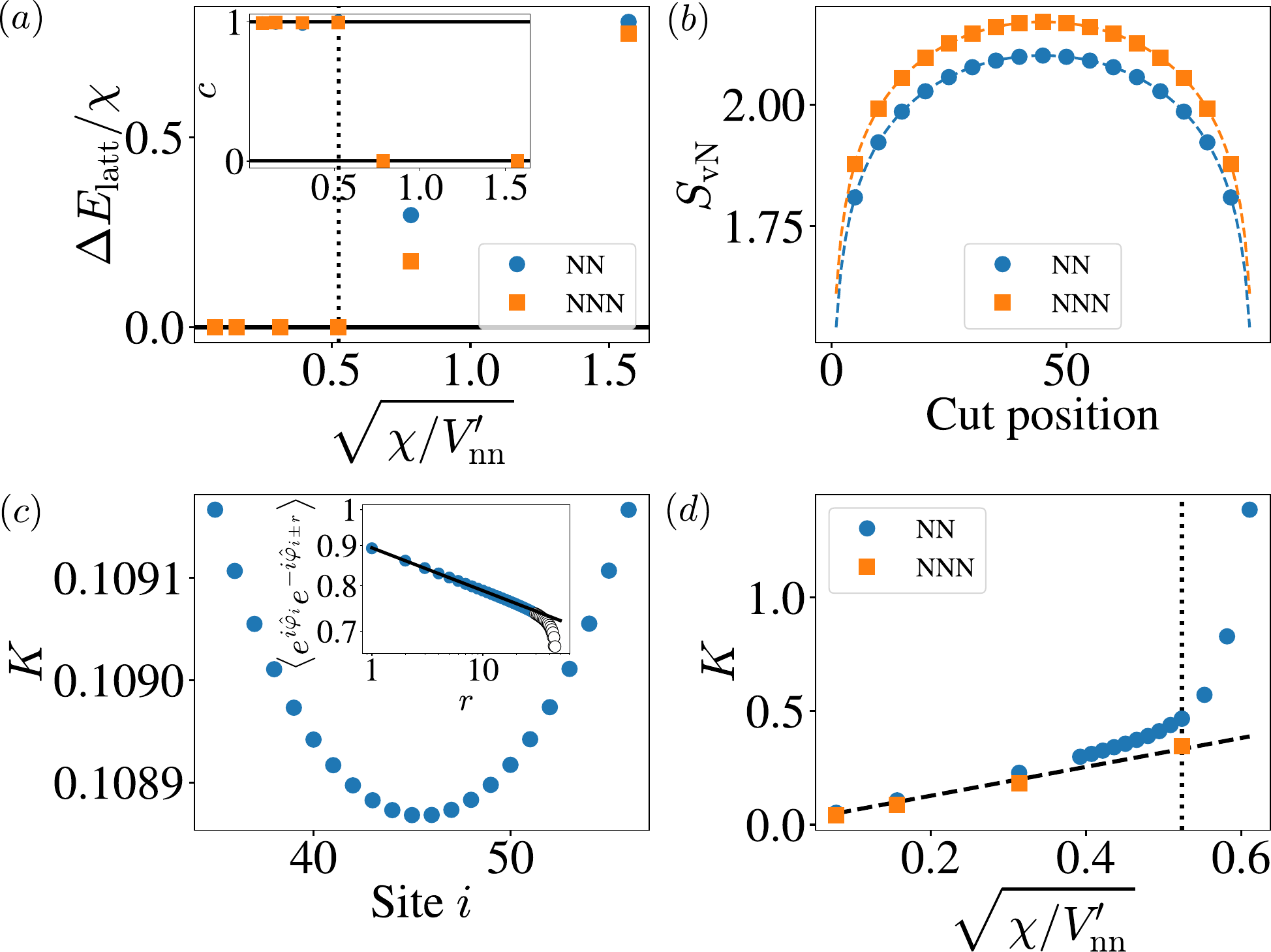}
    \caption{\textbf{Conformal field theory and impact of next-nearest-neighbor interaction.} (a) Energy gap in lattice units $\Delta E_\text{latt}$ and (inset) central charge $c$ as a function of the microscopic parameters ratio $\sqrt{\chi/V_\text{nn}'}$ for nearest-neighbors only (NN, blue dots), and next-nearest-neighbors interaction (NNN, orange squares).   The vertical dotted line denotes the regime in which the lattice Hamiltonian correctly simulates the CFT ($\sqrt{\chi/V_\text{nn}'} \lesssim \pi/6$). (b) Von Neumann entanglement entropy $S_\text{vN}$ evaluated at several cut positions and fitted with Eq.~\eqref{eq:centralchargeEE} (dashed lines) for $N=90, J=18, \sqrt{\chi/V_\text{nn}'}=0.157$. (c) (Inset) Vertex-vertex connected correlation function $\langle \hat{J}^{(i)}_+ \hat{J}^{(i\pm r)}_-\rangle_\text{c} /\sqrt{J(J+1)}$ against the separation $r$ between sites $i(=N/2)$ and $i+r$. We use the same parameters as in (b). (Main) From the slope of the log-log fit we determine the Luttinger parameter $K$ at site $i$ for the NN case. (d) Extrapolated value of $K$ for NN (blue dots) and NNN (orange squares) compared to the prediction from the microscopic parameters Eq.~\eqref{eq:K_microscopic} (dashed black line).}
    \label{fig:AppCFT}
\end{figure}
In Fig.~\ref{fig:AppCFT}(a) we compare the continuum results for both the energy gap $\Delta E_\text{latt}$ (in lattice units) and the central charge $c$ in two cases: for nearest-neighbors only, and also the next-nearest-neighbor term of the dipole-dipole interaction. We observe that higher orders in the dipole-dipole interaction only modify the value of the mass gap $\Delta E_\text{latt}$ in the gapped region (where the lattice model does not correctly simulate the CFT), while the central charge is only minimally modified in the gapless phase.
We thus expect that the long-range interaction does not extend the simulability region of the model, yet it modifies the value of the Luttinger parameter $K$. 
Note that the central charge $c$ is extrapolated upon fitting with Eq.~\eqref{eq:centralchargeEE} the von Neumann entanglement entropy $S_\text{vN}$ at different cut positions along the chain [see Fig.~\ref{fig:AppCFT}(b)]. The continuum result is then determined from a linear fit in $1/J(J+1)$ and a subsequent fit in $1/N$, as described in App.~\ref{sec:AppendixFittingDiscussion}. 

We extrapolate the value of the Luttinger parameter $K$ from the connected vertex-vertex correlator $\langle \hat{J}_+^{(i)} \hat{J}_-^{(i+r)} \rangle_\text{c} /J(J+1)$ by linearly fitting the logarithm of the correlator against the logarithm of the separation $r$ between the two sites considered according to Eq.~\eqref{eq:Luttinger_power_law}. We choose the fitting interval in such a way to avoid boundary effects (we neglect the 15 sites closest to both ends), and to fit over linear regions only [see inset in Fig.~\ref{fig:AppCFT}(c)]. The Luttinger parameter $K$ displays a tiny dependence on the site $i$ at which it is calculated due to finite-size effects; therefore, we consider as our best guess the half-chain result $i=N/2$.
In Fig.~\ref{fig:AppCFT}(d) we compare the numerical results for the Luttinger parameter $K$ with the theoretical prediction Eq.~\eqref{eq:K_microscopic}, and notice that it displays some renormalization w.r.t. the microscopic mapping Eq.~\eqref{eq:K_microscopic}. We expect the renormalization to emerge due to the truncation of the cosine term arising from the dipole-dipole interaction. As expected [see App.~\ref{sec:AppsG_renormbeta}], the renormalization increases for decreasing $V_\text{nn}'$, but it is partially compensated by the NNN interaction term. Indeed, the value $K$ obtained upon considering also next-nearest-neighbors terms is very close to the microscopic prediction Eq.~\eqref{eq:K_microscopic} in the simulated regime.

\section{\label{sec:AppendixEquilibriumSineGordon} Equilibrium properties of the sine-Gordon model}

In this Appendix we discuss additional results regarding the equilibrium properties of the sine-Gordon model.
In App.~\ref{sec:AppsG_compare} we discuss the impact of neglected next-nearest-neighbors dipole-dipole interaction and Ising terms on the simulated sG model. In App.~\ref{sec:AppsG_renormbeta} we discuss the role of the neglected higher-order terms in the cosine expansion and how they impact the renormalization of the parameter $\beta^2$. In App.~\ref{sec:AppsG_renormM0} we illustrate two methods to determine the renormalization of the mass parameter $M_0'$. In App.~\ref{sec:AppsG_conjecture} we discuss the Lukyanov-Zamolodchikov conjecture.

\subsection{\label{sec:AppsG_compare} Neglected terms in the dipole-dipole interaction}
The sG model can be simulated with the 1D lattice Hamiltonian $\hat{H}_\text{latt}$ [Eq.~\eqref{eq:LatticeHamiltonian}], obtained from the Rydberg Hamiltonian $\hat{H}_\text{Ryd}$ [Eq.~\eqref{eq:RydbergLatticeHamiltonian}] by neglecting higher-order dipole-dipole interaction term $V_{ij}$ ($j>i+1$) and the Ising contribution $\hat{J}_z^{(i)} \hat{J}_z^{(j)}$. The latter approximation is justified in the asymptotic limit $J\rightarrow \infty$: using the identification Eq.~\eqref{eq:fields_to_spins} between lattice spin operators and the continuum QFT operators, as well as by introducing the parameter $V_{ij}'=V_{ij}J(J+1)$, the term $\hat{J}_z^{(i)} \hat{J}_z^{(j)}$ is rescaled by the factor $1/J(J+1)$, which makes the Ising term vanish in the large-spin limit $J \rightarrow \infty$ [see App.~\ref{sec:AppendixMappingHamiltonian}].

\begin{figure}
    \centering
    \includegraphics[width=\linewidth]{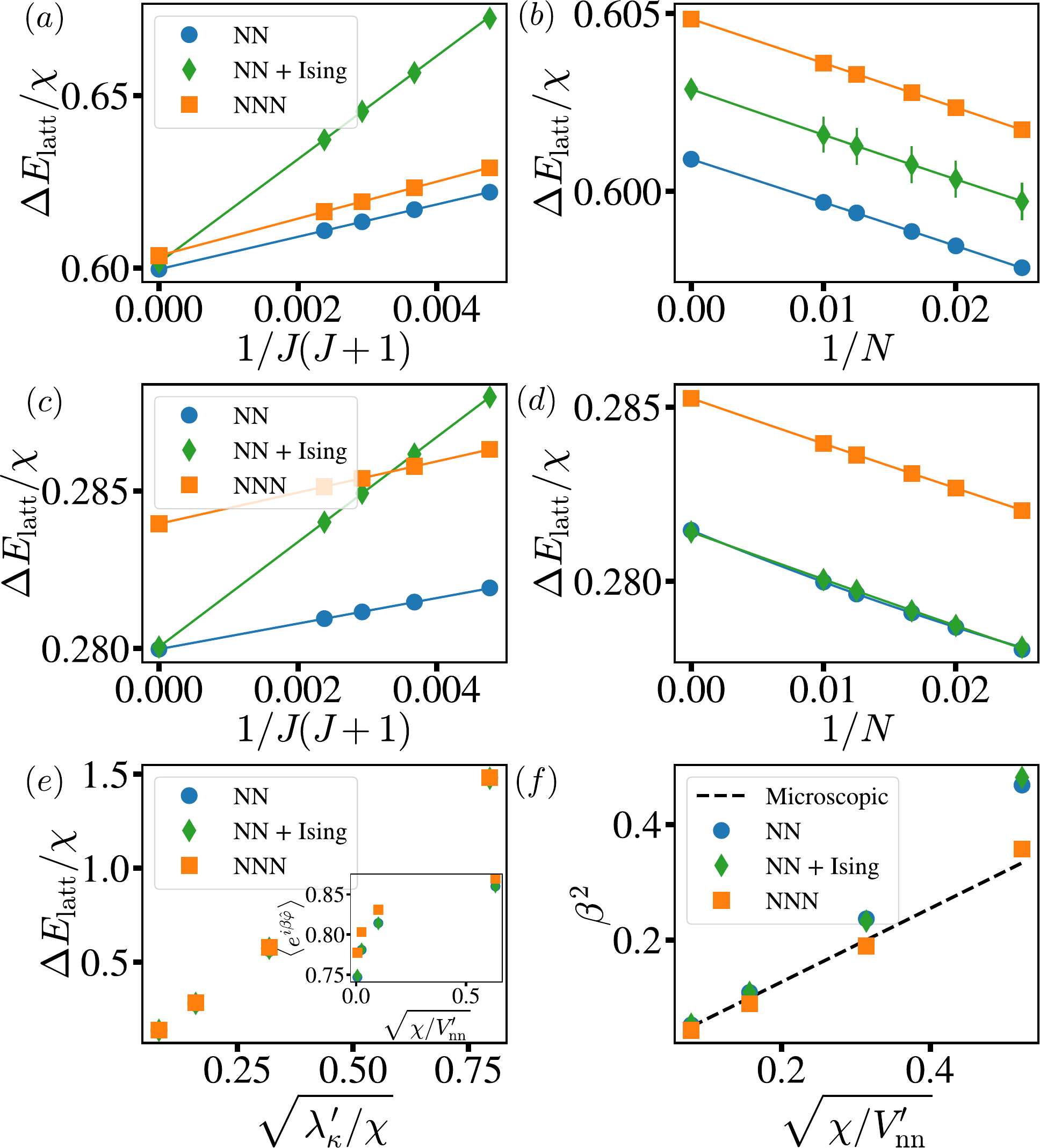}
    \caption{\textbf{Effect of neglected interactions on the sG model simulation}. Comparison of numerical results obtained from DMRG calculations of the lattice model $\hat{H}_\text{latt}$ upon considering only nearest-neighbor dipole-dipole interaction (NN, blue dots), by adding the Ising term $\hat{J}_z^{(i)} \hat{J}_z^{(i+1)}$ (NN + Ising, green diamonds), or by adding next-nearest-neighbor interactions (NNN, orange squares). We choose $\kappa=1$ and unless varying over the corresponding parameters, we consider $N=100,\lambda_\kappa'=1/\pi^2,  \chi=1$ and $V_\text{nn}'=400/\pi^2$ in (a)-(b) resp. $V_\text{nn}'=100/\pi^2$ in (c)-(e). (a), (c) Energy gap in lattice units $\Delta E_\text{latt}$ and corresponding fit in $1/J(J+1)$ for the different interaction terms. (b), (d) The asymptotic value extracted in the large-spin length limit $J\rightarrow \infty$ is plotted for different system sizes $N$ and fitted against $1/N$.
    (e) Continuum results for the lattice gap  $\Delta E_\text{latt}$ and (inset) the vertex operator $\langle e^{i\beta \opphi} \rangle = \langle \hat{J}_+ \rangle /\sqrt{J(J+1)}$ as a function of the microscopic parameter ratio $\sqrt{\lambda_\kappa'/\chi}$. (f) Renormalized $\beta^2$ value against the microscopic ratio $\sqrt{\chi/V_\text{nn}'}$. The dashed black line represents the microscopic prediction according to  Eq.~\eqref{eq:parameter_ident}.
     }
    \label{fig:AppsG_compare}
\end{figure}

To numerically assess the impact of both Ising and NNN interaction terms on the continuum results, we begin by investigating the mass gap $\Delta E_\text{latt}$ (in lattice units). Here we consider $J=14,\dots, 20$ and $N=40, \dots, 100$. The numerical data $\Delta E_\text{latt}$ obtained upon considering also the Ising term converge toward the results for only NN in the limit $J \rightarrow \infty$ [see Fig.~\ref{fig:AppsG_compare}(a),(c)], as expected from the above discussion. 
Nonetheless, in the limit of small $\beta^2$, i.e., large $V_\text{nn}'$ [see Fig.~\ref{fig:AppsG_compare}(b)], we observe some deviations between the energy gaps in the two cases (NN and NN+Ising) even after the extrapolation in $J$. We attribute these deviations to arise from imperfect convergence of the observable in the spin length in the small $\beta^2$ regime. Simulating the model at larger $J$, although computationally more expensive, could provide even more accurate extrapolations.
If we now consider instead the next-nearest-neighbors (NNN) interaction term, we observe an offset with respect to the standard case of only NN interactions, independently of the $\beta^2$-value considered, both after the extrapolation in $J$ and in $N$. 

In Fig.~\ref{fig:AppsG_compare}(e) we compare the continuum results of both the mass gap and the vertex operator at $V_\text{nn}'/\chi=100/\pi^2$ and for different $\sqrt{\lambda_\kappa'/\chi}$ (i.e.,  $M_0'$). For the gap the data points of the three cases agree with each other, while for the vertex operator we observe some deviations between NN and NNN, which increase for decreasing $M_0'$.
This deviation leads to different renormalized $\beta^2$ values [see Fig.~\ref{fig:AppsG_compare}(f)], which, as discussed in the main text, we extrapolate from the log-log fit of mass gap vs. vertex operator.
We notice that the NNN interaction reduces the renormalization  [see Fig.~\ref{fig:AppsG_compare}(f)], in agreement with the CFT case [see App.~\ref{sec:AppendixCFT}]. This behaviour indicates a counteracting effect of the long-range dipole-dipole interaction on the neglected terms in the expansion of the cosine [see App.~\ref{sec:AppsG_renormbeta} for further details].

To allow for a quantitative analysis, we compare the numerical values of the mass gap and the vertex operator in the continuum limit with respect to the results obtained for NN only.
Upon considering also the Ising term, for the mass gap we obtain a maximal relative deviation of up to 3.3\% for $M_0'\leq 0.2$ at $\beta^2=\pi/20$, while when considering the NNN interaction, the maximal error amounts to 2\%.
For the vertex operator, the maximal deviation for the Ising term amounts to 0.9\%, while the deviation for the NNN interaction increases with $V_\mathrm{nn}'$ up to 9\% for the largest values considered.

\subsection{\label{sec:AppsG_renormbeta} \texorpdfstring{$\beta^2$-renormalization and $\kappa$-dependence}{beta-squared-renormalization and kappa-dependence}}
Here we explicitly discuss the neglected terms in the Taylor expansion of the cosine term, which we expect to be responsible for the renormalization of the $\beta^2$ parameter. We further illustrate the dependency of the neglected terms on the microscopic parameter $\kappa$.

To engineer the spatial derivative term $[\partial_x \opphi(x)]^2$ in the sG model, we need to Taylor expand the cosine term $-V_\text{nn}'/2 \cdot \cos(\vphi_{i} - \vphi_{i+1}) = -V_\text{nn}'/2 \cdot \cos(\delta \vphi_{i})$  of the lattice QFT Hamiltonian Eq.~\eqref{eq:Ham_lattice_QFT}, which arises from the dipole-dipole interaction term $- \frac{1}{4} \left( \hat{J}_{+}^{(i)}\hat{J}_{-}^{(j)} + \mathrm{H. c.} \right) $. We thus obtain 
\begin{align}
    -\frac{V_\text{nn}'}{2}\cos(\delta \vphi_{i})& = \\  -\frac{V_\text{nn}'}{2}&\left(1 - \frac{(\delta \vphi_i)^2}{2} + \sum_{n=2}\frac{(\delta \vphi_i)^{2n}}{(2n)!}(-1)^n \right),
\end{align}
where we need to neglect higher-order term $(n\geq 2)$ to retrieve the sG model.
Upon multiplying the Hamiltonian with $\beta^2/(2\chi\kappa^2 a)$, mapping $\kappa  \hat{\vphi}_\text{latt} \rightarrow \beta \hat{\vphi}_\text{cont}$, and using the relation $V_\text{nn}'/\chi=4\kappa^4/\beta^4$, as we do in App.~\ref{sec:AppendixMappingHamiltonian} to obtain the continuum sG model, the neglected terms become 
\begin{equation*}
 \sum_{n=2} \left(\frac{\beta}{\kappa}\right)^{2n-2}\frac{1}{a}\frac{(\delta \vphi_i)^{2n}}{(2n)!}(-1)^n.
\end{equation*}
We believe these terms to renormalize the parameter $\beta^2$, and from the above equation we expect this effect to increase for increasing $\beta^2$ (i.e.,  decreasing $V_\text{nn}'$), but to decrease with larger $\kappa$.
This dependence agrees with the numerical observations discussed in the main text and in the previous appendices.

\subsection{\label{sec:AppsG_renormM0} \texorpdfstring{$M_0'$}{M0'} renormalization}
\begin{figure}
    \centering
    \includegraphics[width=0.8\linewidth]{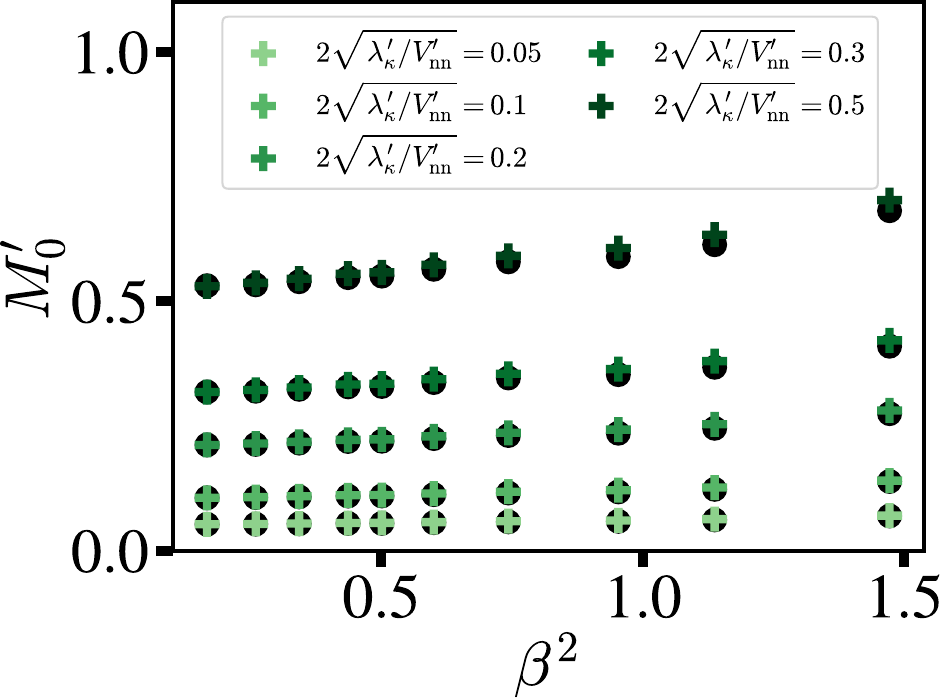}
    \caption{\textbf{Extrapolation of the mass $M_0'$.} The black dots represent the values $M_0'$ determined according to Eq.~\eqref{eq:M0_renormalization} from the effective $\beta^2$ parameter. The colored crosses are instead the values extrapolated upon inverting Eq.~\eqref{eq:breathermass}. Different shaded of green corresponds to different ratios $2\sqrt{\lambda_\kappa'/V_\text{nn}'}$, i.e., to different $M_0'$ according to the naive microscopic mapping Eq.~\eqref{eq:parameter_ident}.} 
    \label{fig:AppMassExtrapolation}
\end{figure}

In the main text we discussed the extrapolation of the effective $\beta^2$ parameter of the simulated sG model. We further observed that the value of the bare mass $M_0'$ also differs from the microscopic identification Eq.~\eqref{eq:parameter_ident}, and that it can be fixed in dependence of the effective $\beta^2$ value according to Eq.~\eqref{eq:M0_renormalization}.
An additional check of the validity of this approach is the alternative extrapolation of $M_0'$ from the numerically determined mass gap $\Delta E_\text{sG}=\beta^2/(2\kappa^2 \chi a) \cdot \Delta E_\text{latt}$ (in sG units). By inverting the theoretical prediction for the breather mass $m_1$ [Eq.~\eqref{eq:breathermass}], it is possible to obtain an expression for $M_0'$ as a function of the mass gap $\Delta E_\text{sG}=m_1$ and the $\beta^2$ parameter. As shown in 
Fig.~\ref{fig:AppMassExtrapolation}, the two extrapolated values of $M_0'$ display a perfect agreement at small $\beta^2$ and $\lambda_\kappa'/V_\text{nn}'$, i.e.,  in the continuum limit, with small deviations at larger ratios $\lambda_\kappa'/V_\text{nn}'$ and for larger $\beta^2$. 

\subsection{\label{sec:AppsG_conjecture} Lukyanov-Zamolodchikov conjecture}
In the main text we introduce the expectation value $\langle e^{i n \beta \hat{\varphi}}\rangle$ and discuss the agreement of numerical results with the unproven conjecture formulated by Lukyanov and Zamolodchikov in Ref.~\cite{lukyanov1997exact}. According to their work, the expectation value $\langle e^{i\sqrt{8\pi} \gamma\hat{\varphi}} \rangle$ is conjectured to be \cite{daviet2019nonperturbative}
\begin{multline}
    \langle e^{i \sqrt{8\pi} \gamma \hat{\varphi}} \rangle = (b\Lambda)^{2\gamma^2}\left( \frac{\pi\Gamma{(1-K)}}{2\Gamma{(K)} (b\Lambda)^2}\frac{(aM_0)^2}{\beta^2}\right)^{\gamma^2/(1-K)}
    \times \\ \exp \left\{ \int_0^\infty \frac{{\rm d}t}{t}\left[\frac{\sinh^2{(2 \gamma \sqrt{K} t )}}{2\sinh{(Kt)}\sinh{(t)}\cosh{((1-K)t)}} 
   \right. \right. \\
   \left.  \left. -2\gamma^2 e^{-2t} \vphantom{\exp \left\{ \int_0^\infty \frac{{\rm d}t}{t}\left[\frac{\sinh^2{(2 \gamma \sqrt{K} t )}}{2\sinh{(Kt)}\sinh{(t)}\cosh{((1-K)t)}} 
   \right. \right.} \right] \right\} ,
\end{multline}
with $K=\beta^2/8\pi$ and is valid for $\beta^2 < 8\pi$, $|\Re \gamma| < \sqrt{8\pi}/(2\beta)$.
For $\gamma=\sqrt{K}=\beta/\sqrt{8\pi}$ the above expression agrees with the theoretical prediction for the vertex operator of the sG model Eq.~\eqref{eq:vertextheory} \cite{lukyanov1997exact}. 

\section{\label{sec:AppSolitonPrepTimeevol} Semi-classical soliton preparation and dynamics}
In this Appendix we provide additional details regarding the preparation of the semi-classical soliton, as well as its dynamics. This Appendix is structured as follows: 
in App.~\ref{sec:AppSol_preparation} we discuss the prepared soliton state, and we show that the two-point correlators remain untouched, in App.~\ref{sec:AppSol_StaticEvol} resp. App.~\ref{sec:AppSol_MovingEvol} we focus on the time evolution of the static resp. moving soliton, in App.~\ref{sec:AppSol_semi-classical} we discuss in more detail the semi-classical model, while in App.~\ref{sec:AppSol_Fluctuations} we illustrate the relation between the $zz$-correlator of the quantum ground state and the standard deviation of the velocity of the semi-classical model.

\subsection{\label{sec:AppSol_preparation} State preparation and correlation functions}

In the main text we discuss how, by applying two sets of gates $\{ \hat{U}_\text{p}^{(i)}, \hat{U}_\text{m}^{(i)}\}$ on the ground state of the sG Hamiltonian, we prepare the system in a state whose expectation values of the phase $\langle \opphi \rangle=\varphi_\text{s}$ and of the momentum $\langle \hat\pi \rangle=\pi_\text{s}$  agree with those of a classical soliton. 
Here, we further investigate the prepared state with respect to one- and two-point correlation functions and show that the latter are left untouched by our protocol.

\begin{figure}
    \centering
    \includegraphics[width=\linewidth]{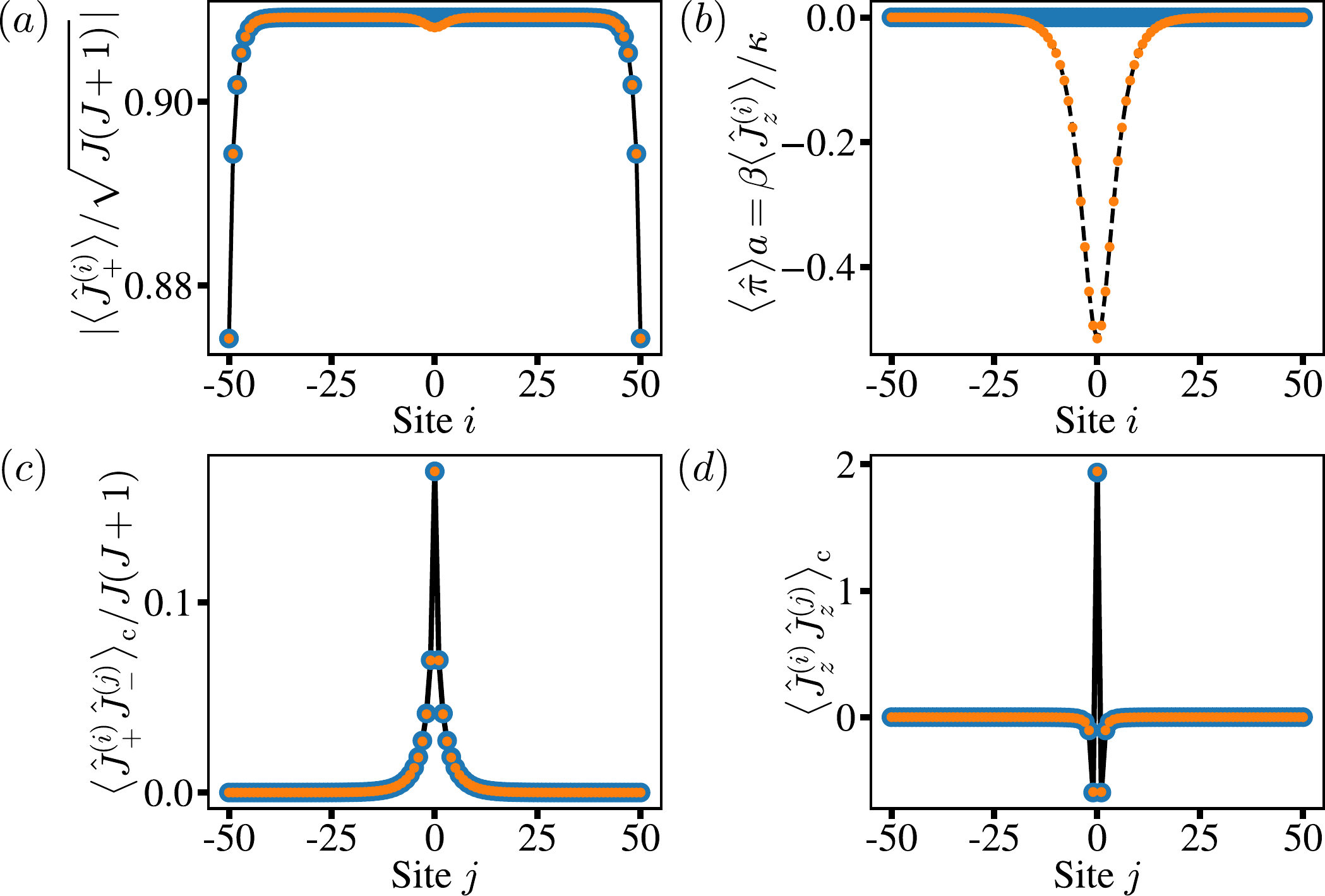}
    \caption{\textbf{One- and two-point correlation functions of the prepared soliton state.} Comparison of selected expectation values for different initial states: the solid black line represents the ground state of the quantum sG model, the blue circles the semi-classical static soliton $v=0$, the orange dots the semi-classical moving soliton at $v=0.5$, the dashed black line the theoretical prediction for the momentum of the classical moving soliton. 
    (a) Absolute value of the expectation value of the vertex operator $|\langle \hat{J}_+ ^{(i)}\rangle/\sqrt{J(J+1)}|$ and (b) the momentum operator $\langle \hat{\pi}\rangle = \beta\langle \hat{J_z}^{(i)}\rangle/\kappa a$ as a function of the site $i$ along the chain. (c) Vertex-vertex connected correlation function  $\langle \hat{J}_+^{(i)} \hat{J}_-^{(j)} \rangle_\text{c}/(J(J+1)) = [\langle \hat{J}_+^{(i)} \hat{J}_-^{(j)} \rangle - \langle \hat{J}_+^{(i)} \rangle \langle \hat{J}_-^{(j)} \rangle]/(J(J+1))$ and (d) $zz$-connected correlation function $\langle \hat{J}_z^{(i)} \hat{J}_z^{(j)} \rangle_\text{c}= \langle \hat{J}_z^{(i)} \hat{J}_z^{(j)} \rangle - \langle \hat{J}_z^{(i)} \rangle \langle \hat{J}_z^{(j)} \rangle$ as a function of the site $j$ for $i=51$ (half-chain). In the plot we consider $N=101, J=20, \chi=\kappa=1, V_\text{nn}'=400/\pi^2, \lambda'_\kappa=6.25/\pi^2$. }
    \label{fig:AppCorrelationSoliton}
\end{figure}

In Fig.~\ref{fig:AppCorrelationSoliton}(a) we plot the absolute value of the vertex operator $|\langle \hat{J}_+^{(i)} \rangle/\sqrt{J(J+1)}|$, which displays a perfect overlap between the ground state of the quantum sG model and the prepared static soliton, while for the moving soliton a small dip emerges in the center of the chain, right in the middle of the transition region. We attribute these deviations to the non-commutativity of the two sets of unitary rotations. 
Moreover, note that the absolute value of the vertex operator $|\langle e^{i\beta \opphi}\rangle|$, i.e., the normalized ``classical" spin length, is never 1 [see Fig.~\ref{fig:Equilibrium}(c)]. Therefore, when determining the rotation angle $\theta^{(i)}$ of the gate $\hat{U}_\text{m}^{(i)}$ for the momentum imprinting, we need to take this local value into account, such that the magnitude of the momentum expectation value, which is proportional to the projection of the spin on the $z$-axis $|\langle \hat\pi \rangle|\propto|\langle \hat{J}_z \rangle|$, agrees with the theoretical prediction [see  Fig.~\ref{fig:AppCorrelationSoliton}(b)]. 

We consider now the two-point functions. Both the vertex-vertex connected correlation function $\langle \hat{J}_+^{(i)} \hat{J}_-^{(j)} \rangle_\text{c}/(J(J+1))= [\langle \hat{J}_+^{(i)} \hat{J}_-^{(j)} \rangle - \langle \hat{J}_+^{(i)} \rangle \langle \hat{J}_-^{(j)} \rangle]/(J(J+1))$ and the $zz$-connected correlation function $\langle \hat{J}_z^{(i)} \hat{J}_z^{(j)} \rangle_\text{c}= \langle \hat{J}_z^{(i)} \hat{J}_z^{(j)} \rangle - \langle \hat{J}_z^{(i)} \rangle \langle \hat{J}_z^{(j)} \rangle$ agree with each other for all the three different states: quantum ground state, semi-classical static and moving soliton [see Fig.~\ref{fig:AppCorrelationSoliton}(c)-(d)]. Note that in the vertex-vertex correlator we need first to account for the correct phase of the spin operator $\hat{J}_\pm\rightarrow \hat{J}_\pm e^{\mp i\beta \vphi_\text{s}}$.
We can thus conclude, that our preparation protocol allows us to impose the desired one-point function, while keeping the two-point correlation functions untouched.

\subsection{\label{sec:AppSol_StaticEvol} Time evolution of the static soliton}
Here we provide additional details regarding the time evolution of the static soliton. In the following, we study the soliton dynamics for $N=101, J=\{16,18,20\}, \beta^2=\{\pi/20, \pi/13, \pi/10\},$ and $M_0'=\{0.1, 0.15, 0.2, 0.25, 0.3\}$. According to Ref.~\cite{kruckenhauser2022highdimensional}, these parameter regimes can be experimentally investigated.

We first investigate the topological charge density $\rho(x,t)=\beta \partial_x \vphi(x,t)/2\pi$, which we determine using $i\beta \partial_x \varphi=e^{-i\beta\varphi}\partial_x e^{i\beta\varphi}$, and approximating the partial derivative with the central finite difference 
\begin{equation}
    \partial_x e^{i\beta\varphi (x)}\approx\frac{ e^{i\beta\varphi(x+\Delta x)}-e^{i\beta\varphi(x-\Delta x)}}{2a},
\end{equation} thus obtaining
\begin{equation}
    \langle \rho(x=i \cdot a,t)\rangle=\Im\left[\frac{\langle\hat{J}_-^{(i)}\hat{J}_+^{(i+1)}-\hat{J}_-^{(i)}\hat{J}_+^{(i-1)}\rangle(t)}{2\pi \cdot 2a\cdot J(J+1)}\right].
\end{equation}

\begin{figure}
    \centering
    \includegraphics[width=\linewidth]{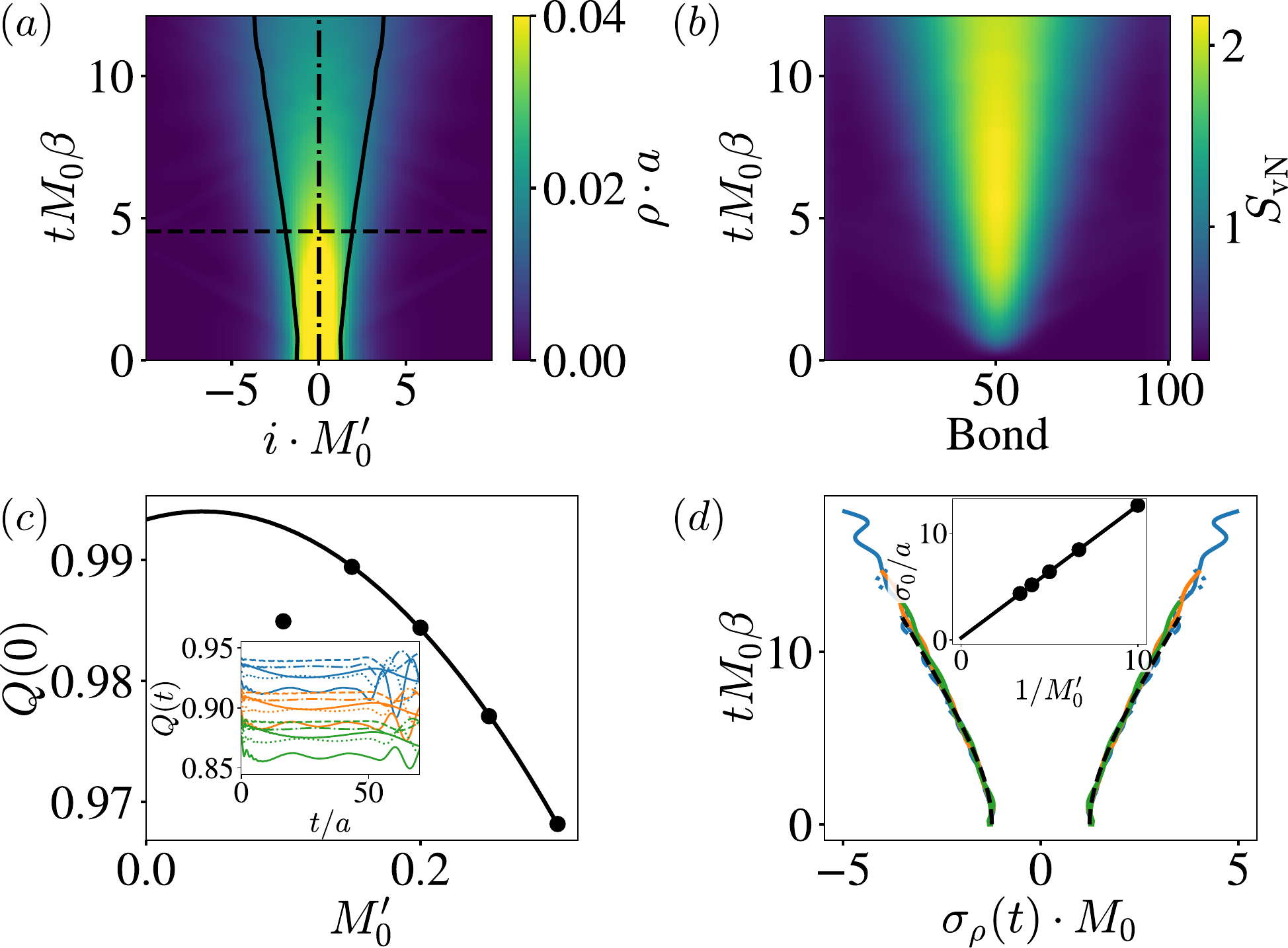}
    \caption{\textbf{Time evolution of the static soliton.} (a) Topological charge density $\rho(x,t)$ in rescaled time and space units. The solid line represents the width of the Gaussian fit, the dash-dotted the mean value of the Gaussian and the dashed one the ``decay'' time. Here we consider $N=101, J=20, \beta^2=\pi/20, M_0'=0.2$. For the same parameters we plot in (b) the von Neumann entanglement entropy $S_\text{vN}$ at each bond along the chain. (c) Large-spin  ($J \rightarrow \infty$) and classical ($\beta^2 \rightarrow 0$) limit value of the topological charge $Q(t=0)$ against the mass $M_0'$ and corresponding fit with a quadratic function. (Inset) The topological charge $Q(t)$ in the large-spin limit $J\rightarrow \infty$ is plotted as a function of time $t$ for several choices of the parameters $M_0'$ and $\beta^2$ (blue lines correspond to $\beta^2=\pi/20$, orange to $\beta^2=\pi/13$, green to $\beta^2=\pi/10$).  (d) Large-spin value of the width $\sigma_\rho(t)$ of the Gaussian fit for several $\beta^2$ and $M_0'$ in rescaled space and time units. The collapse of the numerical data is captured by the function $\sigma_\rho(t)=\sqrt{\sigma_{\rho, 0}^2 + (v_\text{spr}t)^2}$ (dashed). The initial width $\sigma_{\rho, 0}=1.237/M_0'$ is inversely proportional to the mass term $M_0'$ [see inset] and independent of $\beta^2$, while the prefactor of the spreading velocity $v_\text{spr}=0.273\beta$ is determined from the smallest $M_0'=0.15$ reliably simulated.}
    \label{fig:AppSol_StaticEvol}
\end{figure}

As observed in Fig.~\ref{fig:AppSol_StaticEvol}(a)-(b), both the topological charge density  $\langle \rho(x,t)\rangle$ and the von Neumann entanglement entropy $S_\text{vN}$ spread in time, while the soliton remains centered at half-chain. Moreover, for both quantities there are some free excitations propagating with speed $v\approx c$, which bounce and get reflected at the walls at $t\approx 50-55$.

Another interesting quantity is the topological charge $Q(t)$, which we determine by numerically integrating in space the topological charge density $\langle \rho(x,t) \rangle$.
The large-spin limit value ($J\rightarrow \infty$) of the topological charge in the classical limit $\beta^2 \rightarrow 0$ and at $t=0$ shows a quadratic dependence on $M_0'$, converging towards the classical theoretical prediction ($Q=1$) in the continuum limit $M_0'\rightarrow 0$ [see Fig.~\ref{fig:AppSol_StaticEvol}(c)]. We expect the quadratic dependency on $M_0'$ to arise due to the fact that the transition between the two degenerate vacua becomes sharper for increasing $M_0'$, thus making it more challenging  to approximate the soliton profile on a discrete lattice and leading to larger deviations from the theoretical prediction.
Instead, we expect the deviation at $M_0'=0.1$ to arise from the large transition region of the corresponding soliton that cannot be fully captured on a lattice of $N=101$ sites. We will neglect this value in the following. 
Over time the charge is approximately conserved up to  $t\approx 50-55$, when larger deviations appear, which we interpret as a consequence of the interaction of the free excitations with the wall [see inset in Fig.~\ref{fig:AppSol_StaticEvol}(c)].

In Fig.~\ref{fig:AppSol_StaticEvol}(d) we instead plot the large-spin limit ($J\rightarrow \infty$) value of the fitted Gaussian width $\sigma_\rho$ of the topological charge density $\rho$ for several $\beta^2$ and $M_0'$ as a function of time $t$. We observe a good agreement of the numerical results with the theoretical prediction $M_0 \sigma_\rho(t)\approx\sqrt{(M_0\sigma_{\rho, 0})^2 + (M_0 v_\text{spr}t)^2} \approx \sqrt{1.237^2 + (0.273 M_0 \beta t)^2}$ (dashed black line),  where the prefactor of the spreading velocity $v_\text{spr}\approx 0.273\beta$ is determined from the smallest $M_0'=0.15$ value we can reliably simulate. 
The initial width $\sigma_{\rho,0}$ of the Gaussian can be extrapolated by fitting the $\beta$-independent asymptotic value of the width against $1/M_0'$.

In the main text we discussed the results of the time evolution of the static soliton, numerically determined using the TEBD algorithm with time step $\tau=0.01$ and maximal bond dimension $\chi=128$. Note that the time step $\tau=\tau_\text{latt}$ is given in lattice units, and the time in lattice units is related to the time in sG units according to $t_\text{sG}=t_\text{latt} \cdot 2\chi \kappa^2 a/\beta^2$. Here, we compare the numerical results obtained with other parameter choices: $(\tau=0.0025, \chi=128)$ up to $t_\text{latt}=14$, and $(\tau=0.01, \chi=256)$ up to $t_\text{latt}=2.7$ (for reference, in the plots we display numerical results $t_\text{latt}=12$). For both choices we take the largest absolute deviation with respect to $(\tau=0.01, \chi=128)$ and then compute the relative error.
For $(\tau=0.0025, \chi=128)$ we get a maximal error of 0.27\% for the half-chain absolute value of the vertex operator, of 0.7\% for the half-chain von Neumann entanglement entropy and of order $10^{-5}$ for the energy.
For $(\tau=0.01, \chi=256)$ we are restricted to much smaller simulation times and we get a maximal error of 0.4\% for the half-chain absolute value of the vertex operator, of 0.5\% for the half-chain von Neumann entanglement entropy and of order $10^{-6}$ for the energy.
In both cases we expect the errors to increase over time.

\subsection{\label{sec:AppSol_MovingEvol} Time evolution of the moving soliton}
\begin{figure}
    \centering
    \includegraphics[width=\linewidth]{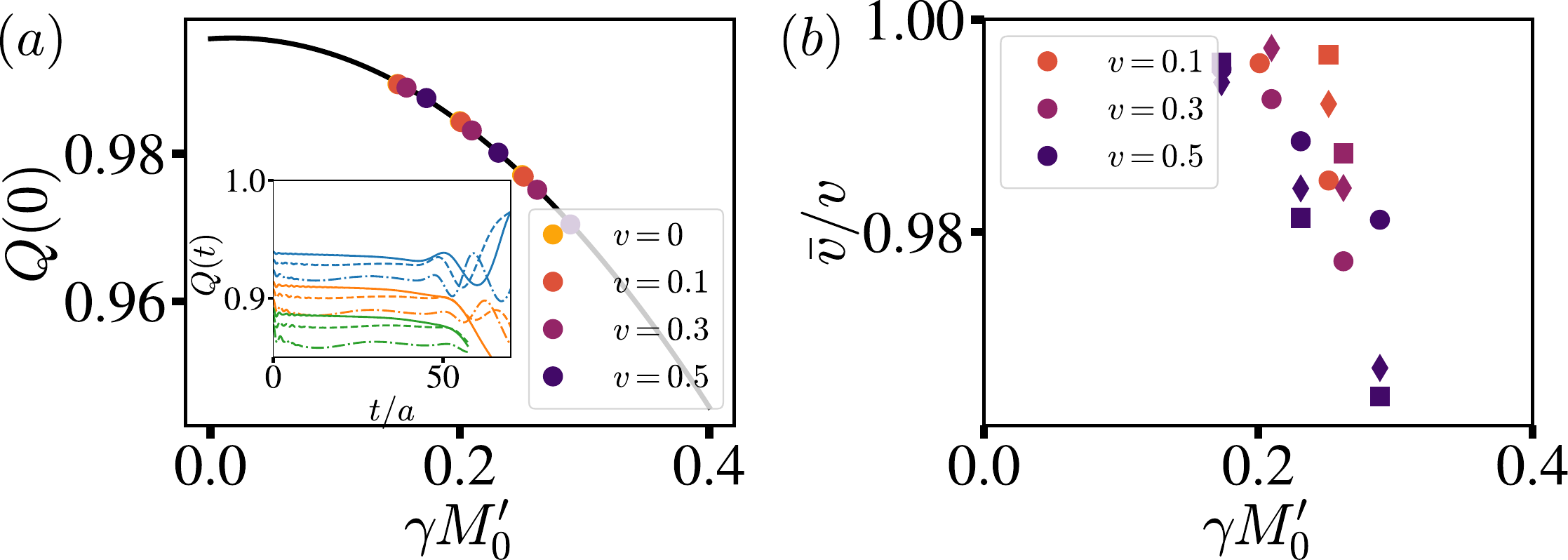}
    \caption{\textbf{Properties of the moving soliton.} (a)  Large-spin ($J\rightarrow \infty$) and classical  ($\beta^2 \rightarrow 0$)  limit values of the topological charge $Q(t=0)$ against $\gamma M_0'$ and fitted with a quadratic function. (Inset) The time-dependent topological charge $Q(t)$ is plotted as a function of time $t$ for several parameters $\beta^2$ (blue lines correspond to $\beta^2=\pi/20$, orange to $\beta^2=\pi/13$, green to $\beta^2=\pi/10$) and $M_0'$ at $J=20, v=0.5$. (b) Ratio of the effective velocity $\bar{v}$ and the imposed classical value $v$ in the large-spin limit against $\gamma M_0'$ for $\beta^2=\pi/20$ (circles), $\beta^2=\pi/13$ (diamonds) and $\beta^2=\pi/10$ (squares).}
    \label{fig:Fig14_AppMovingSoliton}
\end{figure}

We now instead investigate the quench dynamics of the moving soliton ($v\neq 0$) for $N=101, J=\{16,18,20\}, \beta^2=\{\pi/20, \pi/13, \pi/10\},$ and $M_0'=\{ 0.15, 0.2, 0.25\}$. Similarly to the static soliton, in Fig.~\ref{fig:Fig14_AppMovingSoliton}(a) we observe that the topological charge at time $t=0$ follows a quadratic dependence in $\gamma M_0'$ in the large-spin ($J \rightarrow \infty$) and classical ($\beta^2\rightarrow 0$) limit. Moreover, the topological charge is conserved over time up to $t\approx 50$, the collision time of the free excitations with the wall. 
In Fig.~\ref{fig:Fig14_AppMovingSoliton}(b) we plot instead the large-spin $(J\rightarrow \infty)$ value of the effective soliton velocity $\bar v$, extrapolated by fitting with a linear function the position of the quasi-particle determined according to Eq.~\eqref{eq:quasi-particle_position}. The ratio between the effective velocity $\bar{v}$ and the imposed one $v$ displays a dependence on $\gamma M_0'$, and converges to values close to 1 in the continuum limit. Again, possible reasons for the observed behavior include the challenging preparation of solitons with sharp transition regions $\sim 1/\gamma M_0'$ at large $\gamma M_0'$ values.

In the main text we discussed the time evolution of the moving soliton performed with TEBD using the time step $\tau=0.01$ and the maximal bond dimension $\chi=128$.  To imprint the momentum, we use the naive gate $\hat{U}_\text{m}^{(i)}=\exp{( -i \theta^{(i)} \hat{J}_\varphi^{(i)})}$,
with the angle $\theta^{(i)}=\arcsin{(\pi^{(i)}_\text{s} \kappa a/(\beta J))}$, i.e., we assume the classical spin length to be equal $J$.
We now compare the results for $(\tau=0.01, \chi=128)$ to two other sets of data obtained for the parameter choices $(\tau=0.0025, \chi=128)$ up to $t_\text{latt}=7$, and $(\tau=0.01, \chi=256)$ up to $t_\text{latt}=5.5$ (for reference in the main text we consider a maximal time $t_\text{latt}=9$). For both choices we take the largest absolute deviation with respect to $(\tau=0.01, \chi=128)$ and then compute the relative error.
For $(\tau=0.0025, \chi=128)$ we get a maximal error of 0.7\% for the minimum of the absolute value of the vertex operator (analog of the half-chain value in the static case), of 3.5\% for the von Neumann entanglement entropy along the whole chain and of order $10^{-5}$ for the energy.
For $(\tau=0.01, \chi=256)$ we are restricted to way smaller simulation times and we get a maximal error of 1\% for the minimum of the absolute value of the vertex operator, of 4.6\% for the half-chain von Neumann entanglement entropy and of order $10^{-6}$ for the energy.
In both cases we expect the errors to increase over time.

\subsection{\label{sec:AppSol_semi-classical} Semi-classical phenomenological model}

As discussed in the main text, we can engineer a semi-classical phenomenological model, which captures the numerically observed spreading dynamics. In this section we provide additional details.

\subsubsection{Theoretical model} 
\begin{figure}
    \centering
    \includegraphics[width=\linewidth]{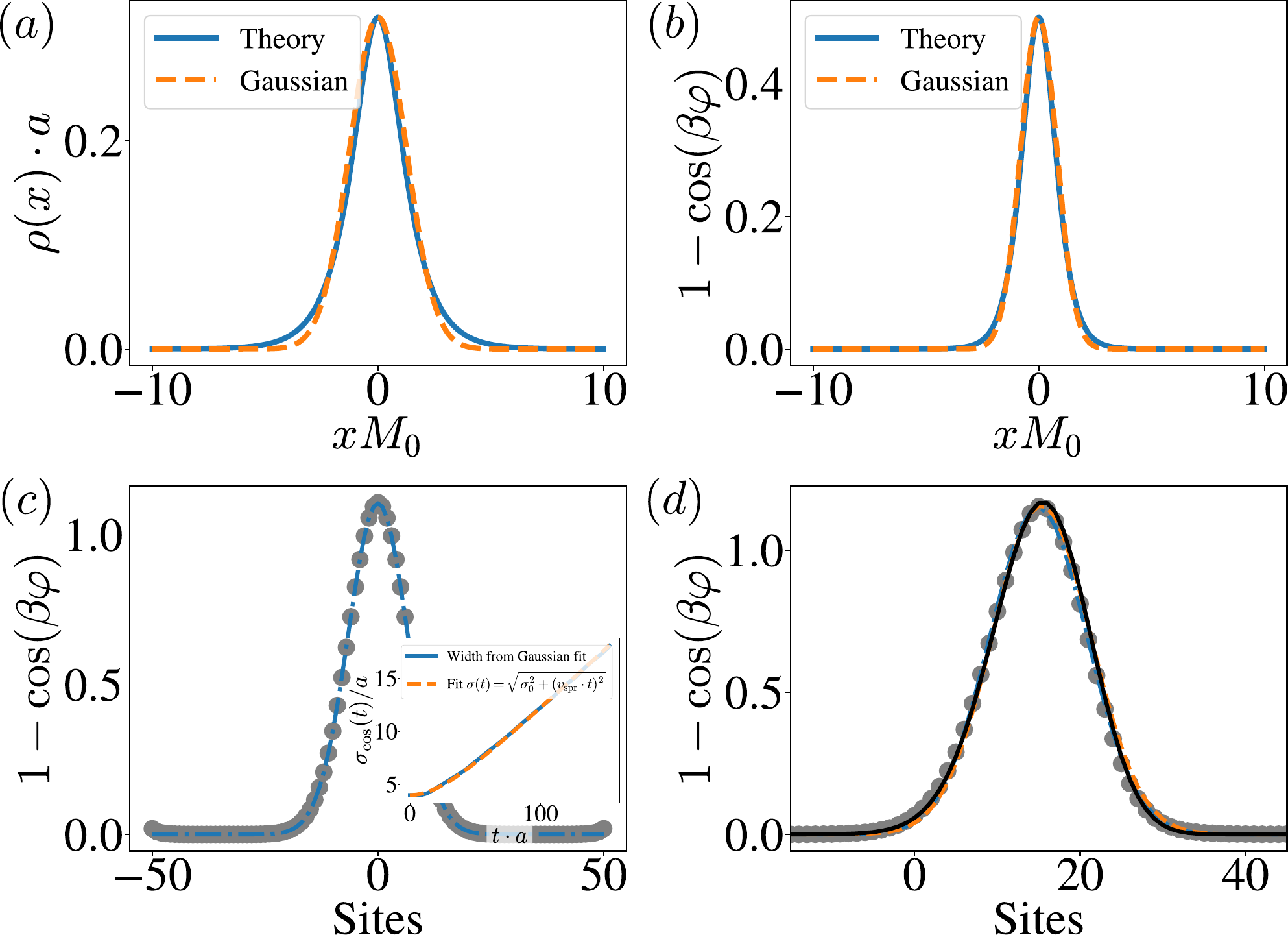}
    \caption{\textbf{Approximation with Gaussian functions.} (a) Topological charge density $\rho(x)$ of the classical soliton at $t=0$ Eq.~\eqref{eq:App_theory_rho} (solid blue line) and comparison with the Gaussian approximation Eq.~\eqref{eq:AppGaussianRho} (dashed orange line). (b) $1-\cos(\beta\opphi)$ of the classical soliton at $t=0$ Eq.~\eqref{eq:App_theory_cos} (solid blue line) and comparison with the Gaussian approximation Eq.~\eqref{eq:AppGaussianCos} (dashed orange line). In both plots we rescale the space by the mass term $M_0$. (c) Cosine interaction term $1-\cos(\beta \opphi)$ of the semi-classical static soliton  ($v=0$)  at $N=101, J=20, \beta^2=\pi/20, M_0'=0.2, t=44.44$: gray dots represent the numerical results, the dash-dotted blue line the fit of the data with the Gaussian function Eq.~\eqref{eq:AppGaussianCos}. (Inset) Gaussian width $\sigma_\text{cos}(t)$: the solid blue line represents the width extrapolated from the previous Gaussian fit, the orange dashed line is an additional fit of the obtained width with the function $\sigma(t)=\sqrt{\sigma_0^2 + (v_\text{spr}\cdot t)^2}$. 
    (d)  For the same parameters, cosine interaction term $1-\cos(\beta \opphi)$ for the semi-classical moving  soliton ($v=0.3$): gray dots represent the numerical results, the dash-dotted blue line the fit with the Gaussian function Eq.~\eqref{eq:AppGaussianCos}, the dashed orange resp. the solid black line the Gaussian function in Eq.~\eqref{eq:AppGaussianObservableRelativistic} neglecting resp. considering the position-dependent shift $\bar x \sigma_v v \gamma$ of the width Eq.~\eqref{eq:AppGaussianWidthRelativistic}. For the last two lines we use the Gaussian fit parameters extracted from the static soliton in (c).}
    \label{fig:AppGaussianApprox}
\end{figure}

We start by considering a classical soliton $\varphi_\text{s}$, for which the charge density $\rho_\text{s}(x,t)$ and the interaction term are given by
\begin{align}
    \rho_\text{s}(x,t)=\frac{\gamma M_0}{\pi}\frac{1}{\cosh(\gamma M_0(x-v t + \delta))} , \label{eq:App_theory_rho}
    \\
    1-\cos{(\beta \vphi_\text{s}(x,t))}=\frac{2}{\cosh^2(\gamma M_0(x-v t + \delta))}. \label{eq:App_theory_cos}
\end{align}
By integrating the charge density over the spatial dimension we determine the topological charge $Q(t)=\int_{-\infty}^{+\infty} {\rm d}x \rho(x,t)$ (for a soliton $Q_\text{s}(t)=1$), while for $1-\cos(\beta \varphi_\text{s})$ we get $\int_{-\infty}^{+\infty} {\rm d}x (1- \cos(\beta \varphi_\text{s}(x,t)))=4/\gamma M_0$. We can normalize the latter to 1 and approximate both quantities with a Gaussian distribution 
\begin{align}
    \rho_\text{s}(x,t)&\approx e^{-\frac{(x-vt+\delta)^2}{2\sigma^2_\rho}}/(\sqrt{2\pi}\sigma_\rho),
    \label{eq:AppGaussianRho}
    \\
    \frac{\gamma M_0}{4}(1-\cos{(\beta \vphi_\text{s}(x,t))})&\approx e^{-\frac{(x-vt+\delta)^2}{2\sigma^2_\text{cos}}}/(\sqrt{2\pi}\sigma_\text{cos}),
    \label{eq:AppGaussianCos}
\end{align}
with $\sigma_\rho=\sqrt{\pi/2}/(\gamma M_0)\approx 1.253/(\gamma M_0)$ and $\sigma_\text{cos}=\sqrt{2/\pi}/(\gamma M_0)\approx 0.798/(\gamma M_0)$ [see Fig.~\ref{fig:AppGaussianApprox}(a)-(b)]. The prefactors of the approximated widths are  in good agreement with the values obtained in the main text by Gaussian fitting the numerical data. 
In the following, we will refer to both quantities generally as the observable $\mathcal{O}_\text{cl}[\varphi_\text{s}(t)]=\mathcal{O}_\text{cl}(x,v,\delta,t)$ approximated by a Gaussian function of width $\sigma_\mathcal{O}$.

In the main text, we approximate time-dependent quantum expectation values of $\mathcal{O}$ as
\begin{align}
    \langle \mathcal{O}(t) \rangle \approx \int D\varphi_\text{s}(0) D\pi_s(0) \, W[\varphi_\text{s}(0), \pi_s(0)] \, \mathcal{O}_\text{cl} \left[\varphi_\text{s}(t)\right],
\end{align}
with $W$ an initial-condition distribution over single solitons, that takes into account different initial phase $\varphi_\text{s}(0)$ and conjugate momentum $\pi_\text{s}(0)$  values. Equivalently, the distribution $W$ corresponds to considering a wave-packet of single solitons with different position shift $\delta$ and velocity $v$, thus capturing initial quantum fluctuations in the two quantities. In particular, we describe quantum fluctuations in $\delta$ and $v$ in the form of normalized Gaussian distributions
\begin{align}
    P_\delta(\mu_\delta, \sigma_\delta)=\frac{e^{-(\delta-\mu_\delta)^2/(2 \sigma_\delta^2)}}{\sqrt{2\pi}\sigma_\delta},\\
    P_v(\mu_v, \sigma_v)=\frac{e^{-(v-\mu_v)^2/(2 \sigma_v^2)}}{\sqrt{2\pi}\sigma_v},
\end{align}    
with $\mu_\delta, \mu_v$ the mean values and $\sigma_\delta, \sigma_v$ the standard deviations.
That is, we assume the distribution $W$ to be well described by two Gaussian distributions in $\delta$ and $v$, and we replace 
\begin{multline}
    D\varphi_\text{s}(0) D\pi_s(0) \, W[\varphi_\text{s}(0), \pi_s(0)] \longrightarrow \\ {\rm d}\delta {\rm d}v P_\delta(\mu_\delta, \sigma_\delta) P_v(\mu_v, \sigma_v).
\end{multline}
The dynamics of the observable $\mathcal{O}(t)$ is then determined by sampling the classical time-dependent expectation values $\mathcal{O}_\text{cl} \left[\varphi_\text{s}(t)\right]=\mathcal{O}_\text{cl}(x,v,\delta,t)$ according to the two Gaussian distributions in $\delta$ and $v$
\begin{multline}
\label{eq:Appsemiclassical_expvalue}
 \tilde{\mathcal{O}}(x,\mu_\delta,\sigma_\delta,\mu_v, \sigma_v, t) =    \langle \mathcal{O}(t) \rangle \approx \\\int D\varphi_\text{s}(0) D\pi_s(0) \, W[\varphi_\text{s}(0), \pi_s(0)] \, \mathcal{O}_\text{cl} \left[\varphi_\text{s}(t)\right]
    \\
   = \int {\rm d}\delta {\rm d}v P_\delta(\mu_\delta, \sigma_\delta) P_v(\mu_v, \sigma_v) \mathcal{O}_\text{cl}(x,v,\delta,t).
\end{multline}

In the following, we focus on the case of a static soliton by assuming $\mu_v= 0$ and $\sigma_v \ll 1$, so that we can approximate $\gamma\approx 1$.
Analytical integration of Eq.~\eqref{eq:Appsemiclassical_expvalue} then leads to 
\begin{equation}
    \tilde{\mathcal{O}}(x,\mu_\delta,\sigma_\delta,\mu_v=0, \sigma_v, t)= \frac{e^{-(x+\mu_\delta)^2/(2 \tilde\sigma_\mathcal{O}^2)}}{\sqrt{2\pi}\tilde\sigma_\mathcal{O}},
    \label{eq:AppGaussianFluctuationStatic}
\end{equation}
i.e.,  we obtain another Gaussian distribution with a time-dependent width $\tilde\sigma_\mathcal{O}(t)=\sqrt{(\sigma_\mathcal{O}^2 + \sigma_\delta^2) + (\sigma_v t)^2}$. This width naturally encodes a spreading of the observables in time, where the spreading velocity is given by $\sigma_v$, while $\sigma_\delta$ simply modifies the initial width $\tilde\sigma_\mathcal{O}(0)=\sqrt{\sigma_\mathcal{O}^2 + \sigma_\delta^2}$. 
We expect $\sigma_\delta$ to be negligible, as this term would simply modify the $t=0$ value of the observables, which we instead impose to agree with the classical predictions. 
Note that the results obtained so far are valid only in the case of a static soliton ($\mu_v=0$), as we neglected the $\gamma$-dependency.

The case of a moving soliton $v=\mu_v$ can be understood as a static soliton ($\mu_v=0$) in an inertial frame moving with velocity $v$ with respect to the observer's inertial frame. We thus apply the Lorentz transformation
\begin{equation}
    x \rightarrow \gamma(x-v\cdot t), \quad t  \rightarrow \gamma(t-x\cdot v),
\end{equation}
where we assume $c=1$. Furthermore, we assume $\mu_\delta=0$ and we investigate the dynamics of the observable with respect to the new spatial coordinate $\bar{x}=x-v\cdot t$, such that the soliton center is always located at $\bar{x}=0$.
After the Lorentz transformation and the change of coordinate, the observable $\Tilde{\mathcal{O}}$ expressed in $\bar x$ is 
\begin{equation}
     \tilde{\mathcal{O}}(\bar{x},\sigma_\delta,v, \sigma_v, t)= \frac{e^{-\gamma^2 \bar{x}^2/(2 \tilde\sigma_\mathcal{O}^2)}}{\sqrt{2\pi}\tilde\sigma_\mathcal{O}},
     \label{eq:AppGaussianObservableRelativistic}
\end{equation}
and the width $\tilde\sigma_\mathcal{O}(\bar{x},t)$ displays an additional dependence on $\bar{x}$
\begin{equation}
    \tilde\sigma_\mathcal{O}(\bar{x},t)=\sqrt{(\sigma_\mathcal{O}^2 + \sigma_\delta^2) + (\sigma_v t/\gamma - \bar{x}\sigma_v v \gamma)^2}.
         \label{eq:AppGaussianWidthRelativistic}
\end{equation}
Upon neglecting the additional contribution $- \bar{x}\sigma_v v \gamma$, which is relevant only at short time and large distances $\bar{x}$, we retrieve 
\begin{align}
    \tilde{\mathcal{O}}(\bar{x},\sigma_\delta,v, \sigma_v, t)= \frac{e^{-\gamma^2 \bar{x}^2/(2 \tilde\sigma_\mathcal{O}^2)}}{\sqrt{2\pi}\tilde\sigma_\mathcal{O}/\gamma}, \\
    \tilde\sigma_\mathcal{O}(t)=\sqrt{(\sigma_\mathcal{O}^2 + \sigma_\delta^2) + (\sigma_v t/\gamma)^2},
         \label{eq:AppGaussianRelativistic2}
\end{align}
where the Lorentz factor $1/\gamma$ in the denominator of $\tilde{\mathcal{O}}$ is necessary to guarantee the normalization.
Comparing it to the standard Gaussian distribution $p_1 e^{-(x-p_3)^2/(2 p_2)^2}/(\sqrt{2\pi}p_2)$, which we use to fit the data, the effective width is $p_2=\tilde\sigma_\mathcal{O}(t)/\gamma=\sqrt{(\sigma_\mathcal{O}^2 + \sigma_\delta^2) + (\sigma_v t/\gamma)^2}/\gamma$, in agreement with the rescaling  $p_2 \rightarrow p_2 \gamma$ and time $t \rightarrow t/\gamma$ performed in the main text, which leads to a collapse
[see Fig.~\ref{fig:TimeEvolFig}(e)].

\subsubsection{Numerical results}
We now compare the expected theoretical dependence of the observable $1-\cos(\beta \opphi)$ with the numerical results for both the semi-classical static and the moving soliton (we expect similar results for the topological charge density $\rho$). 
We first consider the static soliton by fitting in Fig.~\ref{fig:AppGaussianApprox}(c) the numerical data with a Gaussian distribution $p_1 e^{-(x-p_3)^2/(2 p_2)^2}/(\sqrt{2\pi}p_2)$ resembling Eq.~\eqref{eq:AppGaussianFluctuationStatic}. We observe a good agreement of the time-dependent width $p_2(t)$ [see inset of Fig.~\ref{fig:AppGaussianApprox}(c)] with the theoretical predicted form $p_2(t)=\sqrt{(p_2^1)^2+(p_2^2 \cdot t)^2}$, from which the parameters  $p_2^1, p_2^2$ can be numerically extrapolated.

We then investigate the moving soliton case and compare the numerical data to the theoretical predictions
Eqs.~\eqref{eq:AppGaussianObservableRelativistic},\eqref{eq:AppGaussianWidthRelativistic} [see Fig.~\ref{fig:AppGaussianApprox}(d)]. We find again a good agreement when inserting in the theoretical prediction the fit results  $p_1, p_2^1, p_2^2$ of the static soliton, as well as the numerically extrapolated speed $v$ of the moving soliton. 
We notice that even upon neglecting the position-dependent contribution $\bar{x}\sigma_v v \gamma$ in Eq.~\eqref{eq:AppGaussianWidthRelativistic} a good agreement persists, especially for the width, with only some small deviations emerging at large $|\bar{x}|$, as expected.

\subsection{Velocity fluctuations from ground-state quantum fluctuations \label{sec:AppSol_Fluctuations}}

Here we illustrate the relation between the variance of ground-state spin operators and the standard deviation of the velocity $\sigma(v)$. The resulting linear dependence of $\sigma(v)$ on $\beta$ explains the numerically observed spreading of observables in time.

We start by investigating the ground state of the quantum sG model. We look at the $zz$-connected correlator $\langle\hat{J}_z^{(i)}\hat{J}_z^{(j)}\rangle_\text{c}$ and consider a finite interval $F$ containing a fixed number $N_F$ of sites:  for a fixed $\kappa$ we obtain
\begin{equation}
\sum_{i,j\in F}\langle\hat{J}_z^{(i)}\hat{J}_z^{(j)}\rangle_\text{c} \propto \frac{N_F M_0'}{\beta^2}.
\label{eq:AppZZCorr}
\end{equation}
We further recall, as discussed in App.~\ref{sec:AppSol_preparation}, that the two-point connected correlators of the static and moving soliton are completely determined by the ground state of the quantum sine-Gordon model.

To relate the above sum to the standard deviation of the velocity, we first point out that the spatial integral of the classical soliton conjugate momentum $\pi_\text{s}$ is proportional to the soliton velocity: $\int_{-\infty}^{\infty}{\rm d}x \;\pi_\text{s}(x)=-2\pi v/\beta$. Upon truncating the infinite integration domain to a finite interval $F=[-f/(\gamma M_0'), f/(\gamma M_0')]$, whose length is proportional to the soliton transition region $\sim 1/(\gamma M_0')$, we retrieve the same result up to an error inversely dependent on $f$. 
We can physically interpret the truncation using the relation between conjugate momentum and charge density $\pi_\text{s}\propto \rho_s$ (with $Q_s=\int_{-\infty}^{+\infty} {\rm d}x \rho_s(x)=1$ for a soliton) and the duality between sG model and massive Thirring model. The topological charge density $\rho_\text{s}$ in the sG model is well-approximated by a Gaussian distribution in space [see Fig.~\ref{fig:AppGaussianApprox}(a)] and is dual to the fermion probability density in the Thirring model: truncating the integration domain of $\pi_\text{s}$ thus corresponds to integrate only over some part of the probability density (thanks to the Gaussian profile, in our case almost over the whole). In the following, we choose $f=4$, such that $Q_F=\sum_{i\in F} \rho_\text{s}(x=i\cdot a)\approx 0.98$ for the investigated mass values $M_0'=aM_0=0.1, 0.15, 0.2, 0.25, 0.3$.

In our simulation of the sG model, the lattice counterpart to the continuum conjugate  momentum $\hat{\pi}$ is the  $z$-component of the lattice spin operator $\frac{\beta}{\kappa a}\hat{J}_z \rightarrow \hat{\pi}$. Using the above approximations, we can relate the fluctuations of $\hat{J}_z$ to the standard deviation of the soliton velocity $v$. 
In particular, 
\begin{equation}
    \int_{-\infty}^{\infty}{\rm d}x \;\hat{\pi}\approx\int_F {\rm d}x\;\hat{\pi} = \sum_{i\in F} a \frac{\beta}{\kappa a}\hat{J}_z^{(i)},
\end{equation}
and 
\begin{align}
    \frac{2\pi}{\beta}\sigma(v)&=\sigma\left(\int_{-\infty}^{\infty}{\rm d} x\; \hat{\pi}\right)
    \\&\approx\sigma\left(\int_F {\rm d}x\; \hat{\pi}\right)  = \sigma\left(\sum_{i\in F} \frac{\beta}{\kappa}\hat{J}_z^{(i)}\right)
    \\&=\sqrt{\left\langle\left(\sum_{i\in F} \frac{\beta}{\kappa}\hat{J}_z^{(i)}\right)^2\right\rangle-\left\langle\left(\sum_{i\in F} \frac{\beta}{\kappa}\hat{J}_z^{(i)}\right)\right\rangle^2}\\
    &=\sqrt{\left(\frac{\beta}{\kappa}\right)^2\sum_{i,j\in F} \langle\hat{J}_z^{(i)}\hat{J}_z^{(j)}\rangle_\text{c}}.
    \label{eq:AppSumCorrelationFormula}
\end{align}
To evaluate this expression we exploit Eq.~\eqref{eq:AppZZCorr}, and use that for a static soliton ($v=0, \gamma=1$) the interval $F$ is $M_0'$-dependent $F=[-f/M_0', f/M_0']$, i.e.,  the number of atoms  $N_F\propto 1/M_0'$. Due to the mutual cancellation of the mass term, we obtain
\begin{equation}
    \sum_{i,j\in F}\langle\hat{J}_z^{(i)}\hat{J}_z^{(j)}\rangle_\text{c} \propto \frac{N_F M_0'}{\beta^2} \propto \frac{M_0'/M_0'}{\beta^2}= \frac{1}{\beta^2}.    
\end{equation}
In particular, for $\kappa=1$  we get (neglecting a subleading $M_0'$-dependence $\approx -1.4 M_0$)
\begin{equation}
\sum_{i,j\in F}\langle\hat{J}_z^{(i)}\hat{J}_z^{(j)}\rangle_\text{c}\approx \frac{5}{\beta^2},
\label{eq:AppSumCorrelationGroundState}
\end{equation}
and thus for the standard deviation of the velocity $\sigma (v)$ we obtain
\begin{equation}
    \frac{2\pi}{\beta}\sigma(v)\approx \sqrt{5} \longrightarrow \sigma(v)\approx 0.35 \beta.
    \label{eq:AppSpreadingVelocityResult}
\end{equation}

According to the semi-classical phenomenological model [see App.~\ref{sec:AppSol_semi-classical}], the spreading velocity equals the standard deviation of the velocity $v_\text{spr}=\sigma(v)$, and thus 
we get a qualitative agreement with the spreading velocity $v_\text{spr}\approx0.273 \beta$ determined in the main text (the prefactors  differ by a factor $\approx 1.3$).
We expect the same argument to hold at $\kappa>1$, with the sum in Eq.~\eqref{eq:AppSumCorrelationGroundState} getting an additional $\kappa^2$ factor, which cancels out the $\kappa^2$ factor in Eq.~\eqref{eq:AppSumCorrelationFormula}. 

The case of a moving soliton ($v \neq 0$) is instead defined on 
the domain $F=[-f/(\gamma M_0'), f/(\gamma M_0')]$ containing $N_F\propto 1/(\gamma M_0')$ sites, and leading to 
\begin{equation}
    \sum_{i,j\in F}\langle\hat{J}_z^{(i)}\hat{J}_z^{(j)}\rangle_\text{c} \propto  \frac{N_F M_0'}{\beta^2} \propto \frac{1}{\beta^2 \gamma}.    
\end{equation}
Note that the mass term $M_0'$ in the numerator does not get an additional $\gamma$ factor, because it is directly set by the quantum ground state for which no velocity is defined.
Therefore, the spreading velocity of the moving soliton, equivalent to the standard deviation of the soliton velocity, is $v_\text{spr}=\sigma(v)\propto \beta/ \sqrt{\gamma}$. This result is consistent with the rescaling used in Fig.~\ref{fig:TimeEvolFig}(e), and for $v=0$ agrees with the static soliton case Eq.~\eqref{eq:AppSpreadingVelocityResult}.

\section{\label{sec:AppScattering} 
Soliton-antisoliton scattering. Error estimation of the position shift}
Here we describe the approach used to estimate the errors in the position shift $\delta x$ after the semi-classical soliton-antisoliton scattering.
As discussed qualitatively in the main text, we linearly fit the 4 regions of linear motion (before scattering/left half, after/left, before/right, after/right), and we determine the position shift as the distance between quasi-particles at the scattering time, i.e., when the linear fits of the trajectories before and after the scattering cross.

More precisely, for each of the 4 regions we fix a maximal time interval where the quasi-particles move linearly, fit increasingly smaller subsets of it, sort the fit results according to the value of the slope, and choose as best guess for the fit parameters the pair containing the median value of the slope distribution. To determine the uncertainty of each fit parameter (slope and offset), we instead sort all the obtained results increasingly and estimate the error as half of the difference between the first value located at more than 0.8415 of the whole distribution and the first value at more than 0.1585 [see App.~\ref{sec:AppendixFittingDiscussion}].

So far we determined the best guess and the uncertainty of the fit parameters for a linear fit in each of the 4 regions. We now need to understand how these uncertainties combine together in the uncertainty of the position shift. 
First, for each region we determine all the possible combinations of slope and offset values by considering the best guesses as well as by adding and subtracting to them the corresponding errors. We then compute the Cartesian product over the 4 regions of these combinations, i.e., we obtain a set of elements, each of which contains possible slope and offset values for all the 4 linear fits. 
To understand how differences in the linear trajectories impact the position shift, we determine the position shift for each element of the above introduced set. To this end, we first compute the time at which the linear trajectories cross in each half-plane, and then the separation between the quasi-particle trajectories at the mean time. Note that quasi-particle trajectories are determined with a linear function by inserting the slope and offset values. After repeating the position shift estimation for each element of the set, our best guess for $\delta x$ is the median value,  while the error is determined from the whole distribution as above (0.8415 - 0.1585 values). Note that the above considered functions are smooth and the errors symmetric around the best guess, thus the median value is very close to the result we would get directly from the best guess of the 4 linear fits. 
Finally, to estimate the error in the large-spin limit $J\rightarrow \infty$, we simply consider the error of the largest $J$, as we are simulating only three different spin lengths $J\in[16,18,20]$.

\end{document}